\documentclass[11pt]{article}

\usepackage[algo2e]{algorithm2e} 
\usepackage{enumitem}
\usepackage{comment}
\usepackage[english]{babel}
\usepackage{amsthm}
\usepackage[utf8]{inputenc}
\usepackage{xcolor}
\usepackage{amsmath}
\usepackage{physics}
\usepackage[margin=1in]{geometry}
\usepackage{titlesec}
\usepackage{natbib}
\usepackage{graphicx}
\usepackage{graphicx} 
\usepackage{float}

\newtheorem{assumption}{Assumption}
\newtheorem{example}{Example}

\newtheorem{observation}{Observation}
\usepackage{caption}
\usepackage{subcaption}
\usepackage{multirow}
\usepackage{algorithm}
\usepackage{algpseudocode}
\usepackage{placeins}
 
\usepackage{amsmath}

\usepackage{xcolor}
\usepackage{hyperref}
\hypersetup{%
  colorlinks=false,
  linkbordercolor=red
}

\usepackage{natbib}
\usepackage{graphicx}

\usepackage{lineno}

\title{
\textbf{   Claim Reserving via Inverse Probability Weighting: \\ A Micro-Level Chain-Ladder Method}
}
\author{
\textbf{Sebastián Calcetero Vanegas*, Andrei L. Badescu, X. Sheldon Lin} \\
Department of Statistical Sciences\\
  University of Toronto\\
  Toronto, Ontario \\
  *\texttt{sebastian.calcetero@mail.utoronto.ca} \\
  }
  
\begin{document}

\maketitle

\begin{abstract}

Claim reserving primarily relies on macro-level models, with the Chain-Ladder method being the most widely adopted. These methods were heuristically developed without   minimal     statistical foundations, relying on oversimplified data assumptions and neglecting policyholder heterogeneity, often resulting in conservative reserve predictions. Micro-level reserving, utilizing stochastic modeling with granular information,    can improve predictions     but tends to involve less attractive and complex models for practitioners. This paper aims to strike a practical balance between aggregate and individual models by introducing a methodology that enables the Chain-Ladder method to incorporate individual information. We achieve this by proposing a novel framework, formulating the claim reserving problem within a population sampling context. We introduce a reserve estimator in a frequency and severity distribution-free manner that utilizes inverse probability weights (IPW) driven by individual information, akin to propensity scores. We demonstrate that the Chain-Ladder method emerges as a particular case of such an IPW estimator, thereby inheriting a statistically sound foundation based on population sampling theory that enables the use of granular information, and other extensions.

\end{abstract}


\section{Introduction}

Claim reserving is a crucial aspect of insurance and risk management, and is vital for ensuring solvency, assessing risk, and setting appropriate premiums. The insurance industry employs several types of reserves but is primarily interested in the reserve for outstanding claims, which cover the estimated costs of unsettled and non-reported claims, representing the insurer's liability for future payments related to already occurred accidents. This reserve can be split into subcomponents depending on the source of the claim i.e., whether it is from a reported claim or not, and it is of interest to create this distinction for accounting purposes. These reserves play a vital role in maintaining financial stability and ensuring the availability of funds for future claim payments.

Reserving in general insurance is one of the most studied problems in actuarial research. See e.g., \cite{schmidt2011bibliography} for an extended list of most of the research covering this topic. \cite{taha2021insurance} and references therein provide a detailed overview of methods used in insurance reserving. Briefly, two primary approaches to reserving, namely micro-level and macro-level approaches, have been widely studied in the actuarial literature. The theoretical foundations of these methods can be found in the literature on stochastic claim reserving, e.g. \cite{wuthrich2008stochastic}.

On the one hand, the macro-level approach to reserving focuses on estimating claim payments at an aggregate level. Among the macro-level approaches, Chain-Ladder-based techniques are widely employed in the insurance industry due to their ease of implementation, interpretation, and reliance on intuitive assumptions. See for e.g.,  \cite{mack1994stochastic}, \cite{quarg2004munich}, \cite{miranda2012double}. These methods avoid the use of very complex mathematical concepts, such as    machine learning-based     models or stochastic processes, instead relying on simple operations that can be implemented using spreadsheets. Additionally, they only require estimating development factors, which can be easily obtained from aggregate data without the need for specialized software. Consequently, the Chain-Ladder method and its variants are favored by insurance companies and regulators, with    approximately     more than 90\% of insurers relying on them as their primary reserving methods,    according to a survey made by      \cite{astinreport}.

However, these aggregate methods overlook the actual composition and heterogeneity of the insurance portfolio. The Chain-Ladder assumes homogeneity among claims within a given portfolio, disregarding valuable insights that can be gained by considering factors, such as attributes associated with the risk of each policyholder \cite{wuthrich2018neural}. The most recent literature on claim reserving (e.g., \cite{crevecoeur2022hierarchical} and literature therein)  highlights the importance of using all the information available (i.e., the granular data) for the estimation of accurate reserves, and how ineffective is ignoring it. Consequently, the Chain-Ladder and similar macro-level models exhibit clear limitations and modest accuracy of estimation of the reserves when compared to models that do account for granular information i.e., micro-level models.

Moreover, the statistical foundation of aggregate models is most of the time questionable. The Chain-Ladder method was conceived as an ad-hoc method without reliance on a solid statistical basis that justifies the methodology's nature. Therefore, the Chain-Ladder has been the target of critics due to the lack of formalization with a well-founded statistical basis.      Although research in stochastic claim reserving over the past decades has aimed at understanding the statistical framework behind the Chain-Ladder method, including the original work by   \cite{mack1994stochastic} and subsequent studies through the decades e.g., \cite{england_verrall_2002}. \cite{miranda2012double}, \cite{taylor2016stochastic} and \cite{Engler_Lindskog_2024}; the answer is not yet clear and the literature is continuously evolving with new perspectives on the Chain-Ladder model. We refer to the introductions of \cite{pittarello2023chain} and \cite{Engler_Lindskog_2024}, and the literature therein,  for a more comprehensive literature review on the understanding of the Chain-Ladder Method, as well as the most up-to-date developments on this matter.

On the other hand, the micro-level approach to reserving involves estimating individual claim payments by considering detailed characteristics, such as policyholder information, claim type, severity, and other relevant factors, e.g. \cite{boumezoued2017individual}. Micro-level reserving methods utilize probabilistic models that mimic the development of the claims, and are designed to directly capture the behavior of policyholders and their impact on reserves, resulting in relatively more accurate forecasts than the macro-level models. See for e.g., \cite{taylor2008individual}, \cite{antonio2014micro}, \cite{wuthrich2018machine}, \cite{fung2021new}  for various modeling examples. 

While the theory behind micro-level models is relatively well understood given the nature of their construction based on statistical approaches, micro-level models pose challenges in terms of complexity, making them difficult to implement by insurance companies. These models incorporate both stochastic and predictive modeling, adding layers of complexity that may hinder their use in practice. Moreover, micro-level models often require assumptions about model components, such as distributions and simplifications of reality, which may be questionable. Consequently, these models are not widely adopted by actuarial practitioners due to the additional implementation effort required, and the lack of consensus on modeling practices from a regulatory standpoint. Indeed, according to     the survey report by \cite{astinreport}    , micro-level reserving methods are virtually absent among insurance companies worldwide, with almost no one utilizing them, either as their primary method or for internal check-up purposes.

A main obstacle to the consideration of micro-level reserving by practitioners and regulators is the significant disparity in methodologies with respect to macro-level models, in addition to the associated effort required for their construction. Macro-level models, such as the Chain-Ladder, differ significantly from micro-level models in terms of how the reserve estimation is derived, as well as their theoretical foundations. Consequently, the transition from a macro-level to a micro-level model represents a substantial and challenging undertaking for any insurance company. Furthermore, regulators face difficulties in validating and accepting a micro-level model when its underlying principles deviate significantly from the familiar idea of Chain-Ladder and the construction of the reserves via development factors. Therefore, the substantial gap between these two key reserving methodologies hinders the adoption of micro-level modeling in the insurance industry.

In this paper, we focus on bridging the gap between macro-level and micro-level models by introducing a novel approach that translates the reserving problem in non-life insurance into a population sampling problem. By treating the reported claims as a sample from a larger population of claims, we develop a statistically sound approach based on an inverse probability weighting (IPW) method. We therefore develop a new methodology that accommodates for the introduction of individual claim information via a predictive model on the sampling probabilities, similar to how it is achieved to propensity scores in other fields. Our methodology enables the use of individual information in the Chain-Ladder method and therefore improves its performance while retaining most of its simplicity and interpretability.

One of the main features of the  IPW estimator is that it exhibits a functional form reminiscent of the Chain-Ladder method and its development factors. However, it distinguishes itself by having claim-specific factors that depend on the attributes of the claims, and also does not require the specification of a certain granularity on the accident or development dates cells. As a result, our methodology can be viewed as a continuous ``micro-level" version of the Chain-Ladder, where the development of each claim up to its ultimate value is performed at the individual level.

 We observe that this similarity is not incidental and carries profound implications within the literature on claim reserving. We formally demonstrate that the Chain-Ladder method can be viewed as an empirical IPW estimator, operating under some specified homogeneity assumptions. Consequently, we assert that the theory on population sampling and IPW estimators establishes a robust and intuitively sound statistical framework for the Chain-Ladder method, distinct from the traditional approaches explored in the stochastic claim reserving literature      that replicates, sometimes artificially, the Chain-Ladder estimate, as we mentioned earlier.     In contrast, the population sampling framework addresses the reserving problem in a well-defined probabilistic setting, in which the Chain-Ladder method and its extensions naturally emerge without artificial constructions. Uncovering this connection introduces fresh perspectives on the Chain-Ladder method, allowing for extensions and a deeper comprehension of its theoretical properties inherited from its newly revealed nature as an IPW estimator.

The IPW approach signifies an enhancement over aggregate claim reserving models rooted in the traditional Chain-Ladder method, offering a cost-effective alternative to conventional micro-level reserving models. One of its primary strengths lies in its distribution-free methodology for estimating reserves, eliminating the need to specify parametric models for claim arrival processes (frequency) or claim amounts (severity). Notably, the IPW estimator only necessitates modeling the development of claims, encompassing reporting and payment delays, mirroring the approach of traditional aggregate models as the Chain-Ladder. Consequently, modeling efforts are streamlined to focus solely on estimating claim-specific inclusion probabilities based on observed delay distributions, simplifying the process compared to other individual reserving techniques. 

The IPW approach integrates individual claims information in a simple, yet statistically justified manner. It maintains the practicality and interpretability characteristic of macro-level models, rendering it a more attractive choice for both practitioners and regulators. This approach may serve as an initial step to encourage practitioners, who typically rely on macro-level models, to explore the potential benefits and insights obtained from incorporating individual information in the reserving process. Ultimately, it paves the way for practitioners and regulators to consider tailored-made models based on micro-level techniques.

This paper is structured as follows: Section \ref{IPW_Section} introduces the reserving problem as a sampling problem, and shows the derivation of the IPW estimator for the reserve of all outstanding claims. Section \ref{Others_section} extends the methodology to consider other types of reserves, such as the incurred but not reported (IBNR) and the reported but not settled (RBNS) reserves, as well as their cumulative payments counterparts.  Section \ref{theoretical} shows how the Chain-Ladder and some extensions can be viewed as IPW estimators, and discusses its implications. Section \ref{Estimation_Section}  discusses how to estimate the inclusion probabilities.  Section \ref{Numerics_Section} provides a numerical study on a real insurance dataset. Lastly, Section \ref{conclusions} provides the conclusion and future research directions.

\section{Claim reserving via inverse probability weighting}
\label{IPW_Section}

In this section, we present the claim reserving problem and demonstrate how it can be effectively tackled using inverse probability weighting methods. Since there are various types of reserves in general insurance, in this section we provide the overall idea of the methodology for the total reserve of outstanding claims only. Section \ref{Others_section} will delve into the specific details of the methodology for the most prevalent and significant reserves in general insurance, namely RBNS and IBNR reserves.

\subsection{The claim reserving problem}
\label{IPW_Section2.1}

Suppose an insurance company is analyzing its total liabilities associated with claims whose accident times occur between $t=0$ and $t=\tau$, where $\tau$ is the valuation time of analysis as defined by the actuary. In general insurance, accidents are often not immediately reported to the insurance company for various reasons, resulting in a significant delay between the occurrence of a claimable accident and the time the insurance company is notified. Therefore, at a given valuation time $\tau$, the insurance company only has information on the claims reported by $\tau$ and is unaware of the unreported claims. Furthermore, the complexity of the problem increases due to another delay in the payment process. When a claim is reported, it is common for it to be paid in several sub-payments over time rather than as a lump sum. This is because the impact of an accident can evolve, requiring additional payments until its associated claim is fully settled. Therefore, at a given valuation time $\tau$, the insurance company is only aware of the claims that were reported on time, and for each one, it may have paid only a partial amount of the associated claim size.

Along those lines, the insurance company is interested in estimating the total claim amount of unreported claims, as well as the remaining payments of the reported claims, to construct the overall reserve of outstanding claims. This reserve is also known in the insurance jargon as the Incurred But Not Settled (IBNS) \cite[p.~433]{asmussen2020risk}, and is usually decomposed into further subcomponents depending on whether the payment is associated with a reported or not reported claim. For simplicity,    we will use the terms IBNS reserve and the outstanding claims reserve interchangeably. Furthermore, in this Section, we consider the IBNS reserve      without referring to the decomposition.

Let us now describe the payment process as follows:

\begin{itemize}

\item Let $N(\tau)$ represent the total number of different payments associated with all the claims whose accident time is before the valuation time $\tau$.

\item Let $Y_{i}$, $i= 1, \ldots, N(\tau)$ denote the sequence of payment amounts. Note that some payments may belong to the same claim/accident, but we will not make any explicit distinction notation-wise here. Similarly, we assume that the indexes $i$ are sorted according to the payment time, for simplicity of exposition. 

\item  Let $T_i$, $i= 1, \ldots, N(\tau)$ denote the sequence of accident times associated with the claim underlying each payment; let $R_i$, $i= 1, \ldots, N(\tau)$ denote the sequence of the associated reporting times; and let $W_i$, $i= 1, \ldots, N(\tau)$ denote the sequence of the associated times in which the payments take place. Clearly, $T_i < R_i < W_i$ and note that the values $T_i, R_i$ would be the same for payments associated with the same claim, but the $W_i$ would differ.

\item Let $U_i = R_i-T_i$, $i= 1, \ldots, N(\tau)$ be the sequence of the reporting delay times associated with the claim underlying each payment,  $V_i = W_i-R_i$, $i= 1, \ldots, N(\tau)$ be the sequence of the associated payment delay time from reporting, and  $Z_i=W_i-T_i=U_i+V_i$, $i= 1, \ldots, N(\tau)$ be the sequence of the associated payment delay time from the accident (i.e., total delay).  Note that $U_i$ is the same for all the payments associated with the same claim.

\item Let $X_i$, $i= 1, \ldots, N(\tau)$ be the sequence of information/attributes of relevance, that is associated with the accident, the type of claim, the policyholder attributes, or the characteristics of the payment itself.

\item Let $N^{P}(\tau)$ the number of payments made by valuation time $\tau$ out of the total $N(\tau)$, i.e., the number of payments made to the claims reported by $\tau$.

\end{itemize}

Along those lines,  the total liability of the insurance company associated with accidents occurring before the valuation time $\tau$, which we will denote as $L(\tau)$, is given by

$$
L(\tau) = \sum_{i=1}^{N(\tau)} Y_{i}.
$$

Similarly,  the portion of liability that is known to the insurance company (i.e., the so-called paid amount) by valuation time $\tau$, which we will denote as $L^{P}(\tau)$, is

$$
L^{P}(\tau) = \sum_{i=1}^{N^{P}(\tau)} Y_{i}.
$$



Lastly, an actuary is interested in estimating the remaining liability    associated with the outstanding claims i.e., the IBNS reserve.     We will denote this quantity as $L^{   IBNS    }(\tau)$, and it is given by just the difference
$$
L^{   IBNS    }(\tau) = L(\tau)-L^{P}(\tau).
$$

This value is what the insurance company requires to set up the reserve for all outstanding claims, either non-reported, non-settled, or both, and is our goal for estimation. For further details of the claim reserving problem, we refer the reader to \cite{wuthrich2008stochastic}.

\subsection{A population sampling framework for claim reserving}
\label{Sampling_section}

Our proposal in this paper is based on a simple yet novel idea that allows us to frame the reserving problem in the context of \emph{population sampling}, enabling us to leverage techniques from this field to our advantage. Population sampling is a statistical technique used to estimate population totals based on a smaller sample,  especially in contexts where data collection from the entire population is impractical. The \emph{sampling design} is the systematic process of selecting individuals or units from the population to be included in the sample. Different sampling methods are used depending on the research objectives and resources available. By using statistical techniques based on the sampling design, researchers can make reliable inferences about the population based on the sample.

Applying this concept to our reserving problem, we can consider all the      $Y_i, ~ i=1, \ldots, N(\tau)$     payments as the population under study, while the current $N^{P}(\tau)$ payments made by the valuation date,      i.e., the  $Y_i, ~ i=1, \ldots, N^P(\tau)$,     serve as the selected sample for understanding this population. It is important to note that the sampling design and the actual sampling process are not determined or performed by the investigator, but are purely driven by the randomness associated with whether a payment is made or not by the valuation date. Thus, the sample is given     according to an unknown mechanism     rather than being selected by the actuary. 

The sampling mechanism based on the payment data can be conceptualized as a two-stage sampling process \cite[p.~171]{thompson2012sampling}. In the first stage, a \emph{Poisson} sampling without replacement is employed to sample the reported claims. This means, for each of the claims in the population, a Bernoulli experiment is conducted, where success is defined as the claim being reported by the valuation time, and failure occurs if it is not reported. Refer to \cite[p.~85]{sarndal2003model} for more details on the Poisson sampling.

Moving to the second stage, we focus on the payments associated with each of the sampled claims from the previous stage (i.e., the reported claims). In this case, another sampling procedure is carried out to determine which payments of a claim are made before the valuation time and which are not. This is also achieved by Bernoulli-like experiments, however, do note that these are not independent because of the ordering of the payments e.g., a second payment of a claim can be sampled as long as the first payment is sampled. 

As a result of the sampling, we can assign a dichotomic random variable $\mathbf{1}_i(\tau)$, $i=1, \ldots, N(\tau)$  with success probability $\pi_i(\tau)$, to each of the payments $Y_i$ in the population. Such a variable takes the value of 1 or 0, indicating whether the payment $Y_i$ belongs to the sample of payments made or not, respectively,  by a given valuation time $\tau$. These variables are referred to as the \textit{membership indicators} of the payments and are determined based on the delay in reporting (for the first stage of sampling) and the delay in payment (for the second stage) by the valuation time. Mathematically,
$$
\mathbf{1}_i(\tau) = \mathbf{1}_{ \{ W_i \le \tau \} }  = \mathbf{1}_{ \{ T_i+ U_i + V_i \le \tau\} } =\mathbf{1}_{ \{ Z_i \le \tau-T_i \} } =\mathbf{1}_{ \{ U_i \le \tau-T_i \} } \mathbf{1}_{ \{ V_i \le \tau-R_i \} }
$$

\noindent where the indicators in the product on the right-hand side are the indicators of the first and second stages of sampling, respectively. The probabilities $\pi_i(\tau)$ are known as \emph{inclusion probabilities} and can be interpreted as the likelihood of payment $Y_i$ belonging to the sample or, equivalently, being paid by the valuation time $\tau$. These probabilities are dependent on the valuation time and vary across payments due to the     heterogeneity of the portfolio as reflected in the different accident times,  different types of accidents and in general, different attributes associated with each payment.  We also remark that it may be the case that these probabilities are also associated with the actual value of the payment sizes $Y_i$, which is the unit of measure in the sampling context.  This potential dependence is known in the literature on population sampling as the \textit{sampling being informative} as the actual values of the payments may be associated with the sampling design.    

Along those lines, the inclusion probabilities are given by 
\begin{equation}
\pi_i(\tau) = P( U_i \le \tau-T_i \vert X_i,    Y_i, T_i    ) \times P( V_i \le \tau-R_i \vert X_i,    Y_i, T_i, R_i,     U_i) = \pi^{U}_i(\tau) \times \pi^{V}_i(\tau)
\label{pi_t}
\end{equation}

\noindent where $\pi^{U}_i(\tau)=P( U_i \le \tau-T_i \vert X_i,    Y_i, T_i) =P( U_i \le \tau-T_i \vert \mathcal{F}_i    )$ and $\pi^{V}_i(\tau)= P( V_i \le \tau-R_i \vert X_i,    Y_i, T_i, R_i,  U_i)=P( V_i \le \tau-R_i \vert \mathcal{F}_i)$     are the inclusion probabilities of the first and second stage of sampling, respectively,     which also depend on the characteristics of the payment.  Here we use $\mathcal{F}_i$ as the set of all characteristics of the $i$-th claim, i.e., the $X_i,  Y_i, T_i, R_i,  U_i$ etc, to simplify the notation of these conditional probabilities.      Note that the value of the second probability depends on the outcome of the first stage of sampling, and so $\pi^{V}_i(\tau)$ is in principle a conditional probability given the realization of $U_i$.    Additionally, it is important to observe that the first probability is contingent solely upon the reporting of a claim, and thus remains consistent for all payments linked to the same claim. In contrast, the second probability is contingent upon the occurrence of the payment itself, hence it may fluctuate from one payment to another owing to the varying attributes/characteristics of each payment.      We will write these probabilities as $\pi_i^U(\tau)$ and $\pi_i^V(\tau)$ to streamline the notation and emphasize that the indexation on $i$ corresponds to the probabilities being specific for each payment as determined based on their attributes. It is important to note that these probabilities are not predefined and are therefore unknown to the investigator. We will delve into their estimation in Section \ref{Estimation_Section}, via the use of an auxiliary model.

Finally, note that the sample size in the design is not a fixed quantity. The sample size, which in our case is equivalent to the number of payments currently made $N^{P}(\tau)$, is a random variable defined as $N^{P}(\tau) = \sum_{i=1}^{N(\tau)} \mathbf{1}_i(\tau)$, which we can identify as the thinning of the counting process of the total number of payments.

\subsection{Point estimation using the Horvitz-Thompson estimator}
\label{HT_Section}
As motivated by the population sampling literature (e.g., \cite[p.~45]{sarndal2003model}), a well-established estimator of the population total, which in our case is the total of payments (i.e., the ultimate $L(\tau)$)  is provided by the \emph{Horvitz-Thompson} (HT) estimator described as follows

\begin{equation}
\hat{L} (\tau) = \sum_{i=1}^{N^{P}(\tau)} \frac{Y_{i}}{\pi_{i}(\tau)},
\label{IPW_ult}
\end{equation}

\noindent and therefore  an estimator of the outstanding claims is  the difference between the estimated total  and the currently paid amount,
 
\begin{equation}
\hat{L}^{IBNS}(\tau) = \hat{L}(\tau)-L^{P}(\tau) = \sum_{i=1}^{N^{P}(\tau)} \frac{Y_{i}}{\pi_{i}(\tau)} - \sum_{i=1}^{N^{P}(\tau)} Y_{i} = \sum_{i=1}^{N^{P}(\tau)} \frac{1-\pi_{i}(\tau)}{\pi_{i}(\tau)} Y_{i}.
\label{IPW_Y}
\end{equation}

The intuition behind the HT estimator lies in the fact that only a portion of all payments $Y_i$ is reported, proportionally to $\pi_i(\tau)$, and so each payment in the sample is ``augmented" by a factor of $1/\pi_i(\tau)$ to approximate the actual total amount.      The HT estimator has been demonstrated to be unbiased for the population total, regardless of the underlying sampling design \cite[p.~69]{thompson2012sampling}, making it a highly practical choice. One of its most attractive features is that this property holds without assumptions on the probabilistic mechanism generating the data. Indeed, most of the inference results for the HT estimator are based on the so-called \emph{design-based} approach (see for e.g. \cite{arnab2017survey}). This approach relies solely on the randomness associated with the sampling process, rather than the \emph{model-based} approach, which relies on a probabilistic model for the population itself. From an actuarial viewpoint, this essentially implies     that the HT estimator is distribution-free, meaning that it leads to an estimation of the reserve that does not require any assumptions on the underlying distribution of the number of claims (frequency) or the distribution of claim sizes (severity), which is a desirable property for reserving. Additionally, we emphasize the fact that even though the estimator is based on a population-level target (i.e., a macro-level scale), the inclusion probabilities are dependent on the individual attributes of policyholders, claims, and payments. Therefore the estimator incorporates granular information as part of the aggregate estimation.

The HT estimator is widely recognized as one of the most influential estimators in the statistics literature, having been extensively studied for over 70 years in population sampling (e.g., \cite{arnab2017survey}). Consequently, the HT estimator has a solid theoretical foundation and possesses numerous desirable properties that directly inherit to the claim-reserving problem, including consistency, unbiasedness, and sufficiency, among others. More recently, it has also been applied in inverse probability weighting (IPW) methods for estimation in causal inference, e.g. \cite{yao2021survey}, including applications in fairness in insurance. The terminology ``IPW estimator" is more extended in and outside the statistics literature, and so we will mostly refer to the estimator of the reserve as the IPW estimator, and reserve the naming of the HT estimator when referring to the general concept.

In our specific context, it is essential to highlight that the estimator presented in Equation (\ref{IPW_Y}) can be evaluated once the inclusion probabilities have been estimated. Consequently, some of the properties of the HT estimator may be marginally influenced by this preliminary estimation. It is noteworthy that situations requiring such estimation are not uncommon in the context of IPW estimators and are indeed prevalent in practice. For example, in the field of causal inference, inclusion probabilities, often referred to as propensity scores, are routinely estimated before the application of IPW estimators (e.g. \cite{salditt2023parametric}).   Other similar examples occur in the context of handling missing data, e.g., Section on \cite{kim2021statistical}.     It is well-documented that, as long as these preliminary estimations are conducted reasonably, the resulting estimator maintains its reliability and properties. References such as \cite{hirano2003efficient} and \cite{fattorini2006applying} provide further insights into the robustness of estimators under such conditions.

A specific case of interest of the IPW estimator arises when we set $Y_{i} = 1$. In this scenario, all the sums above simplify to a count of the number of payments, allowing us to obtain an estimator for the number of the outstanding payments, ${N}^{IBNS}(\tau)$, to be made after the valuation time $\tau$, as

\begin{equation}
\hat{N}^{IBNS}(\tau) := \hat{N}(\tau)-N^{P}(\tau) = \sum_{i=1}^{N^{P}(\tau)} \frac{1}{\pi_{i}(\tau)} - \sum_{i=1}^{N^{P}(\tau)} 1 = \sum_{i=1}^{N^{P}(\tau)} \frac{1-\pi_{i}(\tau)}{\pi_{i}(\tau)}.
\label{IPW_N}
\end{equation}

It is noteworthy that this particular expression coincides, structure-wise, with the one employed by \cite{hiabu2021} and \cite{fung2022fitting}, specifically for estimating the number of incurred but not reported (IBNR) claims. In the former, the author proposed a density estimation of the distribution of claims in the run-off triangle and used an expression similar to the one in Equation (\ref{IPW_N}) to estimate the number of IBNRs. However, this estimator does not account for the heterogeneity of the claims, nor include payment data. In the latter, the authors derived a similar expression to Equation (\ref{IPW_N}) under the assumption that the number of unreported claims follows a geometric distribution, demonstrating its unbiasedness when the overall number of claims is driven by a Poisson process. It is therefore crucial to emphasize that within the framework of the HT estimator, this result generalizes immediately without the assumption of the geometric distribution, and it also accommodates the case of payment data.

\subsubsection*{A ``Micro-level" Chain-Ladder method}

From an actuarial standpoint, our proposed IPW estimator for the ultimate claims can be perceived as an individual-level adaptation of the Chain-Ladder method, by expressing the estimator in Equation (\ref{IPW_ult}) as

$$
\hat{L}(\tau) = \sum_{i=1}^{N^{P}(\tau)} f_i (\tau) Y_{i},
$$

\noindent where, we can interpret $f_i (\tau) := 1/\pi_i(\tau)$ as an individual development factor assigned to each payment $Y_i$. These factors serve to project the payment to its ultimate value which aligns with the fundamental principle of the Chain-Ladder method. As the factors $f_i(\tau) $ are influenced by the policyholder's attributes, we can think of this methodology as a ``micro-level" version of the Chain-Ladder method, as it implies the development factors on an individual level while retaining the essential characteristics of the Chain-Ladder. This analogy provides the IPW estimator with an intuitive and interpretable estimation of the reserve that is already well-established in the actuarial community and makes its use more appealing to practitioners. We will revisit this discussion with more in-depth details in Section \ref{theoretical}.

\subsection{Confidence interval of the estimation}
\label{CI_section}
Confidence intervals for the reserve can be constructed based on the sampling distribution of the HT estimator, as discussed by \cite[p.~70]{thompson2012sampling}. In summary, under minimal regularity conditions, the HT estimator follows approximately a normal distribution under the two-stage sampling design for large populations \cite{chauvet2018consistency}. Thus, an approximate $1-\alpha$ confidence interval can be constructed using normal quantiles. However, as explained by \cite[p.~70]{thompson2012sampling}, the accuracy of the normal approximation relies on the sample size and the distribution of $Y_i$, which tends to exhibit skewness and heavy tails. Consequently, the normal distribution might provide a suboptimal approximation for our reserving application.

Alternatively, one can construct a confidence interval by applying a log transformation to the liability. This approach utilizes the delta method to construct an interval for the logarithm of the liability, which tends to exhibit behavior closer to normality. Subsequently, the interval is transformed back to the original scale using the reverse transformation. This log-transformed confidence interval can be a more appropriate choice, considering the distribution of the data and its potential skewness and heavy-tailed characteristics.

Therefore, an approximate $1-\alpha$ confidence interval for $L^{   IBNS    }(\tau)$ can be constructed as
\small
\begin{equation}
\left( \exp\left( \log( \hat{L}^{IBNS}(\tau) ) - Z_{\alpha/2} \frac{ \sqrt{ Var(\hat{L}^{IBNS}(\tau) ) } }{ \hat{L}^{IBNS}(\tau)  } \right) , \exp\left( \log( \hat{L}^{IBNS}(\tau) ) + Z_{\alpha/2} \frac{ \sqrt{ Var(\hat{L}^{IBNS}(\tau) ) } }{ \hat{L}^{IBNS}(\tau) } \right) \right),
\label{CIHT}
\end{equation} \normalsize
where $Z_{\alpha/2}$ is the $1-\alpha/2$ quantile of the standard normal distribution. As noted by \cite[p.~71]{thompson2012sampling}, the computation of the variance of the HT estimator can be quite laborious. To address this challenge, \cite{berger1998rate} propose the use of a simpler estimator given by:

\begin{equation}
\hat{Var}( \hat{L}^{IBNS}(\tau) ) =  \frac{ \sum_{i=1}^{N^{P}(\tau)} \left( N^{P}(\tau)\frac{1-\pi_{i}(\tau)}{\pi_{i}(\tau)} Y_{i}- \hat{L}^{IBNS}(\tau)  \right)^2 }{N^{P}(\tau)(N^{P}(\tau)-1)}
\label{VAr_HT}
\end{equation}

This formulation is computationally simple to obtain and tends to provide a conservative estimate of the variance \cite[p.~70]{thompson2012sampling}, which is desirable for the claim reserving problem.


Once again, it is crucial to note that the previous procedure for the construction of confidence intervals is conditional upon the estimated inclusion probabilities and, therefore, does not account for parameter uncertainty associated with such estimation. Numerous studies have addressed the distribution of the HT estimation while considering the estimation of the inclusion probabilities, such as \cite{hirano2003efficient}, \cite{tan2006distributional}, \cite{fattorini2009adaptive}, and references therein. In brief, under certain regularity conditions, the HT estimator remains consistent and approximately normally distributed, with variance depending on the actual process used for estimation. Nevertheless, in general practice, it is not uncommon either to neglect the variance stemming from the uncertainty of the parameters for constructing prediction intervals or to adopt more practical approaches. For instance, one can construct confidence intervals employing the bootstrap procedure, as outlined by \cite[p.~636]{arnab2017survey}. This method allows for the incorporation of parameter uncertainty without introducing excessive complexity, and simultaneously adopting a data-driven approach without relying on the normal approximation.

\subsection{Adjustments to the IPW estimate}
   
\label{adjust}
A key issue that makes the claim reserving estimation process to be a very challenging one, is that the inclusion probabilities can vary significantly impacting the stability of the IPW estimator. An extreme case is when the inclusion probability of a claim is close to 0, which mostly occurs when a claim is recently reported, which represents the majority of the claims included in the IBNR reserve. Such circumstances can lead to instability in the estimator, potentially resulting in abnormally high values of the reserve when compared with the experience of previous reserving exercises. This behavior has been widely documented in the population sampling literature of the HT estimator in the context of heterogeneous inclusion probabilities. See for e.g., \cite{hulliger1995outlier}, \cite{improvingweights}, \cite{ma2020robust} and references therein.

A standard approach to address the instabilities of the HT is by trimming the inclusion probabilities. In this approach, if the inclusion probability is too small, it is replaced with a larger value to get rid of the instability (see for e.g. \cite{improvingweights},\cite{ma2020robust}). As a result, the adjusted inclusion probabilities lead to an adjusted IPW estimator that is more robust towards these scenarios. Data-driven methods, such as \cite{ma2020robust} or Algorithm \ref{IHTest}, as proposed by \, cite{zong2018improved} and illustrated below, offer a systematic approach for this task. In the latter case, it can be shown that the mean square error of the adjusted IPW estimator is less or equal to its not adjusted counterpart for several sampling mechanisms, including the Poisson sampling.

\begin{algorithm}
\small
Obtain the ordered inclusion probabilities $\left\{\pi_{(1)}, \pi_{(2)}, \ldots, \pi_{(N^{P}(\tau))}\right\}$ from smallest to largest. \;

\For{$j =1, \ldots, N^{P}(\tau)$}{
  \If{$\pi_{(j)} \leq \frac{1}{j+1}$ and $\pi_{(j+1)}>\frac{1}{j+2}$ }{
     Modify inclusion probabilities as: \;
     $$\{\underbrace{\pi_{(j)}, \ldots, \pi_{(j)}}_{j-1}, \pi_{(j)}, \pi_{(j+1)}, \ldots, \pi_{(N^{P}(\tau))}\}$$ \;
    }
}
\caption{Trimming  of inclusion probabilities for the IPW estimator}
\label{IHTest}
\end{algorithm}

While these adjustments are theoretically sound,  increasing the inclusion probabilities may introduce a downward bias in the estimation of the reserve. Therefore, it is advisable to make adjustments only if the estimation demonstrates sensitivity to changes in the inclusion probabilities. 

Lastly, we note that there is a vast literature on improving the raw estimation of IPW estimators to make them more robust, as seen in, for example, \cite{ma2020robust}. These adjustments are beyond the scope of this paper, yet it is important to acknowledge their existence and potential application in reserving. Nevertheless, it is worth emphasizing that the decision on whether or not to implement these adjustments should be left to the actuary's discretion, based on their expertise.

\section{Calculation of RBNS, IBNR and  incremental claims reserves}
\label{Others_section}

The reserve for outstanding claims, as discussed earlier, accounts for both unreported and partially paid claims. While Equation (\ref{IPW_Y}) provides an estimator for the total reserve, it does not specify the allocation of the reserve to different types of payments. However, for accounting purposes, cash management, and risk assessment, actuaries need to specify the components of the overall reserve, commonly known as the IBNR (Incurred But Not Reported) reserve, the RBNS (Reported But Not Settled) reserve, and incremental payments over specific time periods.

In this section, we present how the population sampling framework can be adapted to decompose the estimation of the total reserve given by Equation (\ref{IPW_Y}) into these sub-components as per the actuary's requirements. To do so, we introduce the ``change of population principle" as a general approach to accomplish these, and possibly other decompositions within the IPW framework, and then demonstrate its application in deriving the aforementioned reserves.

\subsection{ The change of population principle}

In Section \ref{IPW_Section}, we used the fact that the currently paid amount can be considered as a sample of the total amount of payments. As a result, we defined a sampling design within the total amount of payments along with its corresponding inclusion probabilities. However, it is important to recognize a simple yet crucial fact: the currently paid amount can also be regarded as a sample from various sub-populations within the total amount of payments. 

\begin{figure}[h]
\centering
\includegraphics[width=0.8\textwidth]{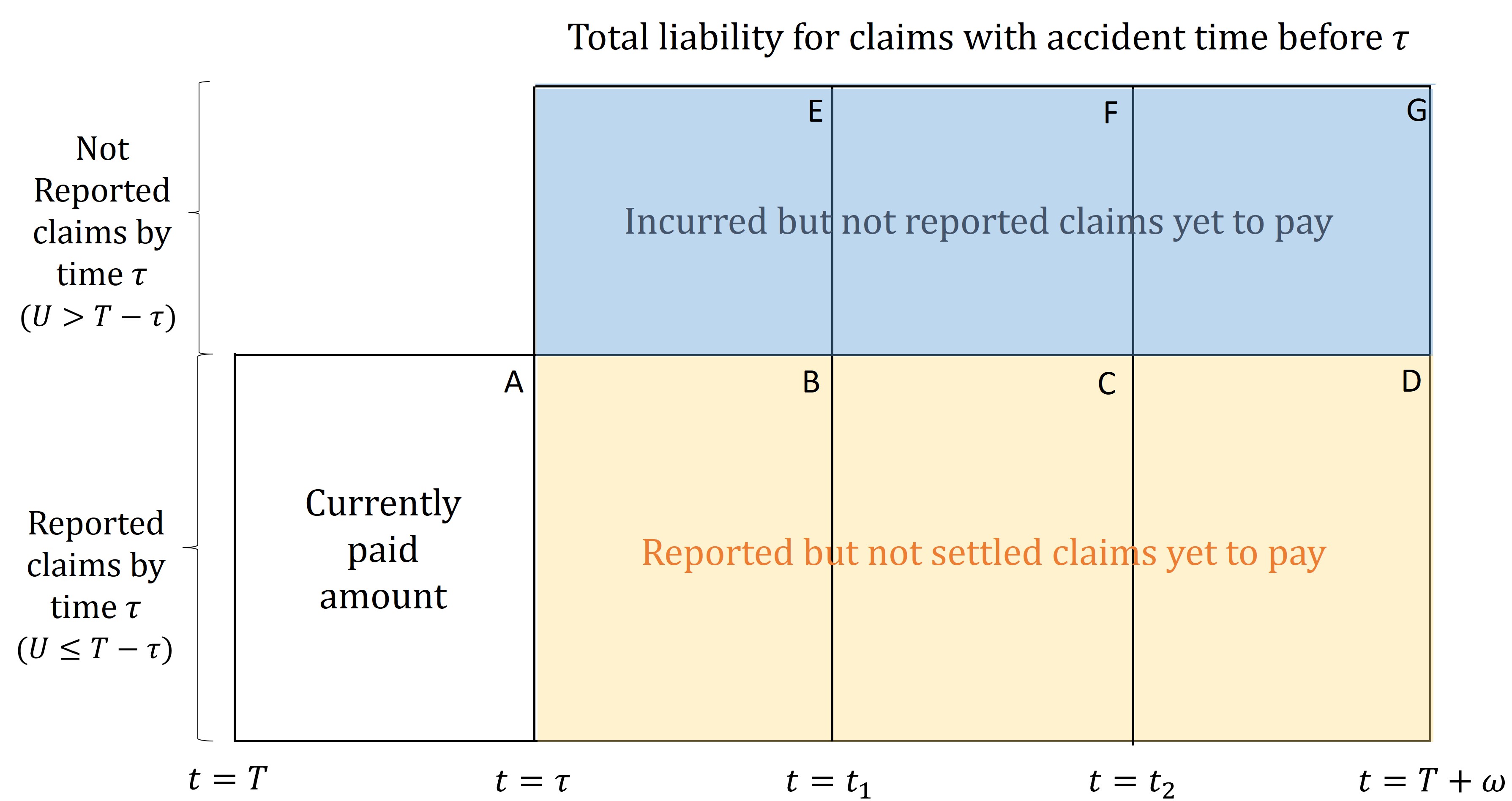}
\caption{Decomposition of the total liability as of valuation time $\tau$ into sub-populations}
\label{subpopu}
\end{figure}

Figure \ref{subpopu} demonstrates a method of partitioning the total liability at a given valuation time $\tau$ into sub-populations associated with specific reserves of interest. This figure provides a visual representation, akin to a run-off triangle, distinguishing reported and non-reported payments at $\tau$. The x-axis represents the development time, which goes from $t=T$ (i.e., the accident time)  up to time $t= T+\omega$, being $\omega$ the maximum settlement time of a claim. This figure can be thought of as a screenshot of the classification of all the payments at a given valuation date $\tau$. 

To illustrate the concept, let's examine Figure \ref{subpopu}. Combining all regions ($A$ to $G$) yields the population discussed in Section \ref{IPW_Section}, representing total liabilities. Focusing on the lower half of the figure, regions $A$ to $D$, encompass payments associated with reported claims at $\tau$. Additionally, by narrowing our focus to the lower half of the figure and considering payments up to a specific time, such as $t=t_2$ (the union of regions $A$, $B$, and $C$), we obtain a truncated version of payments. Specifically, this includes total payments made for currently reported claims, excluding those made after $t=t_2$. Note that the current paid amount is a sample from all of this subpopulation.

Adopting this approach, estimating the total liability for a specific subpopulation involves treating it as the main population from which current payments are sampled. Consequently, the IPW estimator, under a different sampling design, can be employed to estimate the liability. We refer to this approach as the ``change of population principle''. The different sample design under the change of population principle leads to different inclusion probabilities. Nevertheless, these probabilities can be easily determined using elementary conditional probability arguments. Specifically, when we limit the analysis to a subpopulation $\mathcal{S}$, we denote the inclusion probability under this restriction as $\pi^{\mathcal{S}}_j(\tau)$. This probability represents the likelihood of payment $Y_j$ being reported at $\tau$, given its membership in subpopulation $\mathcal{S}$. Note that these probabilities differ in meaning and in value from those defined in Section \ref{IPW_Section}. Bayes' rule allows expressing this probability as

$$
\pi^{\mathcal{S}}_{i}(\tau) =  \frac{\pi_{i}(\tau)}{P_i(\mathcal{S})},
$$

\noindent where $P_i(\mathcal{S})$ denotes the probability of payment $j$ being sampled in subpopulation $\mathcal{S}$ according to the original sampling design, which is influenced by factors like reporting delay and claim evolution. It is important to note that in this context, we assume the subpopulation $\mathcal{S}$ is a subpopulation encompassing the current payments (region $A$ in Figure \ref{subpopu}).

We will observe that for the reserves of interest, these probabilities can be straightforwardly expressed in terms of the probabilities associated with the previously defined delay time random variables $U_i$ and $V_i$. Therefore, no additional estimations are necessary.

\subsection{Calculation of the RBNS reserve}
\label{RBNS_Section}
The Reported But Not Settled (RBNS) reserve represents payments that are yet to be made for claims already reported at valuation time $\tau$. This reserve corresponds to the combined regions B, C, and D in Figure \ref{subpopu}.

To estimate the reserve, we apply the change of population principle and define the population $\mathcal{S}$ as the total payments associated with reported claims at $\tau$. This population corresponds to the lower half of Figure \ref{subpopu}, specifically $\mathcal{S}=A \cup B \cup C \cup D$. In reserving terminology, this corresponds to the ultimate incurred losses for claims reported prior to $\tau$.

Next, we determine the inclusion probabilities. The selection of the subpopulation depends on claim reporting, which occurs with probability $P_i(\mathcal{S})=P(U_i \le \tau -T_i \vert X_i)$. Utilizing the previously mentioned result derived from Bayes' rule, the new inclusion probabilities for this population are
$$
\frac{\pi_{i}(\tau)}{P_i(\mathcal{S})}
 = \frac{P(U_i \le \tau -T_i \vert \mathcal{F}_i) \times P(V_i \le \tau -R_i \vert \mathcal{F}_i)}{P(U_i \le \tau -T_i \vert \mathcal{F}_i)} = P(V_i \le \tau -R_i \vert \mathcal{F}_i) = \pi_i^V(\tau).
$$

This result is intuitive since the region only considers claims already reported at $\tau$, and the remaining randomness pertains to the evolution of payment occurrences only. Consequently, we can utilize the IPW estimator to obtain an estimator for the total payments of reported claims as
$$
 \sum_{i=1}^{N^{P} (\tau)} \frac{ Y_i}{\pi_i^V(\tau)}.
$$

Hence, the RBNS reserve of interest can be obtained by subtracting this quantity from the current paid amount:
\begin{equation}
\hat{L}^{RBNS}(\tau) = \sum_{i=1}^{N^{P} (\tau)} \frac{ Y_i}{\pi_i^V(\tau)} -L^{P}(\tau) = \sum_{i=1}^{N^{P} (\tau)} \frac{1-\pi_i^V(\tau)}{\pi_i^V(\tau)} Y_i.
\label{IPW_Y_RBNS}
\end{equation}

\subsection{Calculation of the pure IBNR reserve}

To estimate the Incurred But Not Reported (IBNR) reserve, we cannot directly apply the change of population principle since the current paid amount is not a subpopulation of the not reported claims population (Figure \ref{subpopu}). However, we can easily overcome this by considering the current paid amount as the difference between two populations: the total payments (all regions in Figure \ref{subpopu}) and the total payments of currently reported claims (lower half of Figure \ref{subpopu}). Estimations for the liabilities associated with these populations have been discussed in Sections \ref{IPW_Section} and \ref{RBNS_Section}, respectively. Therefore, the IBNR liability can be estimated as the difference between these two estimations.

\small
\begin{equation}
\hat{L}^{IBNR}(\tau) = \hat{L}^{IBNS}(\tau) - \hat{L}^{RBNS}(\tau) =  \sum_{i=1}^{N^{P} (\tau)} \left( \frac{1-\pi_i(\tau)}{\pi_i(\tau)}  - \frac{1-\pi_i^{V}(\tau)}{\pi_i^{V}(\tau)} \right)Y_i =  \sum_{i=1}^{N^{P} (\tau)} \left( \frac{1-\pi_i^{U}(\tau)}{\pi_i^{U}(\tau)}  \right) \frac{Y_i}{\pi_i^{V}(\tau)}.
\label{IPW_Y_IBNR}
\end{equation}
\normalsize

We would like to highlight that this approach to estimating the IBNR is analogous to the conventional actuarial method using run-off triangles, where the total reserve is estimated using the incurred claims triangle and subtracting the reserve obtained from the paid claims triangle. Unlike aggregate approaches that may yield negative reserve estimates, our method ensures non-negative estimations.

\subsection{Calculation of cumulative and incremental payments}

The estimator we have presented so far provides the ultimate amount of liabilities, but insurance companies require projections of the reserve payments over specific periods. These payments, known as incremental claims, can be estimated within our framework as follows: we utilize the change of population principle to estimate cumulative claims for different periods and then calculate the incremental claims as the difference between these cumulative claims. Notably, our model is continuous rather than discrete, allowing for the accommodation of any desired periodicity for incremental claims. We will illustrate this process for the total reserve only, however, it can be similarly applied to the RBNS.

Let's consider an insurance company assessing claims whose accidents incurred before the valuation time $\tau$ and is interested in estimating the incremental claims associated with a future period between $t_1$ and $t_2$ ($\tau < t_1 < t_2$), denoted as $L(\tau, t_1, t_2)$. Visually, $L(\tau, t_1, t_2)$ corresponds to $C$ and $F$ in Figure \ref{subpopu}.

We start by considering the population of cumulative claims up to time $t_1 \geq \tau$, where only payments made up to $t_1$ are included i.e., $L(\tau, 0, t_1)$ (regions $A$, $B$, and $E$ in Figure \ref{subpopu}). Using the change of population principle, a payment belongs to this population if its payment time is before $t_1$, which occurs with probability $P_i(\mathcal{S}) = \pi_i(t_1)$. Thus, the inclusion probability is:
$$
\frac{\pi_{i}(\tau)}{P_i(\mathcal{S})}
 = \frac{\pi_{i}(\tau)}{\pi_{i}(t_1) } ,
$$

\noindent and so the IPW estimator for cumulative claims is given by
\begin{equation}
\hat{L}(\tau, 0, t_1)= \sum_{i=1}^{N^{P}(\tau)} \frac{\pi_{i}(t_1)}{\pi_{i}(\tau)}Y_{i} ~.
\label{incrementalprobs}
\end{equation}

The incremental claims between $t_1$ and $t_2$ are then given by $L(\tau, t_1, t_2) = L(\tau, 0, t_2) - L(\tau, 0, t_1)$, and so an unbiased estimator for incremental claims is

\begin{equation}
\hat{L}(\tau, t_1, t_2)= \hat{L}(\tau, 0, t_2)- \hat{L}(\tau, 0, t_1)= \sum_{i=1}^{N^{P}(\tau)} \frac{\pi_{i}(t_2)-\pi_{i}(t_1)}{\pi_{i}(\tau)}Y_{i}.
\label{Incre}
\end{equation}

This is a very intuitive expression: the denominator, $\pi_i(\tau)$, scales the observed claims $Y_i$ to the total amount, while the difference in probabilities in the numerator, $\pi_i(t_2) - \pi_i(t_1)$, captures the proportion of the total observed between $t_1$ and $t_2$.

\section{The Chain-Ladder as a Horvitz-Thompson estimator }
\label{theoretical}

This section constitutes a key contribution of the paper, demonstrating how the classical Chain-Ladder method and certain recent extensions inherently manifest as specific instances of the IPW estimator. Consequently, the theory of population sampling offers an alternative and statistically robust framework for understanding the Chain-Ladder method.

To proceed, it is essential to recognize that the Chain-Ladder provides discrete-time estimates of cumulative claims. For the purposes of the discussion that follows, let's assume we have predefined $m$ periods for the development of the lifetime of a claim, from reporting to settlement. We denote these periods as $T = t_0 < t_1 < t_2 < \ldots < t_m = T+\omega$, where $T$ represents the accident year and $\omega$ is the assumed maximum settlement time for a claim. To simplify the proof, we will only focus on the claims that occur in the same accident year $T$ meaning that we will look only at a particular row in the Chain-Ladder triangle, say the $k$-th row, and we will assume that the accident year is $T=0$, so that calendar time and development time match, notationly.  This choice is made without any loss of generality, since each row of the triangle is treated similarly in terms of the development principle, and the total reserve is obtained by adding up the estimates row-wise. Furthermore, we also assume that the $\tau=t_k$ for simplicity.

To demonstrate the desired relationship, it is important to note that the classical Chain-Ladder method implicitly assumes that the development of claims is homogeneous, with all claims sharing the same development factors during the accident years. We formally state this within our context and notation in Assumption \ref{CLassumtion}.

\begin{assumption}{\emph{[Homogeneous development assumption]}}
The evolution of claims with identical accident years in the portfolio constitutes a homogeneous process, indicating that individual characteristics do not influence the patterns of reporting delays and settlement times. Consequently, the inclusion probabilities are the same for all claims with the same accident time, i.e., $\pi_i(\tau) = \pi(\tau)$. \label{CLassumtion}
\end{assumption}

We will show now that this simplified assumption (that we do not have to impose in general) suffices to derive the Chain-Ladder estimator and several of its extensions, as empirical versions of the IPW estimators.

\subsection{Derivation for the Chain-Ladder}

The Chain-Ladder method offers an iterative estimation of cumulative payments up to the ultimate. In line with our notation, we let $\hat{L}_{CL}(\tau, 0, t)$  denote the cumulative claim estimates from the Chain-Ladder method, with payments associated with accidents occurring before the valuation time $\tau$ and paid before time $t$, $t>\tau$. The Chain-Ladder estimates take the form

$$
\hat{L}_{CL}(\tau,0, \tau) = L^{P} (\tau) 
$$
$$
\hat{L}_{CL}(\tau,0, t_n) =  \hat{L}_{CL}(\tau, 0, t_{n-1}) \times f(t_{n-1},t_n), ~ ~ n=k+1, \ldots, m,
$$
where the factors $ f(t_{n-1},t_n)$ are the so-called development factors or link ratios and $L^{P} (\tau)$, represents the total payments made to time $\tau$. These factors are estimated through Equation (\ref{CLfactors}), as outlined in the literature on stochastic claim reserving e.g., \cite{mack1999standard}. In a similar vein, employing the recursion, the ultimate liability is estimated as

$$
\hat{L}_{CL}(\tau) = \hat{L}_{CL}(\tau, 0, t_m) = L^{P} (\tau) \times  f(t_{k},t_{k+1}) \times  f(t_{k+1},t_{k+2})\times \ldots \times f(t_{m-1},t_m),
$$
and the reserve estimation is obtained by subtracting this estimation from the current paid amount.

Now let's proceed to connect this estimation with the IPW estimator. It is noteworthy that under Assumption \ref{CLassumtion}, we can represent the IPW estimator for cumulative claims at any time $t_n$ using Equation (\ref{incrementalprobs}) as

$$
\hat{L}(\tau,0, t_n)= \sum_{j=1}^{N^{P}(\tau)} \frac{\pi_{j}(t_n)}{\pi_{j}(\tau)}Y_{j} = \left( \sum_{j=1}^{N^{P}(\tau)} Y_{j} \right) \frac{\pi(t_n)}{\pi(\tau)} =  L^{P} (\tau) \frac{\pi(t_n)}{\pi(\tau)},
$$

and so the IPW estimators for the cumulative payments  over the periods above would satisfy the recursive relationship

$$
\hat{L}(\tau,0,  t_n) = \hat{L}(\tau, 0, t_{n-1})\frac{\pi(t_n)}{\pi(t_{n-1})} , ~ ~ n=k+1, \ldots, m
$$

This equation provides an estimate of the reserve, following the same recursive principle as the Chain-Ladder method. In fact, one can easily observe that the implied development factors or link ratios of the IPW estimator are given by

\begin{equation}
f(t_{n-1},t_n) = \frac{\pi(t_n)}{\pi(t_{n-1})}.
\label{CLfactor}
\end{equation}

This expression provides an explicit interpretation of the development factors as the ratio of two probabilities. More specifically, it can be interpreted as the inverse of the conditional probability of a payment occurring by time $t_{n-1}$, given it occurs by time $t_n$.    
Although not explicitly stated in the notation (as we assume $T=0$ for simplification purposes, it is worth noting that the development factors might potentially be influenced by the accident time, given that the inclusion probabilities could exhibit such dependence.    .

Observe that the expression above provides a ``theoretical version" of the development factors because the inclusion probabilities are unknown a priori. We therefore continue by demonstrating how the classical Chain-Ladder method naturally provides an estimator for Equation (\ref{CLfactor}).   To accomplish this, it is important to note that in the classical Chain-Ladder method, the development factors remain constant across rows, meaning they do not vary with accident time. Hence, we need to make the assumption outlined below regarding the progression of claims over time, enabling us to utilize the information from all claims collectively in the estimation process, regardless of their accident time. It is worth mentioning that while such an assumption can be relaxed, it would necessitate a more explicit structure outlining the evolution of claims over time.

\begin{assumption}{\emph{[Stationary development assumption]}}
The evolution of claims is a stationary process through time, indicating that the accident and reporting dates do not influence the patterns of reporting delays and settlement times. Consequently, the distribution associated with the inclusion probabilities does not change over time. \label{CLassumtion2}
\end{assumption}


That said, consider a non-parametric approach where we use the membership indicators of the payments as the data, denoted as ${\mathbf{1}}_i(t)$ for $t \in {t_0, \ldots, t_m}$. In our simplified setup, we can interpret ${\mathbf{1}}_i(t_n)$ as the indicator of whether a payment is made or not before the development period $t_n$.    Note that under Assumption \ref{CLassumtion2},  the membership indicators are all Bernoulli random variables with a common mean $\pi(t)$ that do not change through time.  Assumption \ref{CLassumtion2} enables us to use the membership indicators of all claims regardless of the accident time, as they have the same distribution. Along those lines, consider a Weighted Maximum Likelihood Estimation (WMLE) for the inclusion probabilities $\pi(t)$ using the membership indicators along with pre-specified weights $\gamma_i$. It is straightforward to show that for a fixed value $t$ and treating the payments as exchangeable, the WMLE is given by the weighted empirical proportion of payments falling before period $t$ out of the total $ \hat{\pi}(t) = \frac{ \sum_{i=1}^{N(\tau)}  \gamma_i \mathbf{1}_i(t)   }{  \sum_{i=1}^{N(\tau)}   \gamma_i } $. Therefore, by the invariance property, the WMLE  estimators of the development factors are given by    
\begin{equation}
\hat{f}(t_{n-1},t_n) = \frac{\hat{\pi}(t_n)}{\hat{\pi}(t_{n-1})}= \frac{ \sum_{i=1}^{N(\tau)}   \gamma_i  \mathbf{1}_i(t_n)}{ \sum_{i=1}^{N(\tau)}  \gamma_i  \mathbf{1}_i(t_{n-1}) } .
\label{CLfactor_est}
\end{equation}

The choice of weights influences the principles governing the estimation of development factors on run-off triangles. For example, if we define the weights as $\gamma_i = Y_i$, the estimator simplifies to
\begin{equation}
\hat{f}(t_{n-1},t_n) = \frac{ \sum_{i=1}^{N(\tau)}   Y_{i} \mathbf{1}_i(t_n)}{ \sum_{i=1}^{N(\tau)}  Y_{i}\mathbf{1}_i(t_{n-1}) } =\frac{\textrm{Cumulative payments up to time } t_n}{\textrm{Cumulative payments up to time } t_{n-1}},
\label{CLfactors}
\end{equation}
which is the same estimator traditionally employed in the Chain-Ladder method. In the case of $\gamma_i=1$, we revert to the scenario of an unweighted likelihood, and the estimator takes the form:
$$
\hat{f}(t_{n-1},t_n) = \frac{ \sum_{i=1}^{N(\tau)}  \mathbf{1}_i(t_n)}{ \sum_{i=1}^{N(\tau)}  \mathbf{1}_i(t_{n-1}) } =\frac{\textrm{Number of payments up to time } t_n}{\textrm{Number of payments up to time } t_{n-1}},
$$
which coincides with the development factors obtained by applying the Chain-Ladder method to the triangle of the number of claims. Similarly, if we set up $\gamma_i=Y_i^2$, it is not difficult to show that we recover a version of the least squares estimator of the linear regression through the origin approach, akin to a model described in \cite{mack1999standard}. Other choices of weights can be employed to derive different development principles over run-off triangles, as discussed, for instance, in \cite[p. 437]{asmussen2020risk}.

These examples demonstrate that we can derive various types of triangle development models as particular cases of weighted likelihood estimation of inclusion probabilities. As such, the development factors of the Chain-Ladder (and some alternatives) aim to estimate the associated ratio of inclusion probabilities in Equation (\ref{CLfactor}). It is known from the theory of weighted likelihood estimation that, under some regularity conditions on the weights, the aforementioned WMLE of the inclusion probabilities is consistent \cite{wang2004asymptotic}, and so will be the associated estimators of the development factors in Equation (\ref{CLfactor_est}). In practice, one might prefer one weighting approach over another based on the finite sample size properties of the estimator, such as a better bias-variance trade-off or the actuary's experience.      Finally, it is pertinent to note that while we have justified the estimation using maximum likelihood, alternative valid estimation frameworks are available. For example, one could opt for a method of moments approach for the inclusion probabilities. However, it is important to highlight that this would result in estimators that coincide with the ones presented above.    

In conclusion, we observe that the Chain-Ladder method is essentially a particular case of the IPW estimator, where the inclusion probabilities are estimated using a specific empirical weighted proportion. We emphasize that this construction of the Chain-Ladder naturally follows from the probabilistic interpretation of the reserving problem as a population sampling problem, the IPW principle, and Assumptions  \ref{CLassumtion} and \ref{CLassumtion2}, rather than artificially constructing an estimator that replicates the Chain-Ladder estimate.

     Furthermore, it is important to highlight that this derivation of the Chain-Ladder method deviates significantly from previous modeling approaches, as illustrated in e.g., \cite{england_verrall_2002}. \cite{miranda2012double}, \cite{taylor2016stochastic},  \cite{Engler_Lindskog_2024} and subsequent literature. Specifically, we emphasize that the sampling framework offers a distribution-free construction for the Chain-Ladder method, where the assumptions pertain to the delay of payments rather than the claim amounts themselves. It is crucial to emphasize that our intention is not to discredit previous research on stochastic claim reserving, but rather to illustrate that the favorable properties of the Chain-Ladder method extend beyond the necessity for a stochastic model based solely on frequency and severity.


\subsection{Derivations for some extensions of the Chain-Ladder}

Over the past decade, there has been significant research aimed at comprehending the nature of the development factors in the Chain-Ladder model, with the goal of creating generalizations. In this section, we demonstrate how several of these extensions naturally arise in an interpretable fashion from the understanding of IPW estimators. This highlights the versatility and convenience of the population sampling framework for claim reserving.

\subsubsection*{Double Chain-Ladder}

Let's now concentrate on the associated incremental payments for a given time window $(t_1, t_2)$, specifically for claims on a given accident date. Under the homogeneity Assumption \ref{CLassumtion}, we can express the IPW estimator for incremental claims in Equation (\ref{Incre}) as:    
\begin{align*}
\hat{L}(\tau,t_1,t_2)  & = \sum_{i=1}^{N^{P}(\tau)} \frac{\pi(t_2)-\pi(t_1)}{\pi^U(\tau)\pi^V(\tau)}Y_{i} 
= \left( \frac{N^{P}(\tau)}{\pi^U(\tau)} \right) \left(\pi(t_2)-\pi(t_1) \right) \left( \frac{1}{N^{P}(\tau)} \sum_{i=1}^{N^{P} (\tau)} \frac{ Y_i}{\pi^V(\tau)} \right) 
\\ & = \hat{N}(\tau) p(t_1,t_2) \bar{Y} (\tau)
\end{align*}

\noindent where $\hat{N}(\tau) = N^{P}(\tau)/\pi^U(\tau)$ is the IPW estimator of the total number of claims whose accident year is before valuation time $\tau$, $p(t_1,t_2) = \pi(t_2)-\pi(t_1)$ is the expected proportion of claims that will be paid in the interval $(t_1,t_2)$, and $\bar{Y} (\tau)=\frac{1}{N^{P}(\tau)}\sum_{i=1}^{N^{P} (\tau)} \frac{ Y_i}{\pi^V(\tau)}$ is the estimated average claim amount per claim for claims reported by valuation time $\tau$. 

This estimator is, in spirit, similar to the so-called double Chain-Ladder estimator described by \cite{verrall2010including} and \cite{miranda2012double}. The IPW estimator $N^{P}(\tau)/\pi^U(\tau)$ represents the (first) Chain-Ladder estimation of the total number of claims when applied to the run-off triangle of the number incremental claims, and $\sum_{i=1}^{N^{P} (\tau)} \frac{ Y_i}{\pi^V(\tau)}$ is the (second) Chain-Ladder estimation of the total claim amount of reported claims when applied to such run-off triangle, see Section \ref{RBNS_Section}. Therefore, the IPW implicitly leverages information from both the triangles of paid claims and the number of claims, a modeling approach shown to be desirable as described in \cite{verrall2010including}.      Note that the decomposition of the reserve into IBNR and RBNS, which is naturally achieved in the IPW framework, is analogous to the one obtained from the double Chain-Ladder.

It is worth noting that this derivation of the double Chain-Ladder principle differs from other approaches, such as the collective reserving model in \cite{WAHL201934} and subsequent extensions. However, the IPW does so without imposing assumptions on frequency or severity, in contrast to the collective reserving model. Indeed, analogously to the traditional Chain-ladder method, the IPW framework also provides reasoning that leads to a double Chain-Ladder type estimation in a distribution-free fashion.  Thus, the IPW and the collective reserving frameworks are not particular cases of one another but rather complementary from a theoretical standpoint. Lastly, it is important to note that the point estimation between these methods may differ depending on the actual estimator selected for the different components.

\subsubsection*{Continuous Chain-Ladder}

\cite{miranda2013continuous} and \cite{bischofberger2020continuous} introduced what they termed the continuous Chain-Ladder, utilizing data from observations in a continuous time instead of discrete time as given by cell partition of the triangle. They argue that the Chain-Ladder method can be viewed as a structured density approach on the run-off triangle of claims, denoted as $\widetilde{f}_{Z, T}(z, t)$, and derived an estimate for outstanding claims of the form    
$$
\frac{\int_0^{\omega} \int_{\omega-t}^{\omega} \widetilde{f}_{Z, T}(z, t) \mathrm{d} z \mathrm{d} t}{\int_0^{\omega} \int_0^{\omega-t} \widetilde{f}_{Z, T}(z, t) \mathrm{d} z \mathrm{d} t} \sum_{i=1}^{N^{P}(\tau)}  Y_{i}.
$$

We can identify such an estimator as the IPW estimator of the reserve in Equation (\ref{IPW_Y}), under the homogeneity Assumption \ref{CLassumtion}, and with a specific inclusion probability. Their approach implicitly derives an estimator of the   average    inclusion probability by estimating the joint density function of the $(Z, T)$ variables, which is then integrated out on the upper half of the triangle of reported claims, i.e.,   $\pi = \int_0^{\omega} \pi(t) \widetilde{f}_{T}(t) \mathrm{d} t  = \int_0^{\omega} \int_0^{\omega-t} \widetilde{f}_{Z, T}(z, t) \mathrm{d} z \mathrm{d} t $.     This estimation is then incorporated into an estimator of the form   $ \frac{1-\pi}{\pi} \sum_{i=1}^{N^{P}(\tau)} Y_{i}$.    

The authors acknowledge that they do not provide any estimation theory justifying the structure of the estimator above. Nevertheless, note that the framework of IPW estimators might provide such a theoretical foundation. The details, unfortunately, go beyond the scope of this paper.


\subsubsection*{Chain-Ladder via reverse hazard rates}

A recent contribution by \cite{hiabu2017relationship} demonstrated that the estimated development factors of the Chain-Ladder method have a one-to-one correspondence with the estimator of the hazard rate of a counting process defined in reversed time, as shown in the equation below. This correspondence makes it useful to define more general estimators based on the Chain-Ladder principle.

Such correspondence can be easily derived from the IPW framework. In our notation, $Z$ is the random variable associated with the entire development time of the claim from its accident, and we can define the reversed time random variable as $\tilde Z = \omega-Z$. Consequently, the development factors can be expressed as:      
\begin{align*}
f(t_{n-1},t_n) & = \frac{P(Z \le t_n)}{P(Z \le t_{n-1})}   = \frac{P( \tilde Z \ge \omega-t_n)}{P( \tilde Z \ge \omega-t_{n-1})}   = \left(1-\frac{P( \tilde Z \ge \omega-t_n)-P( \tilde Z \ge \omega-t_{n-1})}{P( \tilde Z \ge \omega-t_n)} \right)^{-1}  
\\ & = \left( 1- \delta_n \alpha_{t_{n}}  \right)^{-1}
\end{align*}

\noindent where $\delta_n = t_n-t_{n-1}$ and $\alpha_{t_{n}}
 = \frac{P( \tilde Z \ge \omega-t_n)-P( \tilde Z \ge \omega-t_{n-1})}{\delta_n P( \tilde Z \ge \omega -t_n)} $ is the hazard rate of the process in reverse time $\tilde Z$ over the interval $(\omega-t_n, \omega-t_{n-1})$. 
 
This expression is employed for modelling development factors through a kernel estimator of the hazard rate. Further research also proposes more general estimation approaches for the reversed hazard rate using multiplicative models, as demonstrated in \cite{hiabu2021}. It is noteworthy that such approaches can be viewed as applications of survival models for the estimation of development factors, and, in a very implicit manner, models for the inclusion probabilities.

\subsubsection*{Chain-Ladder via demographic methods}
 Recent preprints, such as \cite{pittarello2023chain} and \cite{hiabu2024machine}, have expanded on the idea of reverse hazard rates and established a correspondence between development factors and central death rates. Such a correspondence can also be derived from the IPW framework perspective. Using the analogy of the reversed time counting process of claims as a pure death process, we have:    
\begin{align*}
f(t_{n-1},t_n) & = \frac{P( \tilde Z \ge \omega-t_n)}{P( \tilde Z \ge \omega-t_{n-1})} = \left(P( \tilde Z \ge \omega-t_{n-1} \vert \tilde Z \ge \omega-t_n) \right)^{-1} = \left( 1- \frac{ m_{ t_{n} }  }{ 1+ m_{ t_{n} } (1- \eta_{t_{n} })} \right)^{-1} \\
& = \frac{ 1+ m_{t_{n} } (1- \eta_{t_{n}})}{ 1- \eta_{t_{n} } m_{t_{n} }  } 
\end{align*}

where $ m_{t_{n} }$ represents the reversed central death rate associated with the process in reverse time over the interval $(\omega-t_n, \omega-t_{n-1})$, and $\eta_{t_{n} }$ is the so-called separation factor, which denotes the average fraction of the interval $(\omega-t_n, \omega-t_{n-1})$ lived through by those who died in the interval, akin to an exposure term. The second-to-last equality is derived from well-known results from life tables relating survival probabilities and central death rates. See for e.g., \cite[p.~53-55]{london1997survival}.

In this manner, mortality models inspired by demography literature can be applied to describe such central death rates and implicitly the development factors of the Chain-Ladder method. See the two references mentioned earlier for illustration. It is crucial to note that such mortality models are specific instances of a survival model and are closely related to the modeling of inclusion probabilities in this paper.

We conclude this section by re-emphasizing that the previous cases represent particular simplifications obtained under the sampling theory framework. In fact, the IPW estimator in its full generality will allow us to further generalize the classical Chain-Ladder method by allowing the inclusion probabilities to be dependent on each policyholder's risk characteristics, as well as other sources of heterogeneity.

\section{Estimation of the inclusion probabilities and model validation}
   
\label{Estimation_Section}
The key element required to implement the IPW estimator is the unknown inclusion probabilities $\pi_i(\tau)$. These probabilities are associated with the evolution of a claim, i.e., reporting and settlement delays, and depend on various attributes of the nature of the claim and the policyholder, denoted as $X_i$, as well as the payment amount $Y_i$ itself. In this section,       we outline briefly the methods and considerations to take into account for estimation.

As explained in Section \ref{IPW_Section},  the inclusion probabilities consist of two separate components: the probability of reporting $\pi_i^{U}(\tau)$, and the probability of payment $\pi_i^{V}(\tau)$. Each one is estimated separately, so we discuss different strategies in Sections \ref{reporting_section} and \ref{Payment_Section}.

For the estimation, it is crucial to consider the possibility of having more than one payment per claim. For the sake of notation, we will index the different claims themselves using the index $j$ and denote with $\mathcal{I}_j$ the set of indexes of the payments $Y_i$ associated with the $j$-th claim. In other words, the sets $\mathcal{I}_j$ create a partition of the entire set of indexes ${ i=1, \ldots, N(\tau) }$. Additionally, following common practice in reserving applications, we will use $\omega$ to denote the maximum settlement time of a claim. This value is defined based on the nature of the contract, such as the final date to file a claim for an accident after its occurrence, or according to the policies of the insurance company.

\subsection{Estimation of the reporting delay times probabilities $\pi_i^{U}(\tau)$ }
\label{reporting_section}

 As mentioned in Section \ref{IPW_Section2.1}, a value $U_i$ may be repeated several times depending on the number of payments per claim. Since reporting occurs only once, we consider only one realization of this variable per claim. Hence, we will use the notation $U_j$ to refer to these unique values, forming the working dataset.

To estimate the inclusion probabilities, we require the cumulative distribution function of the reporting delay times. It is common in reserving literature (see, for example, \cite{verrall2016understanding}) to assume that the reporting delay times $U_j$, conditioned on claim attributes $X_j$, follow a common distribution function $F_{U \vert X_j}(u)$. This distribution function is the target of estimation in this section, using some regression framework to incorporate dependence on covariates. 

To achieve this, we note that the reporting delay time is a time-to-event random variable commonly studied in survival modeling. Therefore, approaches already implemented in statistical software can be utilized to estimate the overall distribution function and the desired probabilities. For instance, hazard-based models, particularly the Cox regression model, are by far the most popular techniques used in survival analysis for this task (e.g., \cite{george2014survival}), and are widely implemented in packages, e.g. \cite{bender2018generalized}. Nevertheless, note that there are alternative approaches that can offer different and flexible structures, as inspired by the machine learning literature. For instance, \cite{fung2022fitting} propose a flexible model based on mixture of experts, with implementations available for insurance data \cite{tseung2021lrmoe}; \cite{wiegrebe2023deep}  consider non-linear regression on covariates via deep learning approaches; and \cite{sonabend2021mlr3proba} propose approaches based on survival trees, both implementable via \texttt{R} packages as in \cite{sonabend2021mlr3proba}. The choice of the model must be achieved in a data-driven fashion aiming for the best fit to the data.

Consequently, the desired inclusion probabilities for  all the payments $i \in I_j$ associated with the $j$-th claim are derived using the relationship:
\begin{equation}
\pi_j^U(\tau) = Pr(U_j \le \tau-T_i \vert X_j)=F_{U \vert X_j}(\tau-T_j) = 
\label{pi_u}
\end{equation}

A crucial aspect in the estimation of the model above is accounting for the right truncation of the data. Indeed, due to the delay in the reporting times, our observations are limited to the conditional random variables: $U_j \vert U_j \le \tau - T_j$, and ignoring this fact would result in a downward bias in the overall distribution. Fortunately, the literature on survival analysis has widely explored this issue and provided solutions that the user can adopt for the estimation of the model, such as back-censoring strategies or weighted likelihood methods. See for e.g., \cite{verbelen2015fitting}, \cite{fung2022fitting}), and \cite{vakulenko2020inverse} for the specific case of Cox regression models.  Moreover, as motivated in Section \ref{theoretical}, one may also include a weighting strategy that accounts for the claim sizes on the likelihood to better resemble the Chain-Ladder estimation. 



\subsection{Estimation of the payment times probabilities $\pi_i^{V}(\tau)$}  
\label{Payment_Section}

Estimating this probability differs significantly from the previous case due to the potential for multiple payment events per claim, unlike the one-time event of claim reporting. The payment process is a \emph{recurrent event process}, requiring an adapted modeling approach. See for e.g., \cite{cook2007statistical} and \cite{andersen2023models}. 

Recurrent events relate closely to counting processes, where the former focuses on event times, and the latter on the number of events. The modeling approach required for this paper involves selecting a suitable counting process (incorporating attributes) to data-driven model the number of payments per claim. The fitted model is then used to determine the desired inclusion probabilities. 

 That being said, assume the arrivals of payments for the $j$-th claim, conditioned on claim and policyholder attributes, are governed by a counting process $\{ M_j(v); v \in (0, \omega) \}$, with times $V_i, \in \mathcal{I}_j$ as the associated occurrence times. For convenience, we use the convention of $v=0$ as the actual time of reporting the claim, which initiates the payment counting process.

To compute the inclusion probabilities,  we first express the distribution of arrival times within the interval $(0, \omega)$ in terms of the process $M_j(v)$. To do this consider the reversed-time version of the process, denoted as $\tilde M_j(v) = M_j(\omega - v)$,  in which the reversed time is defined as $\tilde v = \omega - v$ (e.g.  \cite{klein2003survival}). This reversed-time process     can be seen as a pure death process, where  $\tilde M_j(0) = M_j(\omega)$ plays the role of the initial number of lives and $\tilde M_j(v)$ the number of lives at time $v$.     See for e.g., \cite{hiabu2017relationship}. The reversed times $\tilde V_i = \omega - V_i$ correspond to    the event times of such process. These     times can be seen as realizations from a lifetime random variable, denoted as $\tilde V$, following a specific distribution based on the original counting process's evolution. This random variable defines the distribution of the location of the occurrences within the interval $[0, \omega)$, and therefore, the distribution of the variable $V:= \omega - \tilde V$, denoted as $F_{V\vert X_j, U_j} (v)$, determines the distribution of occurrence times of the original process $M_j(t)$. The distribution of this random variable determines the inclusion probabilities.

Using the analogy with the reversed time counting process, we can easily compute the distribution of the time occurrences as:

\begin{equation}
 \begin{aligned}
F_{V\vert     \mathcal{F}_j     } (v) & = P(V \le v \vert      \mathcal{F}_j    )   = P( \tilde V \ge \omega - v \vert      \mathcal{F}_j    ) = E \left( P( \tilde V \ge \omega - v \vert \tilde M(0),      \mathcal{F}_j    )  \right) \\
& = E \left( \frac{ E( \tilde M_j(\omega-v) \vert \tilde M_j(0),      \mathcal{F}_j     )}{\tilde M_j(0)} \right) =  E \left( \frac{\tilde M_j(\omega - v)}{\tilde M_j(0)} \Bigm\vert      \mathcal{F}_j    \right) =  E \left( \frac{M_j(v)}{M_j(\omega)} \Bigm\vert     \mathcal{F}_j    \right)
\label{probidev}
\end{aligned}
\end{equation}
 
\noindent where the fourth equality holds by the life-table relationship $ _{\omega -v}p_0  = l_{\omega -v}/l_0$    (see for e.g., relationship (3.51) in  \cite[p.~42]{london1997survival}), where $l_{\omega -v}= E( \tilde M_j(\omega-v) \vert \tilde M_j(0),      \mathcal{F}_j     )$ and $l_0=\tilde M_j(0)$;     and the second last equality is the tower property of conditional expectations.

Equation \ref{probidev} provides an intuitive expression for the desired probability linked to the claim's evolution up to settlement. Essentially, the right-hand side of Equation (\ref{probidev}) represents the expected proportion of the number of payments made by time $v$ out of the total number of payments. Another interpretation is as the inverse of a development factor for the number of payments from time $t$ to the ultimate value at time $\omega$. Note that this expression can be analytically computed only for certain processes, such as the Poisson process and some extensions, as illustrated in Example \ref{exampleNHPP} below.

\begin{example}
Suppose $M(t)$ is a non-homogeneous Poisson process with intensity  $\lambda(v)$, then 
\small
 \begin{equation*}
F_{V\vert      \mathcal{F}_j    } (v) =  E \left( \frac{M_j(v)}{M_j(\omega)} \Bigm\vert      \mathcal{F}_j    \right) = E \left( \frac{ E( M_j(v) \vert M_j(\omega),      \mathcal{F}_j     )}{M_j(\omega)} \right) = E \left( \frac{M_j(\omega) \frac{\int_0^v \lambda_{M \vert      \mathcal{F}_j    }(s)ds}{\int_0^\omega \lambda_{M \vert      \mathcal{F}_j    }(s)ds} }{M_j(\omega)} \right) = \frac{\int_0^v \lambda_{M \vert      \mathcal{F}_j    }(s)ds}{\int_0^\omega \lambda_{M \vert      \mathcal{F}_j    }(s)ds}
 \end{equation*}
\normalsize
\noindent where we use that $M_j(v) \vert M_j(\omega) \sim \textrm{Binom} \left( n=M_j(\omega), p=\frac{\int_0^v \lambda_{M \vert X_i, U_i}(s)ds}{\int_0^\omega \lambda_{M \vert X_i, U_i}(s)ds} \right)$ in the third equality. 
\label{exampleNHPP}
\end{example}

As a result, the desired inclusion probabilities can be derived using Equation (\ref{probidev}). In this paper, we shall focus on the case of a non-homogeneous Poisson process, and so the inclusion probabilities will be computed using the expression in Example \ref{exampleNHPP}.

\section{Numerical study with real data}
\label{Numerics_Section}
In this section, we showcase the application of the IPW estimator using a real data set obtained from a European automobile insurance company. The data set comprises information on Body Injury (BI) claims from January 2009 to December 2012. 


In line with our methodology's primary objective of serving as an alternative to traditional macro-level models, while being simpler than fitting a micro-level model, we maintain a simplified approach to emphasize the practicality of the method in real-world applications.

\subsection{Description of the data }
The dataset contains detailed information about claim settlements, policyholder attributes, and automobile characteristics within the aforementioned time period. This information encompasses factors such as car weight, engine displacement, engine power, fuel type (gasoline or diesel), car age, policyholder age, and region (a total of five regions). Furthermore, the dataset includes details related to the accidents themselves, such as the time of occurrence and the type of accident (type 1 and type 2). Additionally, information regarding the progression of claim payments is available, including reporting time, settlement amounts, and the corresponding occurrence times.

\begin{figure}[h]
     \centering
     \begin{subfigure}[b]{0.45\textwidth}
         \centering
         \includegraphics[width=\textwidth]{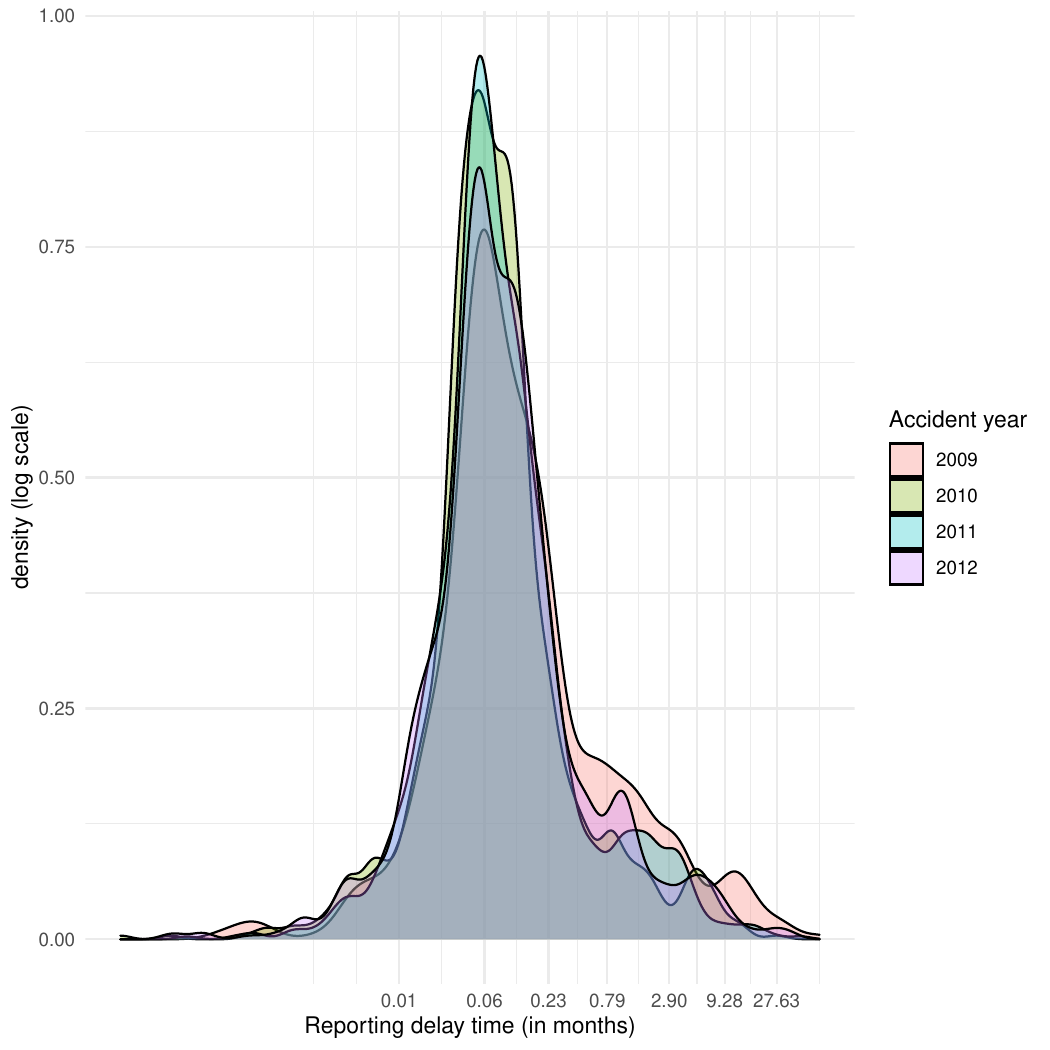}
        \end{subfigure}
     \hfill
     \begin{subfigure}[b]{0.45\textwidth}
         \centering
        \includegraphics[width=\textwidth]{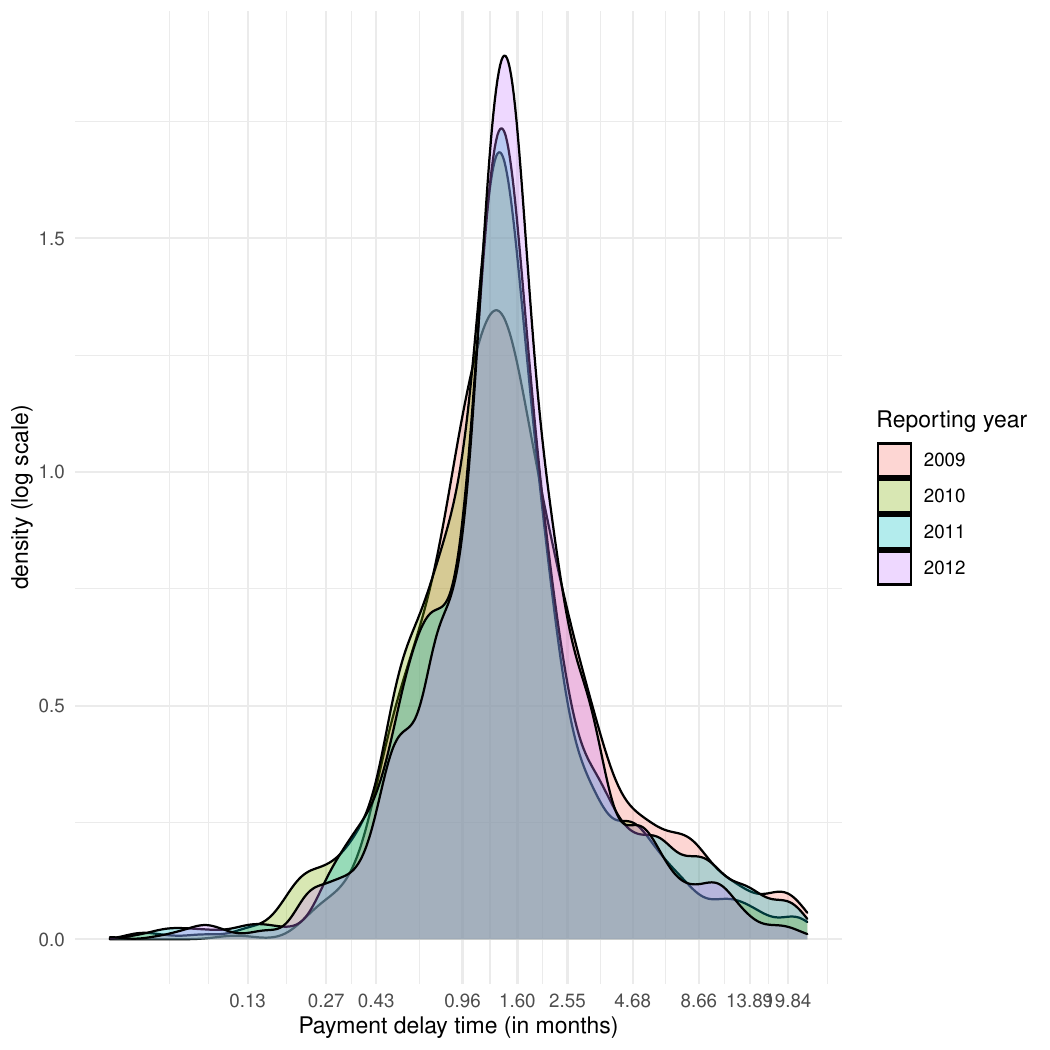}          
     \end{subfigure} 
     \caption{Density functions by year of the reporting delay time (left panel) and the payment time (right panel) in months. Plots are in the log scale.}
     \label{distributions}
\end{figure}

Our statistical modeling study focuses on the evolution of claims, specifically related to reporting delay time and payment times. We illustrate some characteristics of these quantities, such as the distributions in Figure \ref{distributions} and summarize key statistics in Table \ref{summstats}.

\begin{table}[h]
\small
\centering
\begin{tabular}{lcccccccc}
\hline
\hline
\textbf{Variable} & \multicolumn{1}{l}{\textbf{Year}} & \multicolumn{1}{l}{\textbf{Mean}} & \multicolumn{1}{l}{\textbf{Std. Dev.}} & \multicolumn{1}{l}{\textbf{Min.}} & \multicolumn{1}{l}{\textbf{1st Qu.}} & \multicolumn{1}{l}{\textbf{Median}} & \multicolumn{1}{l}{\textbf{3rd Qu.}} & \multicolumn{1}{l}{\textbf{Max.}} \\ \hline
 & 2009 & 1.03 & 4.09 & 0.00 & 0.05 & 0.10 & 0.28 & 43.28 \\
\textbf{Reporting} & 2010 & 0.45 & 1.85 & 0.00 & 0.04 & 0.07 & 0.13 & 25.66 \\
\textbf{delay time} & 2011 & 0.50 & 2.16 & 0.00 & 0.04 & 0.07 & 0.17 & 24.09 \\
\textbf{} & 2012 & 0.57 & 2.34 & 0.00 & 0.04 & 0.08 & 0.18 & 32.69 \\ \hline
\textbf{} & 2009 & 3.32 & 7.04 & 0.10 & 0.93 & 1.43 & 2.52 & 48.05 \\
\textbf{Payment} & 2010 & 2.31 & 4.37 & 0.04 & 0.81 & 1.31 & 1.92 & 35.37 \\
\textbf{delay time} & 2011 & 2.82 & 5.41 & 0.03 & 0.91 & 1.41 & 2.14 & 35.38 \\
 & 2012 & 2.49 & 4.95 & 0.04 & 0.98 & 1.44 & 2.14 & 40.23\\
\hline
\hline
\end{tabular}
\caption{Summary statistics per year of the reporting delay time  and the payment delay time (both in months)}
\label{summstats}
\end{table}

Upon reviewing the information presented in Figure \ref{distributions} and Table \ref{summstats}, it is evident that the reporting delay tends to be relatively short, with an average duration of less than a month. However, there is notable variability in the tail behavior of this variable. In contrast, the progression of claim payments typically spans a few months on average, but there are instances where settlement times can extend over several years. It is important to note that the distributions of these variables exhibit complexities that are challenging to capture using simple parametric models. Specifically, they display significant temporal fluctuations, indicating that historical data may not adequately represent future events. To account for this, we define the maximum development time for future analysis as $\omega=24$ months, with approximately 99.95\% of claims being settled within this timeframe.

\subsection{Model fitting}

To model the reporting delay time, we employ a Cox regression model. Likewise, we adopt a non-homogeneous Poisson process for the recurrent payments process. As the maximum development time is 24 months, then the time window from 2009 to 2010 can be only used for training purposes, while the time window from 2011 to 2012 will be used for testing.

For the sake of illustration, we calculate reserves on a monthly basis i.e., several valuation dates each one month apart from the other. To ensure that the estimation captures the most recent evolution of claims as much as possible, we re-fit the models employing a rolling window approach, where only the last two years of data preceding a valuation date are used for the fitting. This approach aligns with the practices employed by industry professionals in their daily work. We do not consider a time-series model via correlated frailties due to the short time window of the training sets i.e., only 2 years. Although there may be some variations in model parameters across different dates, the overall fit behaves similarly across time. We proceed to present the fitting processes for only one of the valuation dates (September 2012) in the testing period.

\subsubsection*{Reporting delay time}

For the reporting delay time, $U$, we fit a Cox regression model to describe the hazard function $\lambda_{U \vert \mathcal{F} }(u)$ using the attributes of the policyholder (contained in $ \mathcal{F}$)  as covariates as:

\small
\begin{align*}
\log( \lambda_{U \vert \mathcal{F}}(u)  ) =  \log(\lambda_0(u)) & + \beta_1\textrm{Car-Weight}+\beta_2\textrm{Engine-Power}+\beta_3\textrm{Fuel-Type}+\beta_4\textrm{Age}+\beta_5\textrm{Car-Age}\\
&+\beta_6\textrm{Accident-type}+\beta_7\textrm{Claim-Amount}+S_8(\textrm{Accident-day})+\beta_9\textrm{Region}
\end{align*}
\normalsize

The log of the baseline hazard function denoted as $\log(\lambda_0(u))$, is estimated using a B-Spline representation. Additionally, we incorporate non-linear effects for the covariate ``Accident-day" with the term $S_8(\textrm{Accident-day})$, which is also estimated using a B-Spline representation. Note that the non-linear effect associated with calendar time provides the model with a dynamic-like structure, allowing it to account for some temporal changes. Lastly, for the sake of interpretability, we work with the standardized version of the continuous covariates.

\begin{figure}[h]
     \centering
     \begin{subfigure}[b]{0.45\textwidth}
         \centering
         \includegraphics[width=\textwidth]{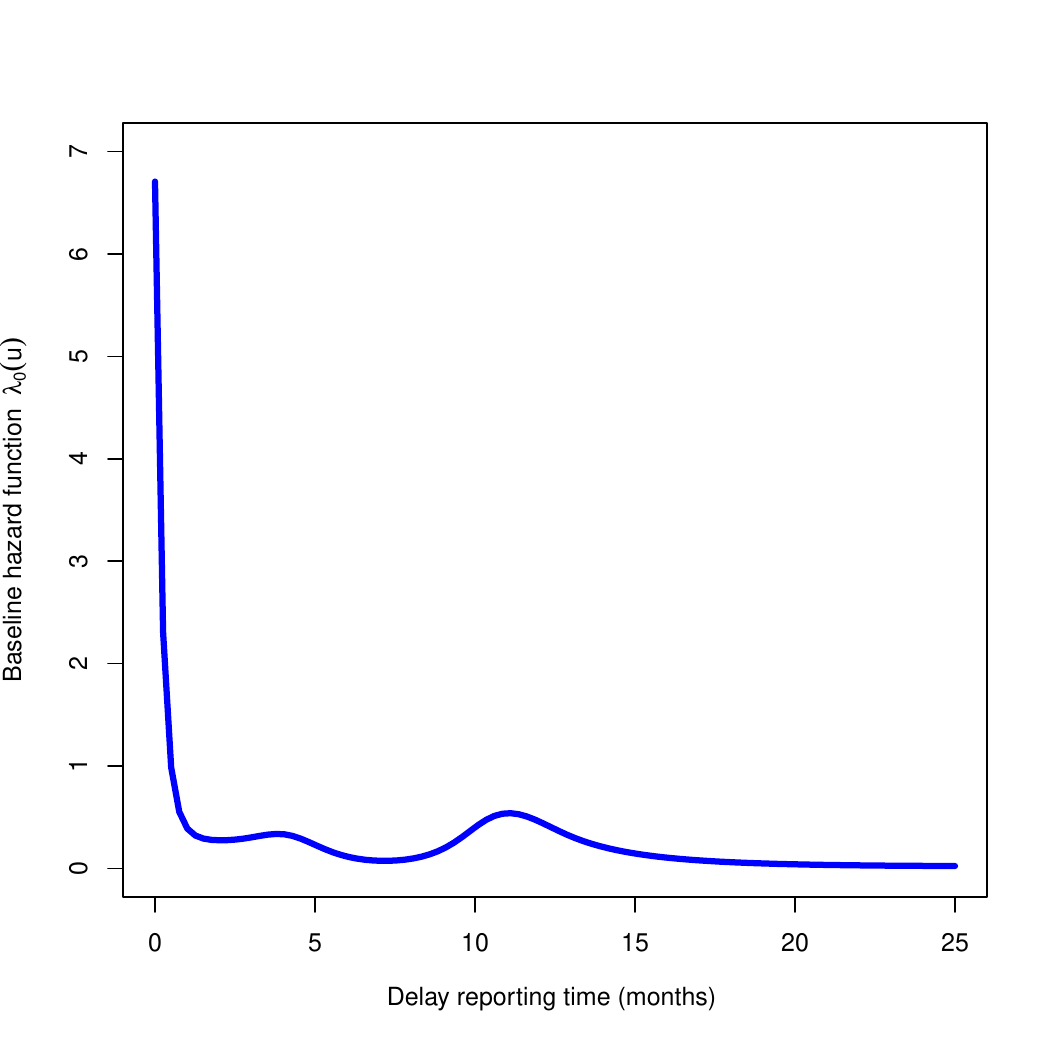}
        \end{subfigure}
     \hfill
     \begin{subfigure}[b]{0.45\textwidth}
         \centering
        \includegraphics[width=\textwidth]{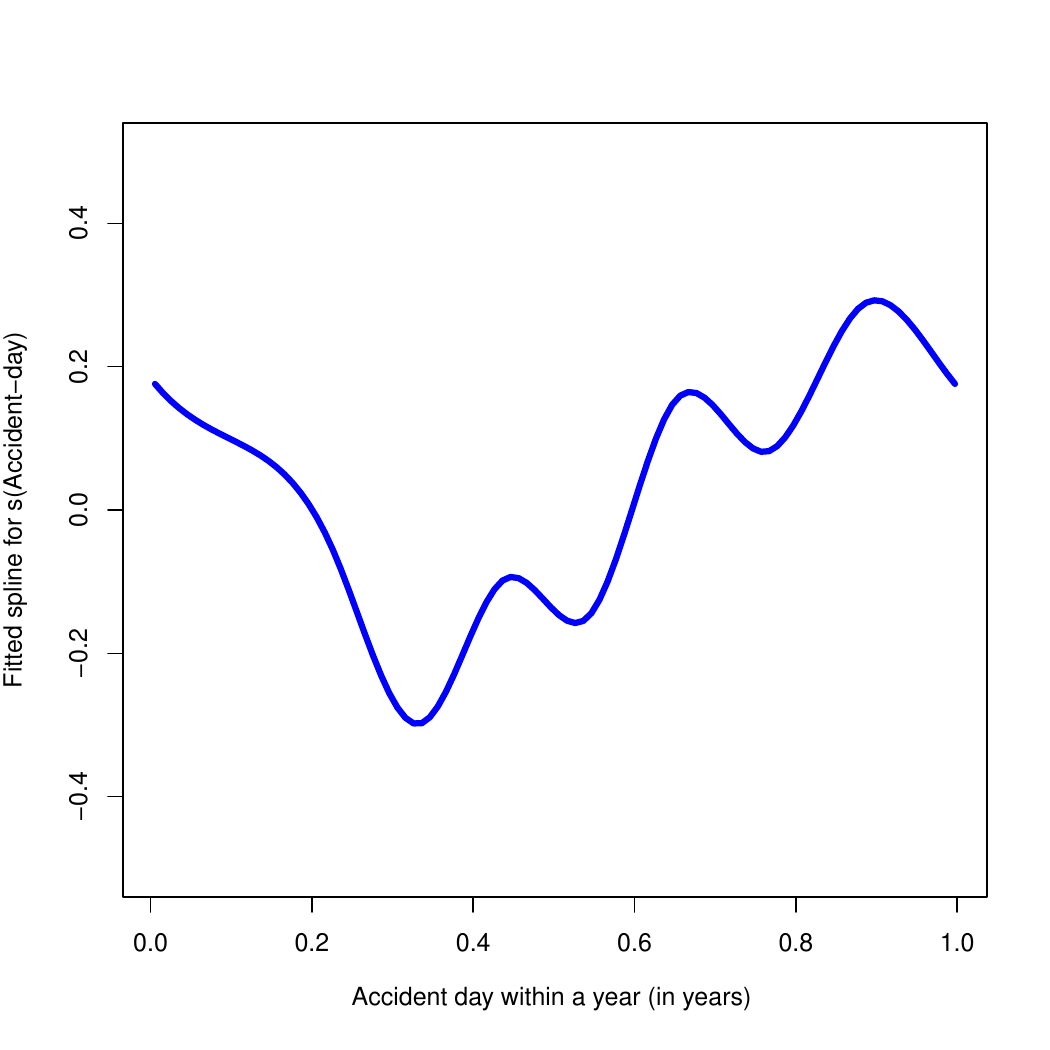}          
     \end{subfigure} 
     \caption{ Fitted baseline hazard function for the reporting delay time  (left panel)  and fitted  effect of the accident day on the hazard of the reporting delay time (right panel) }
     \label{fitted_repo}
\end{figure}

To estimate the parameters while considering the right truncation of the data, we utilize a generalized additive model implementation via the piece-wise exponential modeling approach, as described by \cite{bender2018generalized}. This approach is available in  \texttt{R}  packages such as \texttt{flexsurvreg}, \texttt{pammtools} or \texttt{GJRM}.  Moreover, we use the claim sizes (or case estimates if unavailable) as weights for each observation.  The results of the estimation are presented in Table \ref{regcoef_delay} and Figure \ref{fitted_repo}.

\begin{table}[h]
\small

\centering
\begin{tabular}{cccccccccc}
\hline
\hline 
\textbf{Coefficient}         & \textbf{$\beta_1$} & \textbf{$\beta_2$} & \textbf{$\beta_3$} & \textbf{$\beta_4$} & \textbf{$\beta_5$} & \textbf{$\beta_6$} & $\beta_7$ \\ \hline
Value    &  $0.022$ & $-0.001$ & $0.082$ & $0.074$ & $-0.071$ & $-0.057$ & $-0.069$  \\ 
Std. Err. & $0.0008$ & $0.0007$ & $0.0012$ & $0.0005$            & $0.0005$  & $0.0017$ & $0.0002$  \\
p-val  & $<0.0001$ & $<0.0001$ & $<0.0001$ & $<0.0001$           & $<0.0001$  & $<0.0001$ & $<0.0001$   \\ \hline
\hline
\end{tabular}
\caption{Estimated coefficients of the Cox regression model for the reporting delay time}
\label{regcoef_delay}
\end{table}

Table \ref{regcoef_delay} shows that all the policyholder attributes included in the model are statistically relevant to describe the behaviour of the reporting delay time. The same conclusion applies to the categorical variable ``Region", although the detailed results are not presented in the table due to its numerous categories. The left panel of Figure \ref{fitted_repo} displays the baseline hazard function, indicating a large hazard rate during the initial months, suggesting a concentration of reporting delays within this period. The hazard rate then decreases rapidly but remains nonzero for large delay times, indicating a heavy tail. Furthermore, the right panel of Figure \ref{fitted_repo} illustrates the non-linear effect associated with the accident day within a year. Time 0 is the beginning of the year i.e., January 1  and time 1 indicates the end of the year i.e., December 31.  We can appreciate a seasonal pattern.  Specifically, the hazard rate for the reporting delay time is higher at the end and beginning of the year when compared to the hazard rate at the middle of the year.

To assess the adequacy of the model fit, we employ the so-called normal pseudo-residuals. These are constructed using the probability integral transform \cite{ruschendorf2009distributional}, along with the quantile function of the normal distribution. These quantities should follow a normal distribution if the fitted model resembles properly the distribution of the data. In the case of truncation, these pseudo-residuals are computed as:
$$
r_i^U= \Phi^{-1} \left( \frac{\hat{F}_{U \vert \mathcal{F}_i} (U_i)}{\hat{F}_{U \vert \mathcal{F}_i} (\tau-T_i)} \right)
$$

\noindent 
where $\hat{F}_{U \vert \mathcal{F}_i}$ is the estimated version of ${F}_{U \vert \mathcal{F}_i}$, and $\Phi^{-1}$ is the quantile function of the standard normal distribution.

Along those lines, the left-hand side of Figure \ref{gof_payment} presents a QQ plot comparing these normal pseudo residuals against the theoretical normal distribution. From the plot, it is evident that the normal pseudo residuals exhibit no significant deviations from their expected theoretical counterparts. Hence, there is no evidence suggesting a significant lack of fit in the fitted distribution function.

Lastly, we use the model to obtain the inclusion probabilities, which can be computed by plugging the estimated hazard function in the expression below:

\begin{equation}
\pi_i^U(\tau) = Pr(U_i \le \tau-T_i \vert\mathcal{F}_i)=F_{U \vert\mathcal{F}_i}(\tau-T_j) = 1-\exp( -\int_0^{\tau-T_j} \lambda_{U \vert \mathcal{F}_i}(u)  du).
\end{equation}

\subsubsection*{Payment times}

Next, we present the model for claim evolution utilizing a non-homogeneous Poisson process structure that depends on covariates. In this case, we consider a maximum settlement time of $\omega=24$ months and proceed to model the intensity function of the process via a proportional hazard model (also known as extended Cox regression or as the \emph{Andersen-Gill} model (e.g., \cite{andersen1985counting}))  with intensity described as follows:

\small
\begin{align*}
\log( \lambda_{V \vert \mathcal{F}_i}(v)  ) =  \log( \lambda_0(v)) & + \alpha_1\textrm{Car-Weight}+\alpha_2\textrm{Engine-Power}+\alpha_3\textrm{Fuel-Type}+\alpha_4\textrm{Age}+S_5(v)\textrm{Car-Age}\\
&+\alpha_6\textrm{Accident-type}+\alpha_7(\textrm{Payment-Amount})+S_8(\textrm{Reporting-day})+\alpha_{9}\textrm{Region}\\
&+S_{10}(\textrm{Reporting-delay-time})
\end{align*}
\normalsize

\begin{figure}[h]
     \centering
     \begin{subfigure}[b]{0.45\textwidth}
         \centering
         \includegraphics[width=\textwidth]{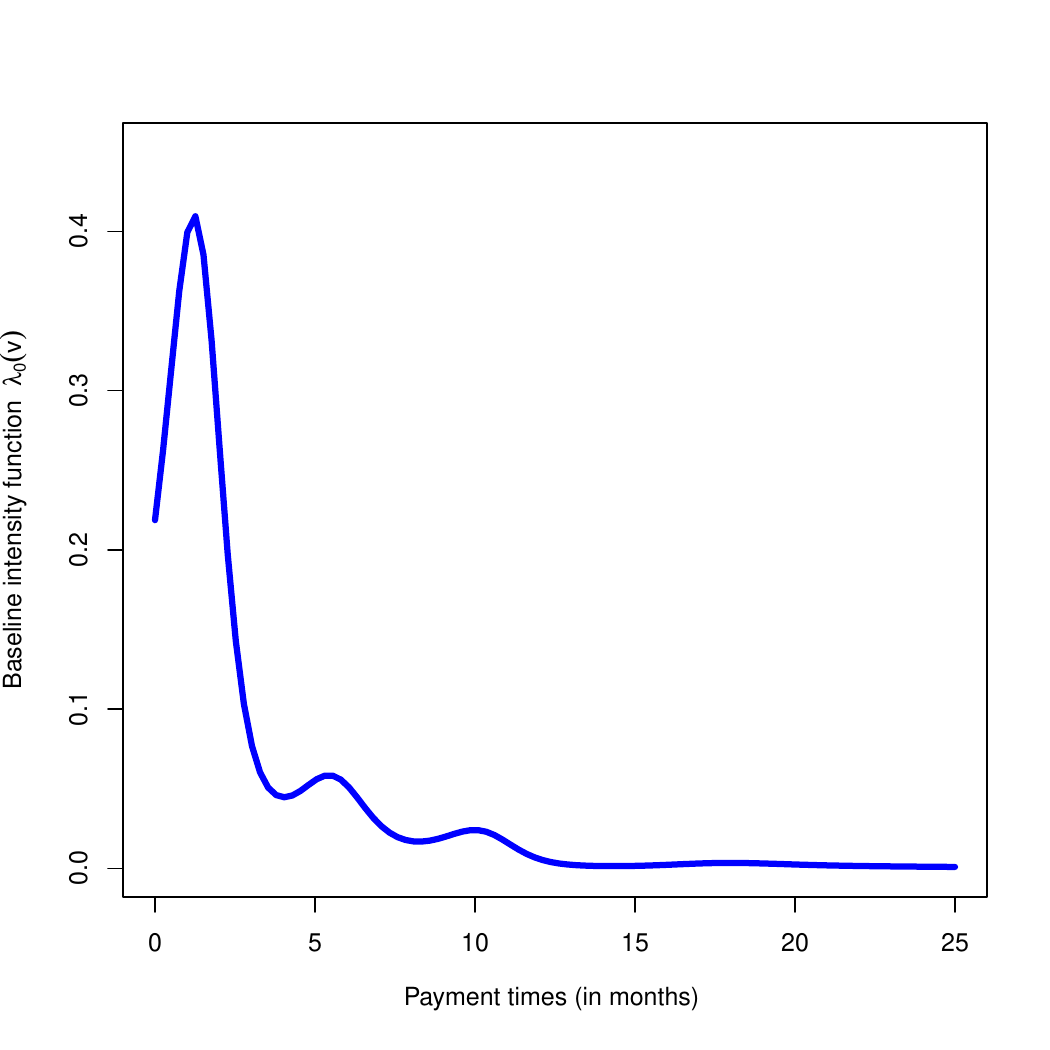}
        \end{subfigure}
     \hfill
     \begin{subfigure}[b]{0.45\textwidth}
         \centering
        \includegraphics[width=\textwidth]{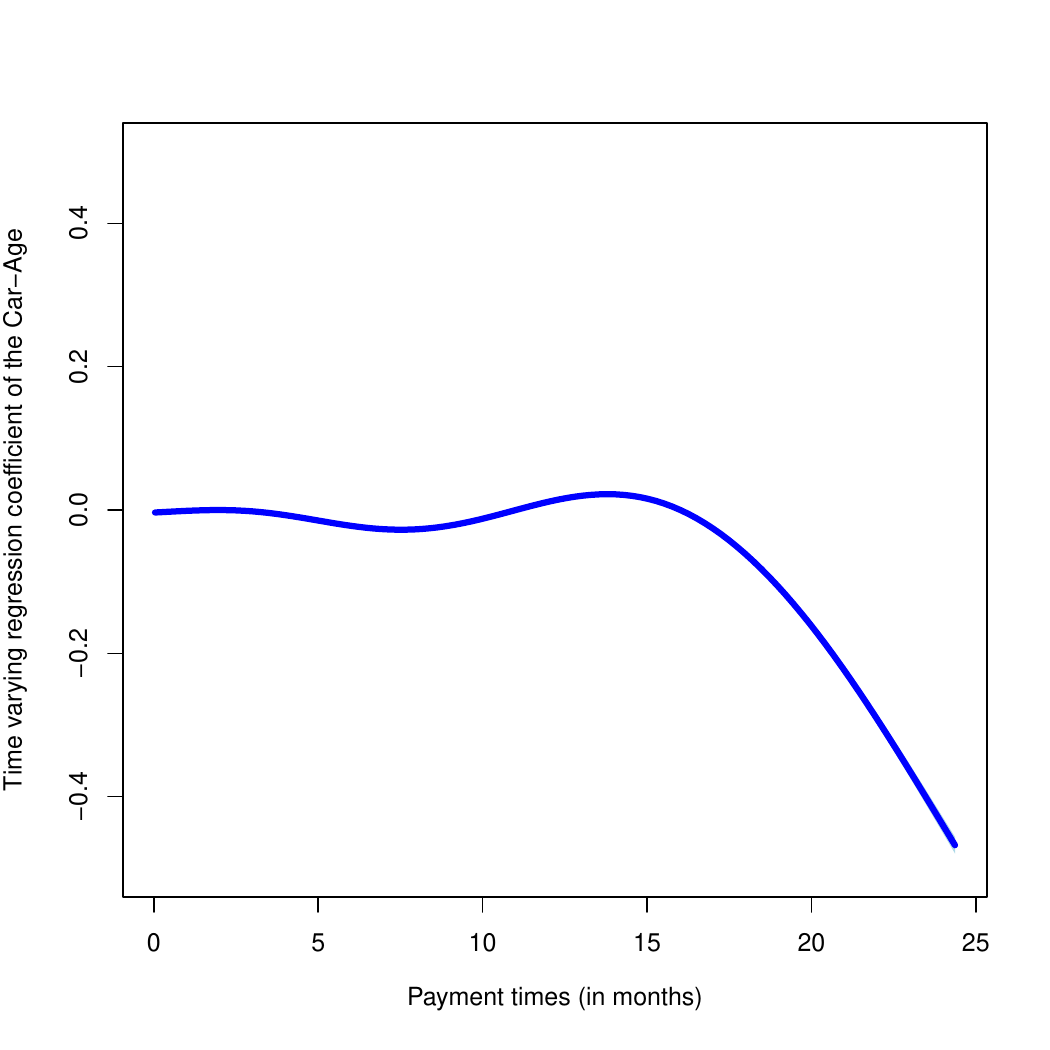}          
     \end{subfigure} 
     \caption{ Fitted baseline intensity function for the payment counting process time (left panel) and fitted effect of the car age on the intensity of the payment counting process(right panel) }
     \label{baseline_payment}
\end{figure}

\begin{figure}[h]
     \centering
     \begin{subfigure}[b]{0.45\textwidth}
         \centering
         \includegraphics[width=\textwidth]{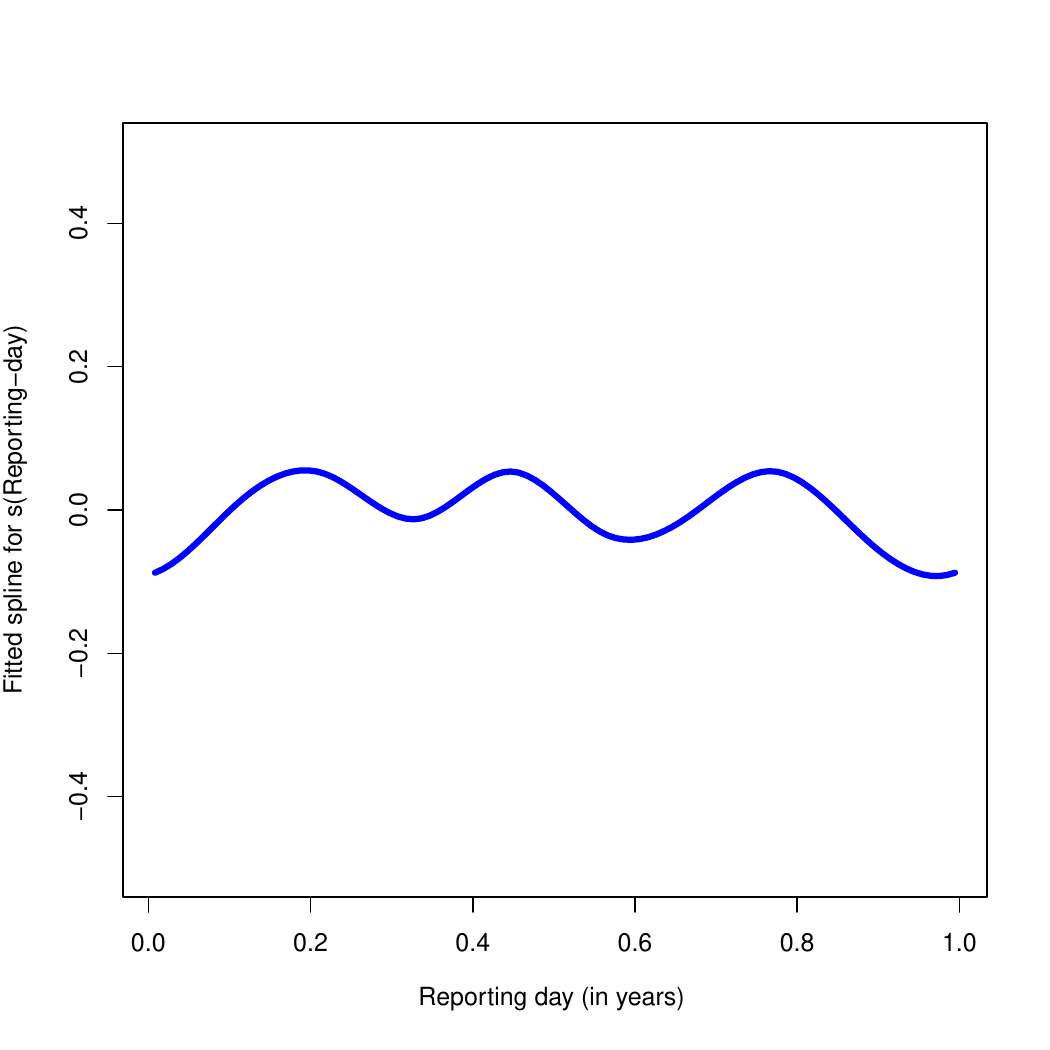}
        \end{subfigure}
     \hfill
     \begin{subfigure}[b]{0.45\textwidth}
         \centering
        \includegraphics[width=\textwidth]{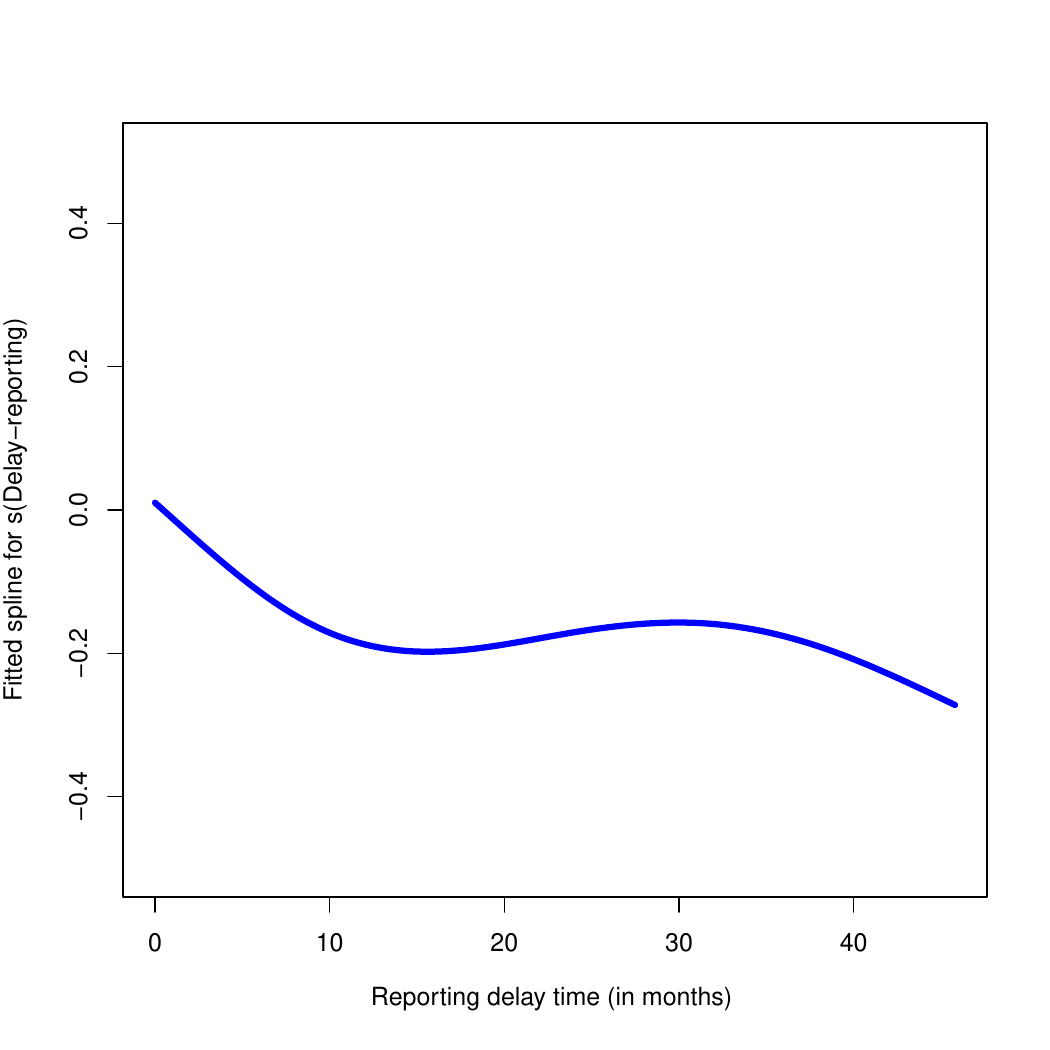}          
     \end{subfigure} 
     \caption{Fitted effect of the reporting day on the intensity function for the payment counting process (left panel) fitted effect of the reporting delay time on the intensity of the payment counting process (right panel) }
     \label{fitted_payment}
\end{figure}

Once again, we work with the standardized version of the continuous covariates for interpretability. In this instance, our model incorporates non-linear effects for the Reporting-day and Reporting-delay-time through the smooth functions $S_8$ and $S_{10}$, respectively, along with a time-varying regression coefficient for Car-Age denoted by $S_5$. These non-linear effects, as well as the logarithm of the baseline intensity function, $\log(\lambda_0(v))$, are estimated using B-Spline representations. The inclusion of these non-linear terms in the regression was based on standard model selection approaches for regression modeling i.e., the significance of regression coefficients, AIC, etc. The estimation is performed with a similar implementation to the previous model. The fitted parameters of the model are detailed in Table \ref{regcoef_payment}, and visual representations are provided in Figures \ref{baseline_payment} and \ref{fitted_payment}.

 This modeling approach mirrors the methodology presented in the previous section, and thus, we refrain from delving into further details. However, we highlight three key distinctions in this regression model. Firstly, the regression structure considers multiple events per claim instead of a single event per claim. Secondly, we incorporate the reporting day as a non-linear effect aiming to capture time-related effects. Thirdly, we include the observed reporting delay time as a covariate, as the counting process is conditional on its realization. 

\begin{table}[h]

\small
\centering
\begin{tabular}{cccccccccc}
\hline
\hline 
\textbf{Coefficient}         & \textbf{$\alpha_1$} & \textbf{$\alpha_2$} & \textbf{$\alpha_3$} & \textbf{$\alpha_4$}  & \textbf{$\alpha_6$} & $\alpha_7$   \\ \hline
Value    &  $-0.005$ & $0.011$ & $-0.007$ & $0.031$  & $0.025$ & $-0.028$  \\ 
Std. Err. & $0.0008$ & $0.0007$ & $0.0011$ & $0.0005$            & $0.0017$  & $0.0002$    \\
p-val  & $<0.0001$ & $<0.0001$ & $<0.0001$ & $<0.0001$           & $<0.0001$  & $<0.0001$   \\ \hline
\hline
\end{tabular}
\caption{Estimated coefficients of the extended Cox regression model for the payment counting process}
\label{regcoef_payment}
\end{table}

The results display similarities to the previous case, as depicted in Table \ref{regcoef_payment}, where all policyholder attributes included in the model are statistically significant in describing the behavior of the number of payments per claim. The left panel of Figure \ref{baseline_payment} illustrates the baseline intensity function of the process. It is apparent that the intensity is initially high during the first couple of months, indicating a significant number of payments occurring within this period. Subsequently, the hazard rate decreases, approaching zero, albeit not rapidly, suggesting the occurrence of some payments after several months from reporting. We also highlight the observation that certain payments tend to occur around 6 months and 1 year, as illustrated in the graph.

The right panel of Figure \ref{baseline_payment} displays the time-varying coefficient associated with the Age of the Car. This effect is practically negligible in the intensity function for the occurrence of payments within approximately the first year, but it negatively influences the frequency of payments after that period. In this scenario, the older the car, the fewer the number of payments after the first year. Similarly, claims from older cars tend to settle more quickly than those from newer cars.

This observation can be explained by factors such as an increased likelihood of total loss events, where major incidents prompt a single lump-sum payment. Additionally, the reduced economic viability of repairing older vehicles and limitations in coverage contribute to insurers opting for settlements over multiple payments. The lower perceived replacement value of older cars further encourages insurers to streamline the process with a lump-sum settlement, minimizing the overall number of payments before claim resolution.

The left panel of Figure \ref{fitted_payment} illustrates the non-linear effect associated with the reporting day, suggesting a quarterly seasonal pattern. In this context, claims reported around the end and beginning of the year tend to be settled with fewer payments than those reported throughout the year. On the other hand, the right panel of Figure \ref{fitted_payment} presents the effect of the reporting delay time on the intensity of the number of payments. In brief, we observe that claims with shorter reporting delay times are slightly associated with claim settlement involving more payments than those with longer delay times. 

Several reasons can explain such a phenomenon. For instance, swiftly reported claims often involve more complex or severe incidents, necessitating prolonged settlement periods for meticulous investigation and processing. In the case of body injury-related claims, an early report might trigger immediate payments for initial expenses, with subsequent adjustments accounting for evolving situations or additional treatments. Additionally, policyholders exhibiting prompt reporting may engage in proactive behavior, contributing to a higher frequency of payments as they actively seek additional coverage or claim multiple aspects of their policy.

\begin{figure}[h]
     \centering
     \begin{subfigure}[b]{0.45\textwidth}
         \centering
         \includegraphics[width=\textwidth]{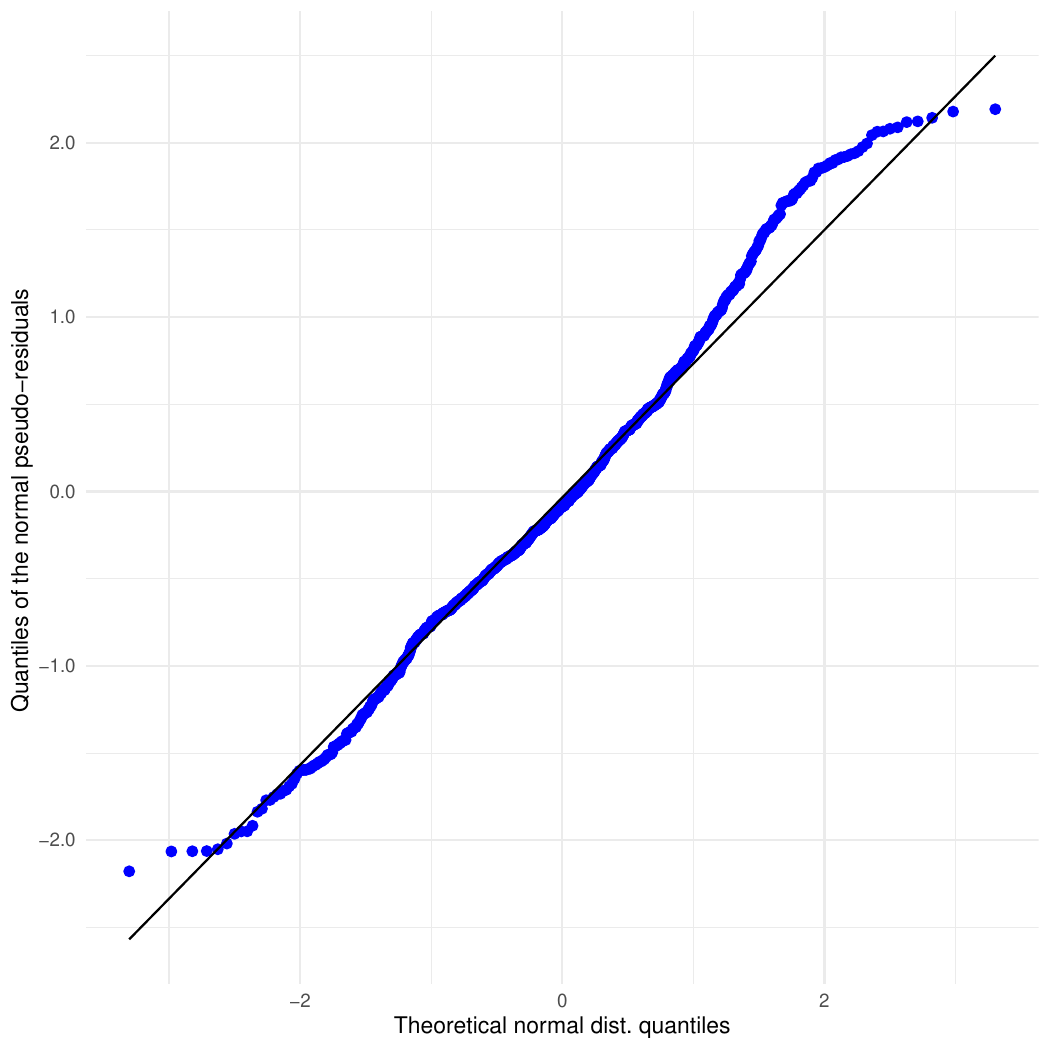}
        \end{subfigure}
     \hfill
     \begin{subfigure}[b]{0.45\textwidth}
         \centering
        \includegraphics[width=\textwidth]{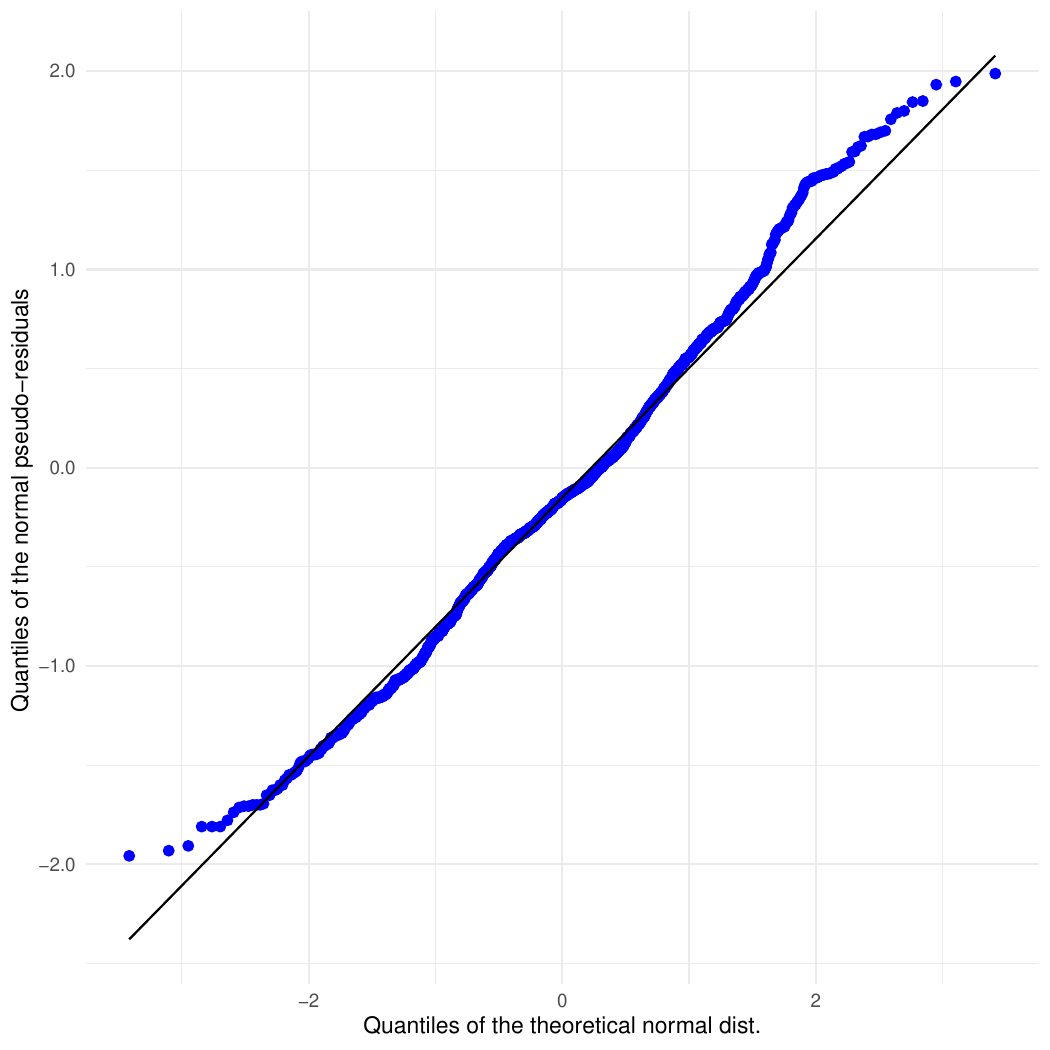}          
     \end{subfigure} 
     \caption{Q-Q plots of the normal pseudo-residuals. The left panel is for the Cox model for the reporting delay time and the right panel is for the extended Cox model for the payment counting process. }
     \label{gof_payment}
\end{figure}

To assess the goodness of fit, we once again employ the normal pseudo residuals (using an analogous expression as the one shown above) and validate them accordingly. The right panel of Figure \ref{gof_payment} presents a QQ plot comparing these normal pseudo residuals against the theoretical normal distribution. Similar to the previous analysis, no significant deviation is observed between the normal pseudo residuals and their expected theoretical counterparts. While the points do not align as closely to the 45-degree line as it was in the previous case, there is no evidence indicating a considerable lack of fit in the fitted distribution function.

Lastly, we use the model to compute the inclusion probabilities  using an expression from Example \ref{exampleNHPP}, as the model is a type of non-homogenous Poisson process:

 \begin{equation}
\pi_i^V(\tau) = P(V_i \le \tau-R_i \vert    \mathcal{F}_i ) = F_{V\vert   \mathcal{F}_i } (\tau-R_i)=  \frac{ \int_0^{\tau-R_i} \lambda_{M \vert  \mathcal{F}_i }(s)  ds  }{\int_0^{\omega} \lambda_{M \vert    \mathcal{F}_i }(s)  ds}
\label{pi_v}
 \end{equation}

\subsection{ Results of the estimation of the reserves for a single date}

Here we show the estimation of the outstanding claims (IBNS), RBNS and IBNR reserves for the same date of the testing period (September 2012). To ease the visualization and comparison, we present the estimation of the reserves in the classical run-off triangle format in Tables \ref{triangletot} and \ref{trianglerbns}, yet however, recall that our method does not rely on a given periodicity for its calculation or the construction of a triangle. 

\begin{figure}[h]
     \centering
     \begin{subfigure}[b]{0.45\textwidth}
         \centering
         \includegraphics[width=\textwidth]{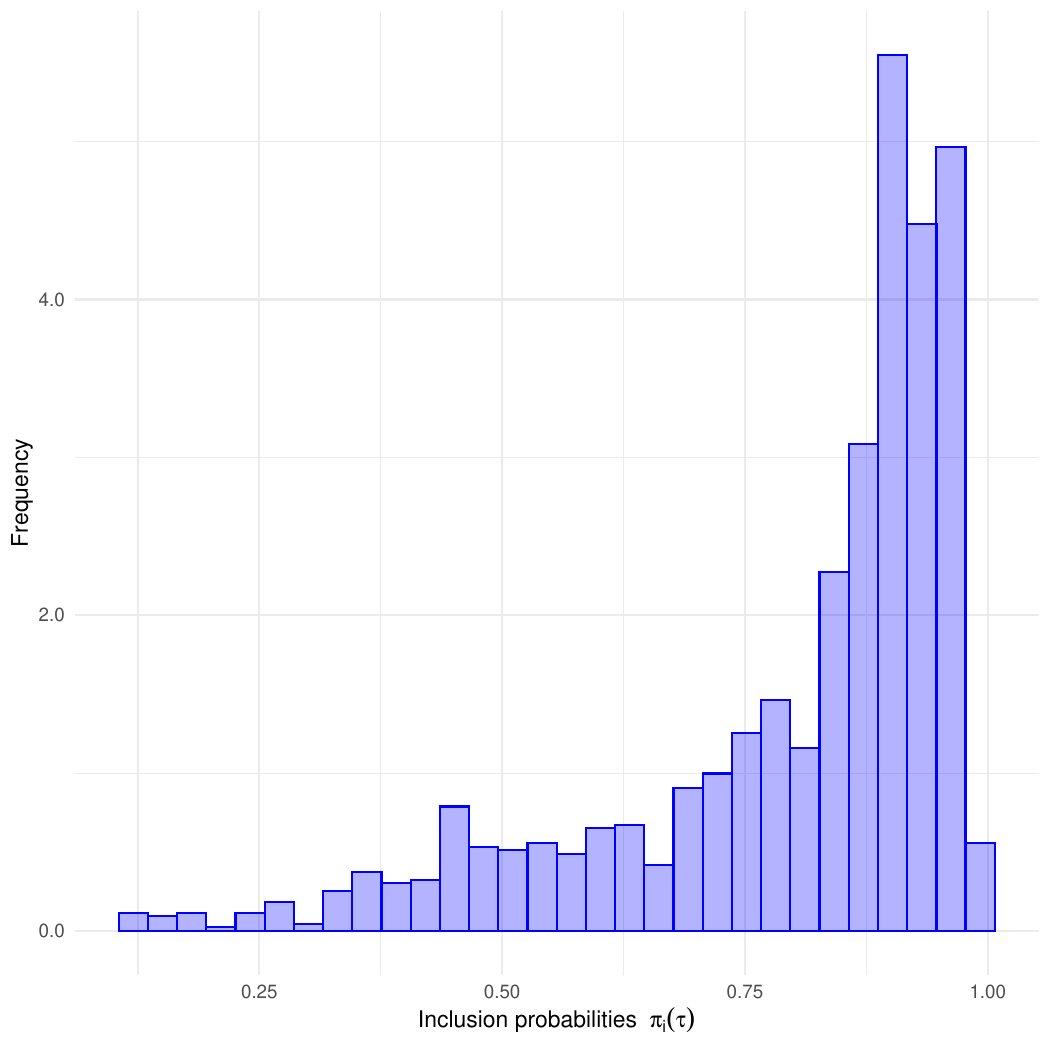}
        \end{subfigure}
     \hfill
     \begin{subfigure}[b]{0.45\textwidth}
         \centering
        \includegraphics[width=\textwidth]{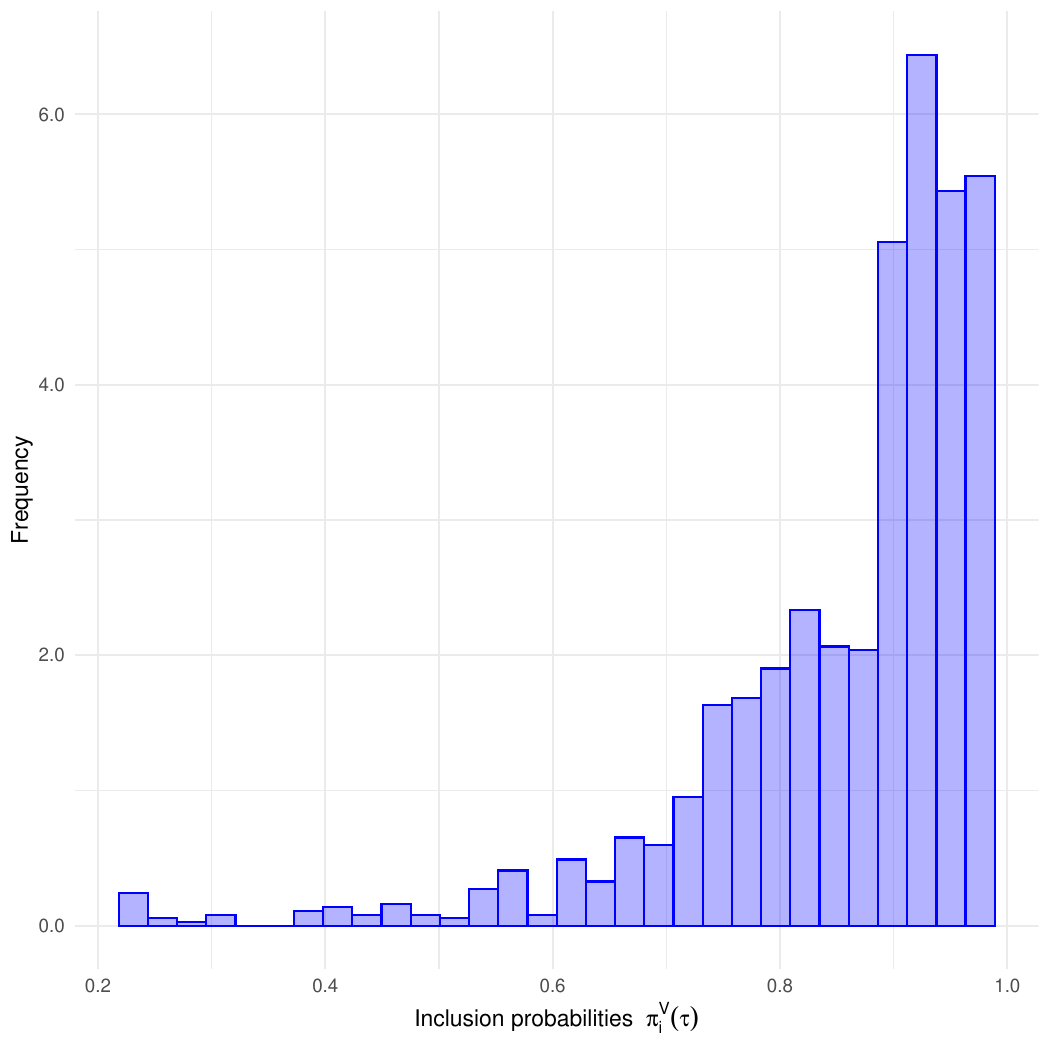}          
     \end{subfigure} 
     \caption{Histogram of the inclusion probabilities used for the IPW estimator of the total reserve (left panel) and the RBNS reserve (right panel)  }
     \label{hist_pi}
\end{figure}

The inclusion probabilities $\pi_i(\tau)$ and $\pi_i^V(\tau)$ for the outstanding not settled claims are estimated directly from the models from the previous section using Equations (\ref{pi_u}), (\ref{pi_v}) and (\ref{pi_t}). Figure \ref{hist_pi} displays the histogram of such probabilities and Table \ref{stats_probs} displays some summary statistics. Briefly, we observe that the inclusion probabilities vary drastically from one claim to another due to the heterogeneity of the claims. Note that the probabilities tend to be closer to 1 than to 0 due to the low average reporting delay and payment time, and therefore only the most recently reported claims have a small probability.

\begin{table}[h]

\centering
\small
\begin{tabular}{lllllll}
\hline
\hline
Probability & Min. & 1st Qu. & Median & Mean & 3rd Qu. & Max. \\ \hline
$\pi(\tau)$ & 0.110 & 0.734 & 0.878 & 0.802 & 0.926 & 0.982 \\
$\pi^V(\tau)$ & 0.239 & 0.806 & 0.901 & 0.858 & 0.943 & 0.984 
\\
\hline
\end{tabular}
\caption{Summary statistics of the inclusion probabilities}
\label{stats_probs}
\end{table}

Tables \ref{triangletot} and \ref{trianglerbns} present cumulative run-off triangles for total outstanding claims and reported but not settled claims, respectively, as of September 2012 The incurred but not reported claims reserve estimation is derived from the difference between these triangles, which is not shown to avoid redundancy. We completed the lower half of the triangles using the true reserve values, the IPW estimator (using Equation (\ref{incrementalprobs})), and the Chain-Ladder method, on a monthly basis, for comparison purposes. To maintain readability and practicality, the table is limited to 13 months, representing approximately 98\% of settled claims within this period.      To differentiate between the RBNS and IBNR claims components in the Chain-Ladder method, we utilize the double Chain-Ladder method by \cite{miranda2012double}.

Analyzing the lower half of the triangles in Tables \ref{triangletot} and \ref{trianglerbns}, we observe that the IPW estimator provides cumulative payment estimations that exhibit similar trends and magnitudes as the actual cumulative claims. No evident patterns of under or over-estimation are observed. Additionally, the IPW estimator exhibits different behavior to the Chain-Ladder method, indicating the impact of using the individual level information in the estimation.

\begin{table}[!ht]
\small
\resizebox{\columnwidth}{!}{%
\begin{tabular}{|c|c|c|c|c|c|c|c|c|c|c|c|c|c|c|}
\hline
\textbf{M} & \textbf{1} & \textbf{2} & \textbf{3} & \textbf{4} & \textbf{5} & \textbf{6} & \textbf{7} & \textbf{8} & \textbf{9} & \textbf{10} & \textbf{11} & \textbf{12} & \textbf{13} & \textbf{ULT} \\ \hline
 &  &  &  &  &  &  &  &  &  &  &  &  &  & 65,812 \\ 
 &  &  &  &  &  &  &  &  &  &  &  &  &  & {\color[HTML]{0000FF}  72,508} \\ 
\multirow{-3}{*}{\textbf{1}} & \multirow{-3}{*}{5,544} & \multirow{-3}{*}{5,544} & \multirow{-3}{*}{5,544} & \multirow{-3}{*}{9,356} & \multirow{-3}{*}{9,356} & \multirow{-3}{*}{9,356} & \multirow{-3}{*}{9,356} & \multirow{-3}{*}{39,787} & \multirow{-3}{*}{39,787} & \multirow{-3}{*}{39,787} & \multirow{-3}{*}{65,812} & \multirow{-3}{*}{65,812} & \multirow{-3}{*}{65,812} & {\color[HTML]{FF0000} 67,687} \\ \hline
 &  &  &  &  &  &  &  &  &  &  &  &  & 408,962 & 464,871 \\ 
 &  &  &  &  &  &  &  &  &  &  &  &  & {\color[HTML]{0000FF}  410,211} & {\color[HTML]{0000FF}  432,108} \\ 
\multirow{-3}{*}{\textbf{2}} & \multirow{-3}{*}{46,283} & \multirow{-3}{*}{187,790} & \multirow{-3}{*}{287,157} & \multirow{-3}{*}{327,979} & \multirow{-3}{*}{331,239} & \multirow{-3}{*}{341,457} & \multirow{-3}{*}{375,575} & \multirow{-3}{*}{386,305} & \multirow{-3}{*}{386,305} & \multirow{-3}{*}{394,726} & \multirow{-3}{*}{408,962} & \multirow{-3}{*}{408,962} & {\color[HTML]{FF0000} 409,390} & {\color[HTML]{FF0000} 427,539} \\ \hline
 &  &  &  &  &  &  &  &  &  &  &  & 471,409 & 471,409 & 521,505 \\ 
 &  &  &  &  &  &  &  &  &  &  &  & {\color[HTML]{0000FF}  473,056} & {\color[HTML]{0000FF}  481,526} & {\color[HTML]{0000FF}  500,189} \\ 
\multirow{-3}{*}{\textbf{3}} & \multirow{-3}{*}{23,813} & \multirow{-3}{*}{218,029} & \multirow{-3}{*}{381,885} & \multirow{-3}{*}{413,203} & \multirow{-3}{*}{457,516} & \multirow{-3}{*}{464,451} & \multirow{-3}{*}{470,453} & \multirow{-3}{*}{471,409} & \multirow{-3}{*}{471,409} & \multirow{-3}{*}{471,409} & \multirow{-3}{*}{471,409} & {\color[HTML]{FF0000} 472,267} & {\color[HTML]{FF0000} 476,769} & {\color[HTML]{FF0000} 491,963} \\ \hline
 &  &  &  &  &  &  &  &  &  &  & 559,824 & 559,824 & 612,687 & 645,614 \\ 
 &  &  &  &  &  &  &  &  &  &  & {\color[HTML]{0000FF}  563,122} & {\color[HTML]{0000FF}  581,329} & {\color[HTML]{0000FF}  598,976} & {\color[HTML]{0000FF}  629,708} \\ 
\multirow{-3}{*}{\textbf{4}} & \multirow{-3}{*}{55,145} & \multirow{-3}{*}{228,820} & \multirow{-3}{*}{387,120} & \multirow{-3}{*}{446,037} & \multirow{-3}{*}{502,665} & \multirow{-3}{*}{511,673} & \multirow{-3}{*}{540,318} & \multirow{-3}{*}{555,795} & \multirow{-3}{*}{555,795} & \multirow{-3}{*}{559,824} & {\color[HTML]{FF0000} 563,068} & {\color[HTML]{FF0000} 577,972} & {\color[HTML]{FF0000} 584,230} & {\color[HTML]{FF0000} 607,539} \\ \hline
 &  &  &  &  &  &  &  &  &  & 403,159 & 407,692 & 407,692 & 409,845 & 412,925 \\ 
 &  &  &  &  &  &  &  &  &  & {\color[HTML]{0000FF}  405,374} & {\color[HTML]{0000FF}  416,024} & {\color[HTML]{0000FF}  430,147} & {\color[HTML]{0000FF}  441,011} & {\color[HTML]{0000FF}  456,839} \\ 
\multirow{-3}{*}{\textbf{5}} & \multirow{-3}{*}{61,068} & \multirow{-3}{*}{167,822} & \multirow{-3}{*}{297,202} & \multirow{-3}{*}{348,347} & \multirow{-3}{*}{370,234} & \multirow{-3}{*}{379,379} & \multirow{-3}{*}{396,235} & \multirow{-3}{*}{396,235} & \multirow{-3}{*}{403,159} & {\color[HTML]{FF0000} 404,157} & {\color[HTML]{FF0000} 417,295} & {\color[HTML]{FF0000} 427,383} & {\color[HTML]{FF0000} 432,219} & {\color[HTML]{FF0000} 447,641} \\ \hline
 &  &  &  &  &  &  &  &  & 492,646 & 492,646 & 492,646 & 550,816 & 551,367 & 554,910 \\ 
 &  &  &  &  &  &  &  &  & {\color[HTML]{0000FF}  496,290} & {\color[HTML]{0000FF}  513,169} & {\color[HTML]{0000FF}  527,684} & {\color[HTML]{0000FF}  553,505} & {\color[HTML]{0000FF}  563,710} & {\color[HTML]{0000FF}  591,803} \\ 
\multirow{-3}{*}{\textbf{6}} & \multirow{-3}{*}{7,925} & \multirow{-3}{*}{123,729} & \multirow{-3}{*}{285,715} & \multirow{-3}{*}{342,347} & \multirow{-3}{*}{446,330} & \multirow{-3}{*}{456,970} & \multirow{-3}{*}{471,253} & \multirow{-3}{*}{492,646} & {\color[HTML]{FF0000} 499,643} & {\color[HTML]{FF0000} 518,644} & {\color[HTML]{FF0000} 534,253} & {\color[HTML]{FF0000} 547,258} & {\color[HTML]{FF0000} 553,134} & {\color[HTML]{FF0000} 573,918} \\ \hline
 &  &  &  &  &  &  &  & 454,720 & 455,515 & 457,373 & 457,737 & 457,737 & 459,130 & 482,794 \\ 
 &  &  &  &  &  &  &  & {\color[HTML]{0000FF}  452,884} & {\color[HTML]{0000FF}  466,792} & {\color[HTML]{0000FF}  475,395} & {\color[HTML]{0000FF}  493,926} & {\color[HTML]{0000FF}  504,702} & {\color[HTML]{0000FF}  511,191} & {\color[HTML]{0000FF}  541,259} \\ 
\multirow{-3}{*}{\textbf{7}} & \multirow{-3}{*}{14,738} & \multirow{-3}{*}{160,911} & \multirow{-3}{*}{314,757} & \multirow{-3}{*}{359,414} & \multirow{-3}{*}{443,251} & \multirow{-3}{*}{447,872} & \multirow{-3}{*}{454,720} & {\color[HTML]{FF0000} 450,614} & {\color[HTML]{FF0000} 464,207} & {\color[HTML]{FF0000} 477,651} & {\color[HTML]{FF0000} 490,427} & {\color[HTML]{FF0000} 501,720} & {\color[HTML]{FF0000} 505,714} & {\color[HTML]{FF0000} 521,377} \\ \hline
 &  &  &  &  &  &  & 336,823 & 336,970 & 345,319 & 345,319 & 345,319 & 345,319 & 367,580 & 382,743 \\ 
 &  &  &  &  &  &  & {\color[HTML]{0000FF}  339,792} & {\color[HTML]{0000FF}  354,857} & {\color[HTML]{0000FF}  363,281} & {\color[HTML]{0000FF}  375,890} & {\color[HTML]{0000FF}  382,666} & {\color[HTML]{0000FF}  390,574} & {\color[HTML]{0000FF}  395,153} & {\color[HTML]{0000FF}  430,756} \\ 
\multirow{-3}{*}{\textbf{8}} & \multirow{-3}{*}{21,450} & \multirow{-3}{*}{129,410} & \multirow{-3}{*}{266,031} & \multirow{-3}{*}{333,922} & \multirow{-3}{*}{334,447} & \multirow{-3}{*}{336,823} & {\color[HTML]{FF0000} 337,586} & {\color[HTML]{FF0000} 347,283} & {\color[HTML]{FF0000} 355,723} & {\color[HTML]{FF0000} 365,547} & {\color[HTML]{FF0000} 375,033} & {\color[HTML]{FF0000} 383,383} & {\color[HTML]{FF0000} 386,419} & {\color[HTML]{FF0000} 397,049} \\ \hline
 &  &  &  &  &  & 346,572 & 382,372 & 383,440 & 388,635 & 390,737 & 390,737 & 391,697 & 480,511 & 506,410 \\ 
 &  &  &  &  &  & {\color[HTML]{0000FF}  353,058} & {\color[HTML]{0000FF}  396,769} & {\color[HTML]{0000FF}  425,225} & {\color[HTML]{0000FF}  452,507} & {\color[HTML]{0000FF}  462,365} & {\color[HTML]{0000FF}  473,363} & {\color[HTML]{0000FF}  482,960} & {\color[HTML]{0000FF}  491,322} & {\color[HTML]{0000FF}  549,739} \\ 
\multirow{-3}{*}{\textbf{9}} & \multirow{-3}{*}{19,856} & \multirow{-3}{*}{214,340} & \multirow{-3}{*}{292,746} & \multirow{-3}{*}{323,878} & \multirow{-3}{*}{346,572} & {\color[HTML]{FF0000} 375,245} & {\color[HTML]{FF0000} 404,109} & {\color[HTML]{FF0000} 419,255} & {\color[HTML]{FF0000} 432,730} & {\color[HTML]{FF0000} 446,150} & {\color[HTML]{FF0000} 458,600} & {\color[HTML]{FF0000} 468,617} & {\color[HTML]{FF0000} 473,299} & {\color[HTML]{FF0000} 489,991} \\ \hline
 &  &  &  &  & 318,371 & 338,935 & 353,534 & 359,803 & 372,754 & 381,612 & 381,612 & 383,137 & 384,757 & 387,255 \\ 
 &  &  &  &  & {\color[HTML]{0000FF}  324,861} & {\color[HTML]{0000FF}  371,492} & {\color[HTML]{0000FF}  437,596} & {\color[HTML]{0000FF}  489,902} & {\color[HTML]{0000FF}  504,456} & {\color[HTML]{0000FF}  516,363} & {\color[HTML]{0000FF}  526,579} & {\color[HTML]{0000FF}  535,969} & {\color[HTML]{0000FF}  598,840} & {\color[HTML]{0000FF}  668,998} \\ 
\multirow{-3}{*}{\textbf{10}} & \multirow{-3}{*}{13,500} & \multirow{-3}{*}{141,237} & \multirow{-3}{*}{252,815} & \multirow{-3}{*}{318,371} & {\color[HTML]{FF0000} 323,699} & {\color[HTML]{FF0000} 380,267} & {\color[HTML]{FF0000} 409,408} & {\color[HTML]{FF0000} 424,897} & {\color[HTML]{FF0000} 438,136} & {\color[HTML]{FF0000} 452,085} & {\color[HTML]{FF0000} 464,319} & {\color[HTML]{FF0000} 474,899} & {\color[HTML]{FF0000} 480,107} & {\color[HTML]{FF0000} 497,165} \\ \hline
 &  &  &  & 268,993 & 340,294 & 356,838 & 372,422 & 376,798 & 378,068 & 379,643 & 383,812 & 383,812 & 383,812 & 438,539 \\ 
 &  &  &  & {\color[HTML]{0000FF}  273,024} & {\color[HTML]{0000FF}  314,104} & {\color[HTML]{0000FF}  387,050} & {\color[HTML]{0000FF}  453,353} & {\color[HTML]{0000FF}  474,677} & {\color[HTML]{0000FF}  487,953} & {\color[HTML]{0000FF}  496,034} & {\color[HTML]{0000FF}  504,104} & {\color[HTML]{0000FF}  547,521} & {\color[HTML]{0000FF}  619,442} & {\color[HTML]{0000FF}  641,400} \\ 
\multirow{-3}{*}{\textbf{11}} & \multirow{-3}{*}{23,920} & \multirow{-3}{*}{138,409} & \multirow{-3}{*}{268,993} & {\color[HTML]{FF0000} 268,971} & {\color[HTML]{FF0000} 306,864} & {\color[HTML]{FF0000} 333,077} & {\color[HTML]{FF0000} 353,658} & {\color[HTML]{FF0000} 364,294} & {\color[HTML]{FF0000} 373,699} & {\color[HTML]{FF0000} 383,949} & {\color[HTML]{FF0000} 393,323} & {\color[HTML]{FF0000} 402,357} & {\color[HTML]{FF0000} 405,363} & {\color[HTML]{FF0000} 417,156} \\ \hline
 &  &  & 82,789 & 227,599 & 279,909 & 289,413 & 304,134 & 307,233 & 309,908 & 309,908 & 309,908 & 309,908 & 309,908 & 314,645 \\ 
 &  &  & {\color[HTML]{0000FF}  73,368} & {\color[HTML]{0000FF}  106,215} & {\color[HTML]{0000FF}  141,850} & {\color[HTML]{0000FF}  173,761} & {\color[HTML]{0000FF}  183,134} & {\color[HTML]{0000FF}  191,238} & {\color[HTML]{0000FF}  195,803} & {\color[HTML]{0000FF}  198,792} & {\color[HTML]{0000FF}  210,286} & {\color[HTML]{0000FF}  242,232} & {\color[HTML]{0000FF}  247,162} & {\color[HTML]{0000FF}  252,446} \\ 
\multirow{-3}{*}{\textbf{12}} & \multirow{-3}{*}{8,104} & \multirow{-3}{*}{82,789} & {\color[HTML]{FF0000} 73,727} & {\color[HTML]{FF0000} 93,682} & {\color[HTML]{FF0000} 103,765} & {\color[HTML]{FF0000} 112,512} & {\color[HTML]{FF0000} 118,300} & {\color[HTML]{FF0000} 121,646} & {\color[HTML]{FF0000} 124,837} & {\color[HTML]{FF0000} 127,785} & {\color[HTML]{FF0000} 131,167} & {\color[HTML]{FF0000} 133,885} & {\color[HTML]{FF0000} 134,909} & {\color[HTML]{FF0000} 138,631} \\ \hline
 &  & 17,219 & 100,679 & 161,887 & 190,028 & 192,544 & 240,792 & 242,969 & 245,873 & 248,874 & 255,900 & 255,900 & 255,900 & 693,148 \\ 
 &  & {\color[HTML]{0000FF}  14,921} & {\color[HTML]{0000FF}  33,036} & {\color[HTML]{0000FF}  51,481} & {\color[HTML]{0000FF}  66,857} & {\color[HTML]{0000FF}  70,311} & {\color[HTML]{0000FF}  74,241} & {\color[HTML]{0000FF}  76,843} & {\color[HTML]{0000FF}  78,126} & {\color[HTML]{0000FF}  83,186} & {\color[HTML]{0000FF}  96,763} & {\color[HTML]{0000FF}  99,325} & {\color[HTML]{0000FF}  100,067} & {\color[HTML]{0000FF}  101,716} \\ 
\multirow{-3}{*}{\textbf{13}} & \multirow{-3}{*}{17,219} & {\color[HTML]{FF0000} 13,220} & {\color[HTML]{FF0000} 38,396} & {\color[HTML]{FF0000} 44,118} & {\color[HTML]{FF0000} 48,006} & {\color[HTML]{FF0000} 50,899} & {\color[HTML]{FF0000} 53,026} & {\color[HTML]{FF0000} 54,332} & {\color[HTML]{FF0000} 55,442} & {\color[HTML]{FF0000} 57,184} & {\color[HTML]{FF0000} 58,512} & {\color[HTML]{FF0000} 59,019} & {\color[HTML]{FF0000} 59,403} & {\color[HTML]{FF0000} 60,765} \\  \hline
\end{tabular}%
}
\caption{Monthly cumulative run-off triangle for all outstanding claims for September 2012. In black is the actual value, in blue is the estimation using the IPW estimator, and in red is the traditional Chain-Ladder}
\label{triangletot}
\end{table}

\begin{table}[h]
\small
\resizebox{\columnwidth}{!}{%
\begin{tabular}{|c|c|c|c|c|c|c|c|c|c|c|c|c|c|c|}
\hline
\textbf{M} & \textbf{1} & \textbf{2} & \textbf{3} & \textbf{4} & \textbf{5} & \textbf{6} & \textbf{7} & \textbf{8} & \textbf{9} & \textbf{10} & \textbf{11} & \textbf{12} & \textbf{13} & \textbf{ULT} \\ \hline
 &  &  &  &  &  &  &  &  &  &  &  &  &  & 65,812 \\ 
 &  &  &  &  &  &  &  &  &  &  &  &  &  & {\color[HTML]{0000FF}  67,577} \\ 
\multirow{-3}{*}{\textbf{1}} & \multirow{-3}{*}{5,544} & \multirow{-3}{*}{5,544} & \multirow{-3}{*}{5,544} & \multirow{-3}{*}{9,356} & \multirow{-3}{*}{9,356} & \multirow{-3}{*}{9,356} & \multirow{-3}{*}{9,356} & \multirow{-3}{*}{39,787} & \multirow{-3}{*}{39,787} & \multirow{-3}{*}{39,787} & \multirow{-3}{*}{65,812} & \multirow{-3}{*}{65,812} & \multirow{-3}{*}{65,812} & {\color[HTML]{FF0000} 65,812} \\ \hline
 &  &  &  &  &  &  &  &  &  &  &  &  & 408,962 & 439,571 \\ 
 &  &  &  &  &  &  &  &  &  &  &  &  & {\color[HTML]{0000FF}  410,211} & {\color[HTML]{0000FF}  428,203} \\ 
\multirow{-3}{*}{\textbf{2}} & \multirow{-3}{*}{46,283} & \multirow{-3}{*}{187,790} & \multirow{-3}{*}{287,157} & \multirow{-3}{*}{327,979} & \multirow{-3}{*}{331,239} & \multirow{-3}{*}{341,457} & \multirow{-3}{*}{375,575} & \multirow{-3}{*}{386,305} & \multirow{-3}{*}{386,305} & \multirow{-3}{*}{394,726} & \multirow{-3}{*}{408,962} & \multirow{-3}{*}{408,962} & {\color[HTML]{FF0000} 409,388} & {\color[HTML]{FF0000} 424,855} \\ \hline
 &  &  &  &  &  &  &  &  &  &  &  & 471,409 & 471,409 & 471,409 \\ 
 &  &  &  &  &  &  &  &  &  &  &  & {\color[HTML]{0000FF}  473,056} & {\color[HTML]{0000FF}  479,076} & {\color[HTML]{0000FF}  491,644} \\ 
\multirow{-3}{*}{\textbf{3}} & \multirow{-3}{*}{23,813} & \multirow{-3}{*}{218,029} & \multirow{-3}{*}{381,885} & \multirow{-3}{*}{413,203} & \multirow{-3}{*}{457,516} & \multirow{-3}{*}{464,451} & \multirow{-3}{*}{470,453} & \multirow{-3}{*}{471,409} & \multirow{-3}{*}{471,409} & \multirow{-3}{*}{471,409} & \multirow{-3}{*}{471,409} & {\color[HTML]{FF0000} 472,266} & {\color[HTML]{FF0000} 476,759} & {\color[HTML]{FF0000} 488,865} \\ \hline
 &  &  &  &  &  &  &  &  &  &  & 559,824 & 559,824 & 612,687 & 645,614 \\ 
 &  &  &  &  &  &  &  &  &  &  & {\color[HTML]{0000FF}  563,122} & {\color[HTML]{0000FF}  578,585} & {\color[HTML]{0000FF}  588,820} & {\color[HTML]{0000FF}  612,891} \\ 
\multirow{-3}{*}{\textbf{4}} & \multirow{-3}{*}{55,145} & \multirow{-3}{*}{228,820} & \multirow{-3}{*}{387,120} & \multirow{-3}{*}{446,037} & \multirow{-3}{*}{502,665} & \multirow{-3}{*}{511,673} & \multirow{-3}{*}{540,318} & \multirow{-3}{*}{555,795} & \multirow{-3}{*}{555,795} & \multirow{-3}{*}{559,824} & {\color[HTML]{FF0000} 561,346} & {\color[HTML]{FF0000} 571,447} & {\color[HTML]{FF0000} 577,620} & {\color[HTML]{FF0000} 596,900} \\ \hline
 &  &  &  &  &  &  &  &  &  & 403,159 & 407,692 & 407,692 & 409,845 & 412,352 \\ 
 &  &  &  &  &  &  &  &  &  & {\color[HTML]{0000FF}  405,374} & {\color[HTML]{0000FF}  416,015} & {\color[HTML]{0000FF}  425,558} & {\color[HTML]{0000FF}  432,532} & {\color[HTML]{0000FF}  446,039} \\ 
\multirow{-3}{*}{\textbf{5}} & \multirow{-3}{*}{61,068} & \multirow{-3}{*}{167,822} & \multirow{-3}{*}{297,202} & \multirow{-3}{*}{348,347} & \multirow{-3}{*}{370,234} & \multirow{-3}{*}{379,379} & \multirow{-3}{*}{396,235} & \multirow{-3}{*}{396,235} & \multirow{-3}{*}{403,159} & {\color[HTML]{FF0000} 403,879} & {\color[HTML]{FF0000} 413,700} & {\color[HTML]{FF0000} 419,768} & {\color[HTML]{FF0000} 424,508} & {\color[HTML]{FF0000} 436,897} \\ \hline
 &  &  &  &  &  &  &  &  & 492,646 & 492,646 & 492,646 & 550,816 & 551,367 & 551,516 \\ 
 &  &  &  &  &  &  &  &  & {\color[HTML]{0000FF}  496,290} & {\color[HTML]{0000FF}  513,169} & {\color[HTML]{0000FF}  525,411} & {\color[HTML]{0000FF}  538,032} & {\color[HTML]{0000FF}  546,020} & {\color[HTML]{0000FF}  564,057} \\ 
\multirow{-3}{*}{\textbf{6}} & \multirow{-3}{*}{7,925} & \multirow{-3}{*}{123,729} & \multirow{-3}{*}{285,715} & \multirow{-3}{*}{342,347} & \multirow{-3}{*}{446,330} & \multirow{-3}{*}{456,970} & \multirow{-3}{*}{471,253} & \multirow{-3}{*}{492,646} & {\color[HTML]{FF0000} 499,467} & {\color[HTML]{FF0000} 517,292} & {\color[HTML]{FF0000} 528,722} & {\color[HTML]{FF0000} 536,975} & {\color[HTML]{FF0000} 542,727} & {\color[HTML]{FF0000} 559,589} \\ \hline
 &  &  &  &  &  &  &  & 454,720 & 454,720 & 456,578 & 456,942 & 456,942 & 458,335 & 462,642 \\ 
 &  &  &  &  &  &  &  & {\color[HTML]{0000FF}  452,884} & {\color[HTML]{0000FF}  466,792} & {\color[HTML]{0000FF}  475,395} & {\color[HTML]{0000FF}  485,127} & {\color[HTML]{0000FF}  494,178} & {\color[HTML]{0000FF}  500,544} & {\color[HTML]{0000FF}  511,826} \\ 
\multirow{-3}{*}{\textbf{7}} & \multirow{-3}{*}{14,738} & \multirow{-3}{*}{160,911} & \multirow{-3}{*}{314,757} & \multirow{-3}{*}{359,414} & \multirow{-3}{*}{443,251} & \multirow{-3}{*}{447,872} & \multirow{-3}{*}{454,720} & {\color[HTML]{FF0000} 450,548} & {\color[HTML]{FF0000} 462,084} & {\color[HTML]{FF0000} 475,002} & {\color[HTML]{FF0000} 484,279} & {\color[HTML]{FF0000} 490,159} & {\color[HTML]{FF0000} 494,052} & {\color[HTML]{FF0000} 506,155} \\ \hline
 &  &  &  &  &  &  & 336,823 & 336,970 & 338,189 & 338,189 & 338,189 & 338,189 & 360,450 & 361,875 \\ 
 &  &  &  &  &  &  & {\color[HTML]{0000FF}  339,792} & {\color[HTML]{0000FF}  354,857} & {\color[HTML]{0000FF}  363,281} & {\color[HTML]{0000FF}  369,985} & {\color[HTML]{0000FF}  376,656} & {\color[HTML]{0000FF}  384,445} & {\color[HTML]{0000FF}  388,954} & {\color[HTML]{0000FF}  397,548} \\ 
\multirow{-3}{*}{\textbf{8}} & \multirow{-3}{*}{21,450} & \multirow{-3}{*}{129,410} & \multirow{-3}{*}{266,031} & \multirow{-3}{*}{333,922} & \multirow{-3}{*}{334,447} & \multirow{-3}{*}{336,823} & {\color[HTML]{FF0000} 337,545} & {\color[HTML]{FF0000} 346,725} & {\color[HTML]{FF0000} 353,592} & {\color[HTML]{FF0000} 362,745} & {\color[HTML]{FF0000} 369,439} & {\color[HTML]{FF0000} 374,010} & {\color[HTML]{FF0000} 376,963} & {\color[HTML]{FF0000} 384,903} \\ \hline
 &  &  &  &  &  & 346,572 & 382,372 & 383,440 & 388,635 & 390,737 & 390,737 & 390,737 & 390,737 & 406,591 \\ 
 &  &  &  &  &  & {\color[HTML]{0000FF}  353,058} & {\color[HTML]{0000FF}  396,769} & {\color[HTML]{0000FF}  419,629} & {\color[HTML]{0000FF}  433,862} & {\color[HTML]{0000FF}  443,238} & {\color[HTML]{0000FF}  453,713} & {\color[HTML]{0000FF}  462,879} & {\color[HTML]{0000FF}  468,211} & {\color[HTML]{0000FF}  481,965} \\ 
\multirow{-3}{*}{\textbf{9}} & \multirow{-3}{*}{19,856} & \multirow{-3}{*}{214,340} & \multirow{-3}{*}{292,746} & \multirow{-3}{*}{323,878} & \multirow{-3}{*}{346,572} & {\color[HTML]{FF0000} 374,215} & {\color[HTML]{FF0000} 401,968} & {\color[HTML]{FF0000} 416,266} & {\color[HTML]{FF0000} 427,811} & {\color[HTML]{FF0000} 440,167} & {\color[HTML]{FF0000} 448,907} & {\color[HTML]{FF0000} 454,714} & {\color[HTML]{FF0000} 459,227} & {\color[HTML]{FF0000} 472,401} \\ \hline
 &  &  &  &  & 318,371 & 338,935 & 353,534 & 359,803 & 372,754 & 381,612 & 381,612 & 383,137 & 384,757 & 387,255 \\ 
 &  &  &  &  & {\color[HTML]{0000FF}  324,861} & {\color[HTML]{0000FF}  367,427} & {\color[HTML]{0000FF}  399,674} & {\color[HTML]{0000FF}  423,228} & {\color[HTML]{0000FF}  435,738} & {\color[HTML]{0000FF}  445,998} & {\color[HTML]{0000FF}  454,836} & {\color[HTML]{0000FF}  462,960} & {\color[HTML]{0000FF}  470,552} & {\color[HTML]{0000FF}  485,388} \\ 
\multirow{-3}{*}{\textbf{10}} & \multirow{-3}{*}{13,500} & \multirow{-3}{*}{141,237} & \multirow{-3}{*}{252,815} & \multirow{-3}{*}{318,371} & {\color[HTML]{FF0000} 323,016} & {\color[HTML]{FF0000} 372,680} & {\color[HTML]{FF0000} 400,357} & {\color[HTML]{FF0000} 414,838} & {\color[HTML]{FF0000} 425,943} & {\color[HTML]{FF0000} 438,909} & {\color[HTML]{FF0000} 447,482} & {\color[HTML]{FF0000} 453,126} & {\color[HTML]{FF0000} 458,089} & {\color[HTML]{FF0000} 471,391} \\ \hline
 &  &  &  & 268,993 & 340,294 & 355,398 & 370,982 & 375,358 & 376,628 & 378,202 & 379,554 & 379,554 & 379,554 & 421,869 \\ 
 &  &  &  & {\color[HTML]{0000FF}  273,024} & {\color[HTML]{0000FF}  312,406} & {\color[HTML]{0000FF}  329,884} & {\color[HTML]{0000FF}  350,764} & {\color[HTML]{0000FF}  367,205} & {\color[HTML]{0000FF}  377,425} & {\color[HTML]{0000FF}  383,668} & {\color[HTML]{0000FF}  389,927} & {\color[HTML]{0000FF}  398,312} & {\color[HTML]{0000FF}  403,495} & {\color[HTML]{0000FF}  413,163} \\ 
\multirow{-3}{*}{\textbf{11}} & \multirow{-3}{*}{23,920} & \multirow{-3}{*}{138,409} & \multirow{-3}{*}{268,993} & {\color[HTML]{FF0000} 267,216} & {\color[HTML]{FF0000} 298,681} & {\color[HTML]{FF0000} 318,792} & {\color[HTML]{FF0000} 337,697} & {\color[HTML]{FF0000} 347,236} & {\color[HTML]{FF0000} 354,739} & {\color[HTML]{FF0000} 363,956} & {\color[HTML]{FF0000} 370,192} & {\color[HTML]{FF0000} 374,876} & {\color[HTML]{FF0000} 377,668} & {\color[HTML]{FF0000} 386,216} \\ \hline
 &  &  & 82,789 & 227,599 & 279,909 & 289,413 & 290,300 & 290,300 & 292,976 & 292,976 & 292,976 & 292,976 & 292,976 & 295,580 \\ 
 &  &  & {\color[HTML]{0000FF}  73,368} & {\color[HTML]{0000FF}  106,206} & {\color[HTML]{0000FF}  116,191} & {\color[HTML]{0000FF}  123,287} & {\color[HTML]{0000FF}  129,959} & {\color[HTML]{0000FF}  135,669} & {\color[HTML]{0000FF}  138,861} & {\color[HTML]{0000FF}  140,972} & {\color[HTML]{0000FF}  143,818} & {\color[HTML]{0000FF}  146,301} & {\color[HTML]{0000FF}  147,963} & {\color[HTML]{0000FF}  150,926} \\ 
\multirow{-3}{*}{\textbf{12}} & \multirow{-3}{*}{8,104} & \multirow{-3}{*}{67,608} & {\color[HTML]{FF0000} 73,340} & {\color[HTML]{FF0000} 90,664} & {\color[HTML]{FF0000} 98,545} & {\color[HTML]{FF0000} 104,954} & {\color[HTML]{FF0000} 110,107} & {\color[HTML]{FF0000} 112,993} & {\color[HTML]{FF0000} 115,499} & {\color[HTML]{FF0000} 118,112} & {\color[HTML]{FF0000} 120,267} & {\color[HTML]{FF0000} 121,588} & {\color[HTML]{FF0000} 122,515} & {\color[HTML]{FF0000} 125,105} \\ \hline
 &  & 17,219 & 97,030 & 137,099 & 163,918 & 166,434 & 183,094 & 185,094 & 185,094 & 188,095 & 194,227 & 194,227 & 194,227 & 194,227 \\ 
 &  & {\color[HTML]{0000FF}  14,921} & {\color[HTML]{0000FF}  32,761} & {\color[HTML]{0000FF}  39,335} & {\color[HTML]{0000FF}  42,195} & {\color[HTML]{0000FF}  44,376} & {\color[HTML]{0000FF}  46,857} & {\color[HTML]{0000FF}  48,497} & {\color[HTML]{0000FF}  49,306} & {\color[HTML]{0000FF}  50,204} & {\color[HTML]{0000FF}  51,129} & {\color[HTML]{0000FF}  52,057} & {\color[HTML]{0000FF}  52,446} & {\color[HTML]{0000FF}  53,276} \\ 
\multirow{-3}{*}{\textbf{13}} & \multirow{-3}{*}{12,814} & {\color[HTML]{FF0000} 12,814} & {\color[HTML]{FF0000} 35,838} & {\color[HTML]{FF0000} 40,351} & {\color[HTML]{FF0000} 43,107} & {\color[HTML]{FF0000} 45,418} & {\color[HTML]{FF0000} 47,226} & {\color[HTML]{FF0000} 48,186} & {\color[HTML]{FF0000} 49,124} & {\color[HTML]{FF0000} 50,346} & {\color[HTML]{FF0000} 50,959} & {\color[HTML]{FF0000} 51,351} & {\color[HTML]{FF0000} 51,683} & {\color[HTML]{FF0000} 52,538} \\  \hline
\end{tabular}%
}
\caption{Monthly cumulative run-off triangle for only reported claims for September 2012. In black is the actual value, in blue is the estimation using the IPW estimator, and in red is the double Chain-Ladder estimation}
\label{trianglerbns}
\end{table}

The overall findings indicate that, at the cell level, the IPW estimator performs similarly to the Chain-Ladder method for this date. We do note that the IPW tends to better capture the changes in the reserve on the most recent dates as a result of accounting for the composition of the portfolio in the estimation. However, it is important to note that the IPW estimator does not consistently outperform the Chain-Ladder method in all cells of the triangle. We emphasize that, even though the IPW estimator can provide such estimations at the cell level, it may not possess the same precision as the estimation of the reserve as a whole. The more granular the desired estimation (i.e., the smaller the subpopulation of interest), the lower the level of accuracy.

Along those lines, instead of focusing on cell-level comparisons, our emphasis lies now on the aggregation of cells to determine the actual reserve value, which is the ultimate objective of estimation. It is noteworthy that the IPW estimator directly provides an estimation of the total reserves using Equations (\ref{IPW_Y}), (\ref{IPW_Y_RBNS}) and (\ref{IPW_Y_IBNR}), eliminating the need for constructing the run-off triangle in comparison to the Chain-Ladder method. Table \ref{tab_err_date} presents the aggregated reserve values obtained by summing the ultimate values for each accident date, along with the corresponding estimation errors. Our findings reveal that the IPW estimator yields reserve values that closely align with their true counterparts for all reserve types, exhibiting significantly lower estimation errors compared to the Chain-Ladder method. 

\begin{table}[h]

\centering
\small
\begin{tabular}{cccccc}
\hline
\hline
\textbf{Reserve Type} & \textbf{Method} & \textbf{Ultimate} & \textbf{Reserve} & \textbf{Error} & \textbf{\% Error} \\ \hline
\multirow{3}{*}{\textbf{   IBNS    }} & True value & 5,871,171& 1,643,872
& - & - \\
& IPW & 5,869,469& 1,642,169
& 1,702& 0.1\%
\\
 & CL & 5,138,422& 911,123
& 732,749& 44.6\%
\\ \hline
\multirow{3}{*}{\textbf{RBNS}} & True value & 5,116,312& 908,598
& & -\\
& IPW & 5,104,503& 896,789
& 11,810& 1.3\%
\\
 & CL & 4,971,626& 763,912
& 144,686& 15.9\%
\\ \hline
\multirow{3}{*}{\textbf{IBNR}} & True value & 754,859& 735,273
& & - \\
& IPW & 764,966& 745,381
& -10,107& -1.4\%
\\
 & CL & 166,796& 147,210
& 588,063& 80.0\%
\\
 \hline
 \hline
\end{tabular}
\caption{Error metrics for the aggregate reserves for September  2012.}
\label{tab_err_date}
\end{table}

\begin{figure}[H]
     \centering
     \begin{subfigure}[b]{0.31\textwidth}
         \centering
         \includegraphics[width=\textwidth]{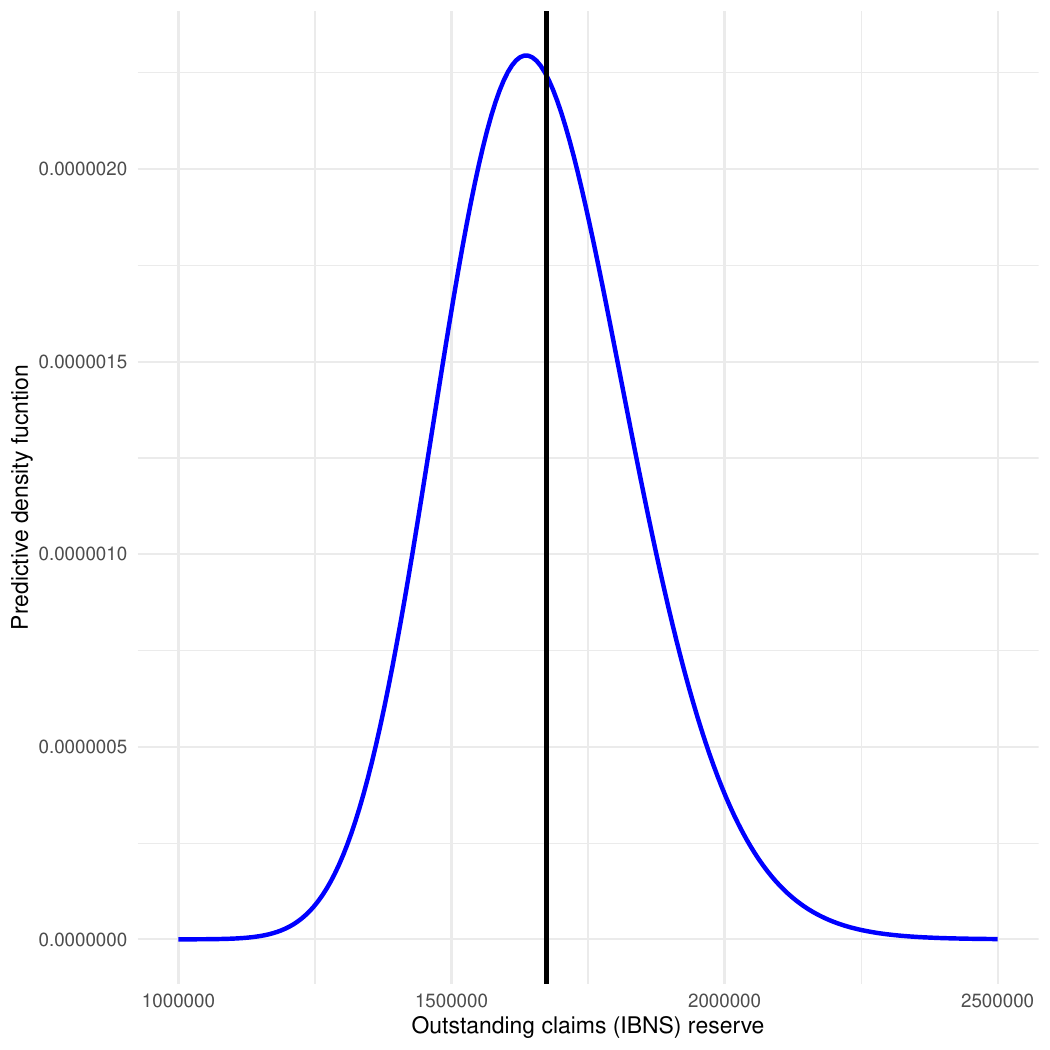}
        \end{subfigure}
     \hfill
     \begin{subfigure}[b]{0.31\textwidth}
         \centering
        \includegraphics[width=\textwidth]{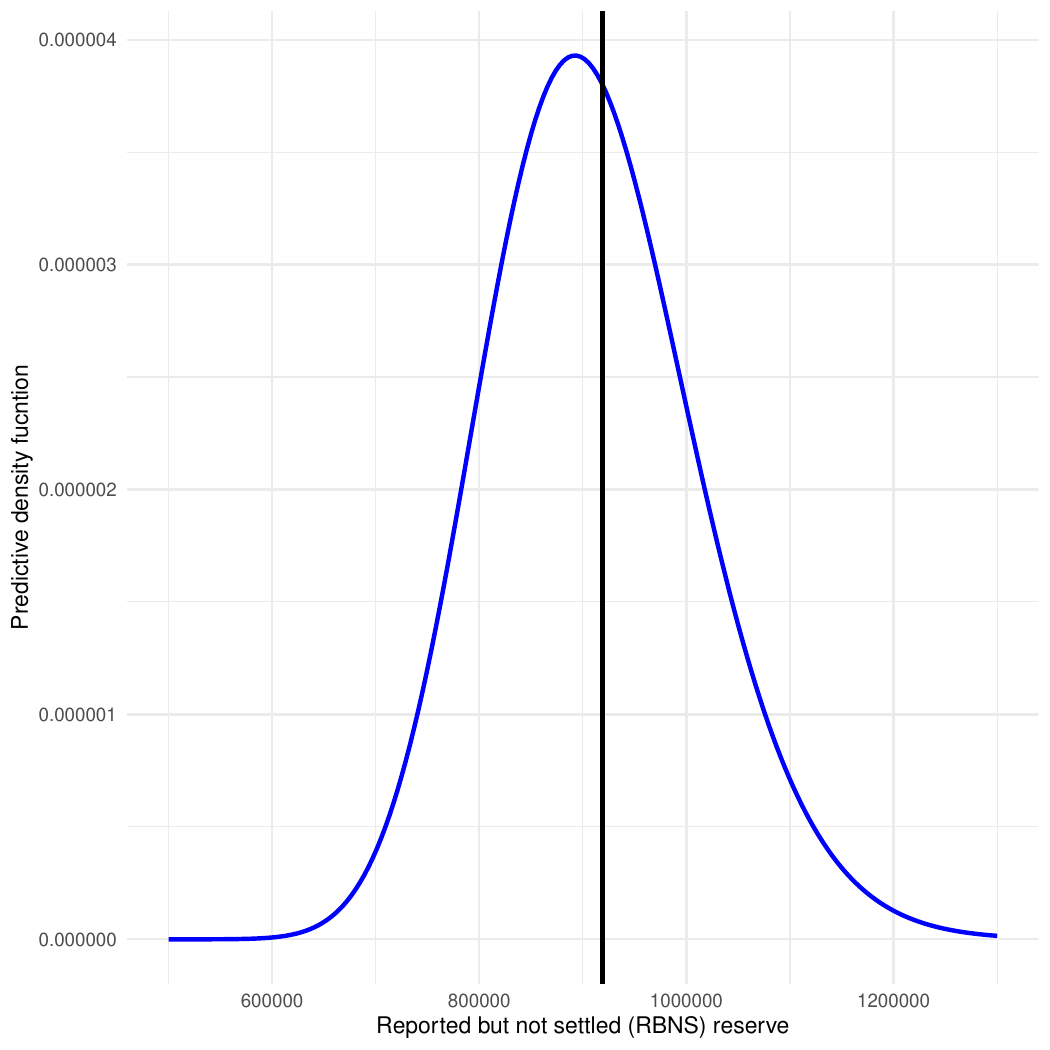}   
     \end{subfigure} 
     \hfill 
     \begin{subfigure}[b]{0.31\textwidth}
         \centering
        \includegraphics[width=\textwidth] {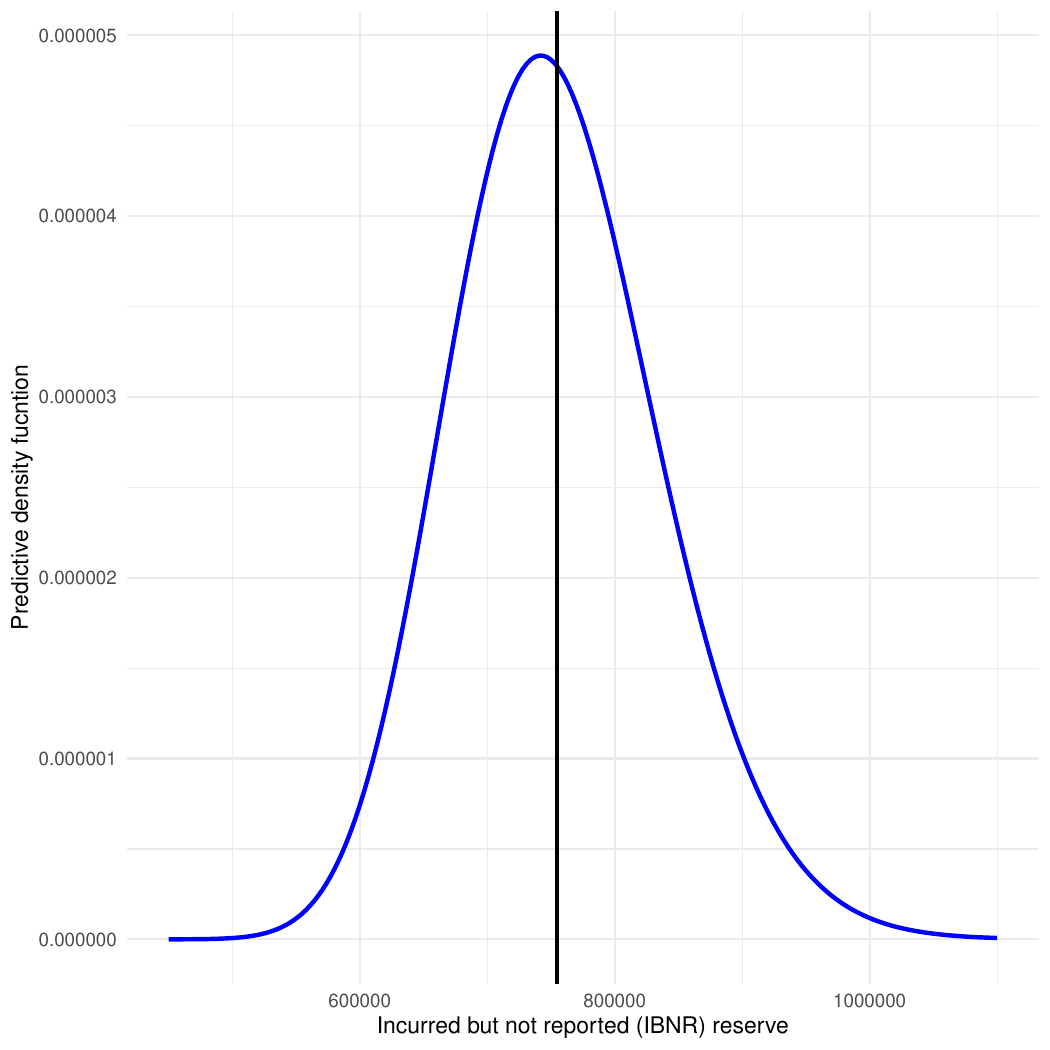}      
     \end{subfigure}
     \caption{Predictive distributions of the IPW method for the IBNS, RBNS and IBNR reserves for September 2012. The vertical line in black is the true observed value. }
     \label{predictive_dist}
\end{figure}

Furthermore, to evaluate the predictive quality of these estimates from a probabilistic standpoint, Figure \ref{predictive_dist} illustrates the predictive distributions of the reserves juxtaposed with the actual observed values.   These are constructed based on the log-transformation, and the approximate sampling distribution of the IPW estimators as discussed in Section \ref{CI_section}    . Notably, we observe that the true values consistently fall within the central region of the distribution, closely aligning with the corresponding modes, which represent the predicted reserve values. Consequently, the IPW-based predictions exhibit consistency with the observed reality.

\newpage

\subsection{Estimation of the reserve for several dates}

\begin{figure}[h]
        \centering
        \includegraphics[width=\textwidth]{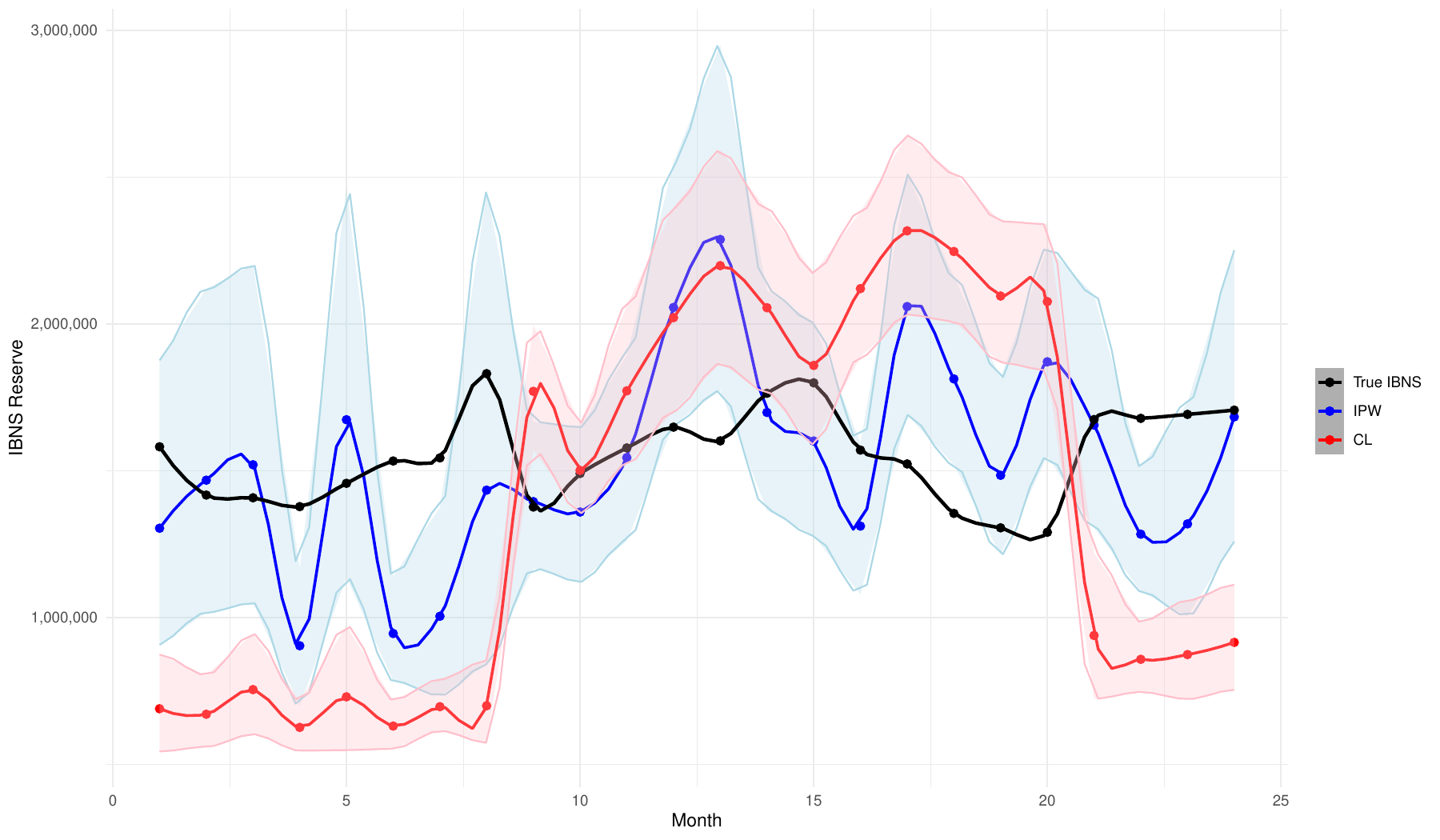}         
     \caption{      Estimation for the IBNS reserve per month }
     \label{plot_tot}
\end{figure}

\begin{figure}[!ht]
        \vspace{0.5cm}
        \centering
        \includegraphics[width=\textwidth]{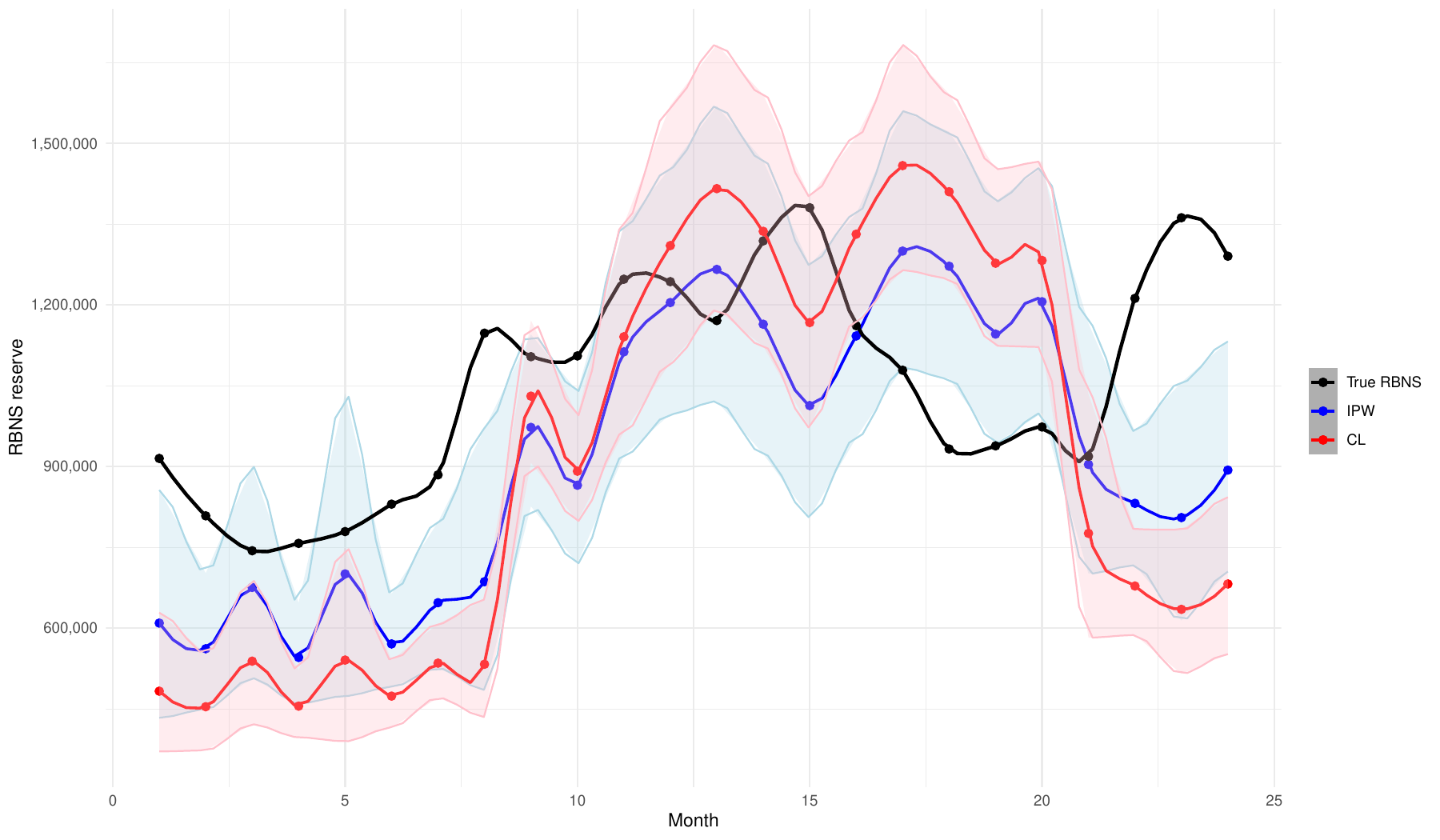}         
     \caption{      Estimation for the RBNS reserve per month }
     \label{plot_rbns}
\end{figure}

\begin{figure}[h]
        \centering
        \includegraphics[width=\textwidth]{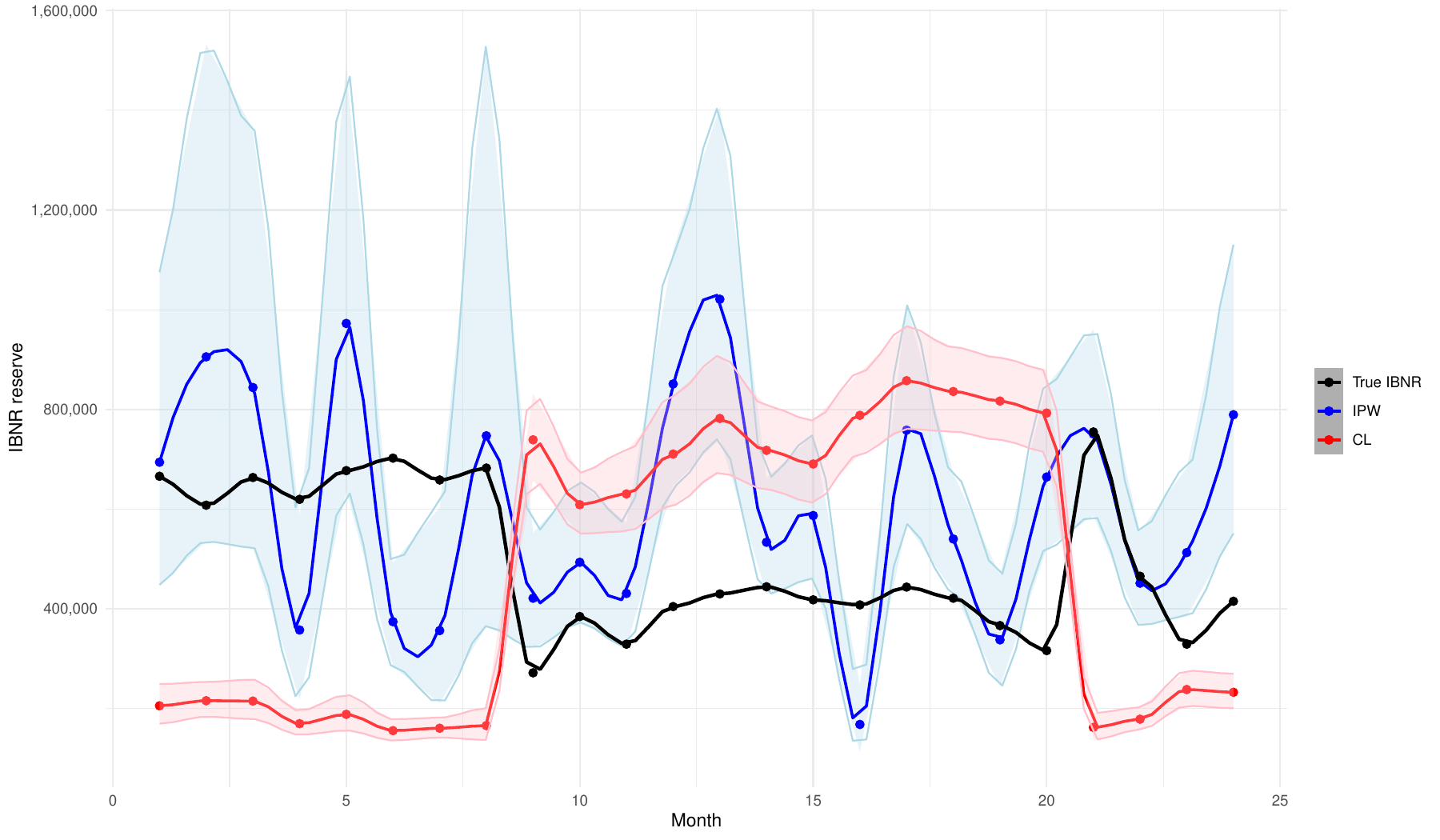}        
     \caption{      Estimation for the IBNR reserve per month }
     \label{plot_ibnr}
\end{figure}

Here we present the estimation of reserves for all 24 months in the testing period. Figure \ref{plot_tot} illustrates the estimations for the IBNS compared to the true value of the reserve at the corresponding month. Additionally, Figures \ref{plot_rbns} and \ref{plot_ibnr} depict the estimation for RBNS and IBNR, respectively. To provide a comprehensive analysis, we include 95\% confidence intervals for the estimations and include the estimates of the Chain-Ladder (CL) method or comparison.      We use the double Chain-Ladder approach to distinguish between IBNR and RBNS, and we use the approach of Section \ref{CI_section} to construct the confidence intervals.     Furthermore, Table \ref{tab_err} presents error metrics to assess the disparities between the estimations across all dates. We would like to note that this kind of temporal analysis is often overlooked in the claim-reserving literature due to its inherent challenges for having consistent estimations. Our aim is not to boast about the complexity of the analysis but rather to show the behavior of the IPW method in distinct valuation dates.

\begin{table}[h]

\centering
\small
\begin{tabular}{cccccc}
\hline
\hline
\textbf{Reserve Type} & \textbf{Method} & \textbf{ME} & \textbf{RMSE} & \textbf{MAE} & \textbf{MAPE} \\ \hline
\multirow{2}{*}{\textbf{   IBNS    }} & IPW & 21,908& 268,386& 292,419& 17\%
\\
 & CL & 170,363& 749,437& 647,976& 49\%
\\ \hline

\multirow{2}{*}{\textbf{RBNS}} & IPW & 133,738& 226,994& 225,052& 23\%
\\
 & CL & 144,503& 305,765& 311,635& 33\%
\\ \hline

\multirow{2}{*}{\textbf{IBNR}} & IPW & -111,830& 182,154& 210,047& 43\%
\\
 & CL & 25,860& 414,344& 386,959& 76\%
\\
 \hline
 \hline
\end{tabular}
\caption{Error metrics for the total of the reserves over the testing period. ME: Mean error, RMSE: Root mean square error, MAE: Mean absolute error, MAPE: Mean absolute percentage error}
\label{tab_err}
\end{table}

With respect to the IBNS i.e., the total reserve, Figure \ref{plot_tot} demonstrates that the IPW estimator produces predictions that closely align with the true value of the reserve for the majority of the observed periods, exhibiting no discernible pattern of under or overestimations. Additionally, the actual reserve value consistently falls within the associated confidence intervals, indicating a consistent fit with the predicted value. Notably, the IPW prediction proves to be more accurate than the traditional Chain-Ladder method on almost all dates of the considered period. This observation is further supported by the results in Table \ref{tab_err}, where the error metrics for the IPW over the 24-month period outperform those of the Chain-Ladder method. Therefore, the IPW along with the use of individual claims information has more predictive power than the macro reserving method.      It is worth noting that the confidence intervals for the IPW method typically tend to be wider than those for the Chain-Ladder method. Essentially, this arises from the bias-variance trade-off, as the IPW method factors in the heterogeneity of the data in its predictions, whereas the Chain-Ladder method does not.   

 With respect to the RBNS, Figure \ref{plot_rbns} reveals that the IPW estimator produces estimations exhibiting a similar pattern to those generated by the Chain-Ladder method. Both methods initially underestimate the reserve for the first year, followed by estimations that fluctuate around the true value of the reserve for the second year. While the estimated patterns align for these methods, the IPW estimator tends to be closer to the true value of the reserves in almost all cases, and when not, the estimation is comparable. Indeed, the IPW consistently outperforms the Chain-Ladder method on average throughout the entire period, as indicated by the lower error metrics in Table \ref{tab_err}. 
 
 From a global perspective, it is noteworthy that the estimations capture the general trend of the true RBNS reserve, demonstrating congruent patterns of fluctuations and variations in a comparable manner with the true value. This is evident, for instance, in the increasing trend during the first year of the reserve and its subsequent decrease in a portion of the second year.

Regarding the IBNR reserve, Figure \ref{plot_ibnr} illustrates that the IPW estimator provides a reasonable estimation throughout the considered period, fluctuating around the true reserve. It is noteworthy that the IPW estimator displays a more variable behavior compared to previous scenarios. This variability is expected due to the relatively low reporting delay time, resulting in the IBNR representing a smaller proportion of the subpopulation. Generally, as the size of the subpopulation decreases, the variance of the IPW increases.

Despite this variability, the IPW estimator consistently outperforms the traditional Chain-Ladder method over the entire 24-month period, as indicated in Table \ref{tab_err}. It is important to highlight that the Chain-Ladder estimations of the reserves follow a consistent trend across all reserves, i.e., underestimation at the beginning of the first year, overestimation at the end of the first year and the beginning of the second, and then underestimation again at the end. In contrast, the IPW estimation does not exhibit the same pattern. This discrepancy is attributed to the fact that the Chain-Ladder assumes a homogeneous portfolio, while the IPW does not.

\begin{observation}
We encountered instability in the behavior of the IPW estimator (i.e., absurdly abnormal large reserves) when performing the estimation on some dates, along the same lines as the behavior described in Section \ref{adjust}. To address this issue, we implemented the adjusted version of the IPW estimator, as described in Algorithm \ref{IHTest}, and compared it to the raw IPW estimator. If the percentual difference between the two estimations exceeded a certain threshold (e.g., more than 3\%), we retained the adjusted estimation. For cases where the difference was not significant, we kept the original estimation. As a result, the estimations become more stable across dates.

\end{observation}

\section{Conclusions}
\label{conclusions}
Macro-level reserving models, particularly the Chain-Ladder method, overlook the underlying heterogeneity within the portfolio of policyholders, treating all claims equally, and providing most of the time modest estimations. Therefore, the estimation of the reserve does not benefit from the use of the individual attributes of the policyholders, which have been shown to provide a significant improvement in the accuracy of these methods in the literature of micro-level reserving. 

In this paper, we address the limitation of macro-level reserving models by proposing a statistically justified macro-level reserve estimator based on Inverse Probability Weighting (IPW). Unlike traditional macro-level models, our method incorporates individual-level information in the weights to improve the accuracy of reserve estimation. Moreover, such incorporation is achieved within a less complex framework compared to micro-level models, in the sense that no explicit assumptions on claim frequency or severity are made.

The IPW estimator serves as a hybrid approach that bridges the gap between macro and micro-level methods. It assigns attribute-driven weights to each claim, allowing for a development factor specific to each claim's settlement, similar to the familiar principles of the Chain-Ladder method when applied at the granular level. This method represents an initial step towards obtaining more precise reserves from macro-level models and serves as an intermediate stage in the development of a customized micro-level reserving model. We hope practitioners find this method appealing as it is a natural extension of the traditional Chain-Ladder method, accounting for portfolio heterogeneity in a statistically justified fashion.

The IPW and population sampling framework offers a new theoretical insight into the Chain-Ladder method, revealing it as a specific case. Several properties of the IPW estimator directly apply to the Chain-Ladder, allowing the derivation of both known and potentially unknown characteristics. For instance, the well-established unbiasedness of the Chain-Ladder stems from the unbiasedness of the IPW estimator and Assumption \ref{CLassumtion}. A similar result applies to the distributional properties of the estimator.  It is worth noting that specific assumptions might differ from those in other frameworks like \cite{mack1999standard}. Although we don't explicitly outline all connections here, recognizing their potential is crucial.

Future research exploring the interplay between claim reserving and population sampling techniques holds promise for advancing reserve estimation. This involves investigating inherited properties of the Chain-Ladder as an IPW estimator and exploring implications. Additionally, other population sampling-motivated methods, like double robust estimators, offer new avenues for addressing the reserving problem.   Indeed, our ongoing work has revealed that population sampling has the potential to offer a broader framework for reserving, encompassing both general macro-level and micro-level models. Additionally, adopting a sampling perspective may yield insights into improving model estimation techniques, thereby striving for enhanced outcomes.

Other research directions should explore alternative approaches for estimation, potentially through tailored models for the development of claims that are specifically designed for the estimation of the inclusion probabilities in Equation (\ref{probidev}). In fact, one may consider alternative inclusion probabilities to use as weights,  aiming for better reserve estimations.        Moreover, the IPW framework and the change of population principle could be applied beyond the discussed reserves. For instance, the incurred but not paid (IBNP) reserve, and the unearned premium reserve (UPR), among others.


\section*{Acknowledgments}
This work was partly supported by Natural Sciences and Engineering Research Council of Canada [RGPIN 284246, RGPIN-2017-06684]. Sebastián Calcetero Vanegas acknowledges the Mountain Pygmy Possum, an endangered species in Australia, for inspiring research on population sampling and also this project. 

\section*{Conflicts of interest or Competing interests }
The authors declare no conflicts of interest or competing interests in this paper, with no financial or personal affiliations that could compromise the objectivity or integrity of the presented work.

\bibliographystyle{apalike}
\bibliography{references}

\end{document}